\documentclass[fleqn,usenatbib]{mnras}

\usepackage{newtxtext,newtxmath}

\usepackage[T1]{fontenc}
\usepackage{ae,aecompl}

\usepackage{graphicx}	
\usepackage{amsmath}	
\usepackage{amssymb}	
\usepackage[flushleft]{threeparttable}

\title[Pattern speeds of MaNGA barred galaxies]{SDSS-IV MaNGA: pattern speeds of barred galaxies}

\author[Guo et al.]{
Rui Guo$^{1,2}$\thanks{E-mail: guorui13@bao.ac.cn},
Shude Mao$^{3,1,4}$,
E.~Athanassoula$^{5}$,
Hongyu Li$^{1}$,
Junqiang Ge$^{1}$,
R.J. Long$^{1,4}$, 
\and
\ Michael Merrifield$^{6}$,
Karen Masters$^{7,8}$
\\
$^{1}$National Astronomical Observatories, Chinese Academy of Sciences, 20A Datun Road, Chaoyang District, Beijing 100012, China\\
$^{2}$University of Chinese Academy of Sciences, Beijing 100049, China\\
$^{3}$Physics Department and Tsinghua Centre for Astrophysics, Tsinghua University, Beijing 100084, China\\
$^{4}$Jodrell Bank Centre for Astrophysics, School of Physics and Astronomy, The University of Manchester, Oxford Road, Manchester M13 9PL, UK\\
$^{5}$Aix Marseille Univ, CNRS, CNES, LAM, Marseille, France\\
$^{6}$School of Physics and Astronomy, University Nottingham, Nottingham NG7 2RD, UK \\
$^{7}$Haverford College, Department of Physics and Astronomy, 370 Lancaster Avenue, Haverford, Pennsylvania 19041, USA\\
$^{8}$Institute of Cosmology and Gravitation, University of Portsmouth, Dennis Sciama Building, Portsmouth, PO1 3FX, UK\\
}

\date{Accepted XXX. Received YYY; in original form ZZZ}

\pubyear{20XX}

\begin{document}
\label{firstpage}
\pagerange{\pageref{firstpage}--\pageref{lastpage}}
\maketitle

\begin{abstract}
The MaNGA project has obtained IFU data for several thousand nearby galaxies, including barred galaxies. With the two dimensional spectral and kinematic information provided by IFUs, we can measure the pattern speed of a barred galaxy, which determines the bar dynamics. We apply the non-parametric method proposed by Tremaine \& Weinberg to estimate the bar pattern speed for 53 barred galaxies, making this the largest sample studied so far in this way. Our sample is selected from the MaNGA first public data release as part of SDSS Data Release 13 according mainly to the axis ratio and position angle difference between the bar and disc, while kinematic data is from the later SDSS Data Release 14. We have used both the photometric position angle from the photometric image and the kinematic position angle from the stellar velocity map to derive the pattern speed. Combining three independent bar length measurements and the circular velocity from Jeans Anisotropic modelling (JAM), we also determine the dimensionless ratio $\cal{R}$ of the corotation radius to the bar length. We find that the galaxy's position angle is the main uncertainty in determining the bar pattern speed. The kinematic position angle leads to fewer ultrafast bars than the photometric position angle, and this could be due to the method of measuring the kinematic position angle. We study the dependence of $\cal{R}$ values on galaxy properties such as the dark matter fraction from JAM modelling and the stellar age and metallicity from stellar population synthesis (SPS). A positive correlation between the bar length and bar strength is found: the longer the bar, the stronger the bar. However, no other significant correlations are found. This may result from errors in deriving the $\cal{R}$ values or from the complex formation and slowdown processes of galactic bars.
\end{abstract}




\section{Introduction}

Barred galaxies are one branch of the Hubble morphological classification. About 25-50\% of nearby disk galaxies observed in optical wavelengths host a bar \citep[e.g.][]{marinova2007, barazza2008, aguerri2009, nair2010, masters2011}. This fraction is even higher when galaxies are observed in near-infrared wavebands \citep[e.g.][]{knapen2000, eskridge2000, menendezdelmestre2007, buta2015}. The bar fraction depends on many galaxy properties, such as Hubble type, stellar mass, galaxy colour and bulge prominence \citep[e.g.][]{aguerri2009, nair2010, masters2011}. Our Milky Way is also a barred galaxy \citep[e.g.][]{devaucouleurs1964, blitz1991}, and several works have tried to estimate its pattern speed \citep[e.g.][]{long2013, antoja2014, portail2015}.

Though bars have a relatively small fraction of the visible mass, they play an important role in disc galaxy evolution and bulge formation. Their strong quadrupole moment allows them to redistribute angular momentum, energy and mass between the galactic bulge, stellar and gaseous discs and dark matter halo \citep[e.g.][]{weinberg1985, debattista1998, debattista2000, athanassoula2003, martinezvalpuesta2006, sellwood2006a, sellwood2006b, villavargas2009, athanassoula2013} . In particular, the amount of angular momentum exchanged is related to galaxy properties, such as the bar mass, halo density, halo velocity dispersion \citep[e.g.][]{athanassoula2003, sellwood2006a, sellwood2006b}, and the central dark matter fraction.

Bars can be described by three important parameters: length, strength and pattern speed. Their evolution depends on the redistribution of angular momentum within the galaxy. Different methods have been proposed to measure these bar parameters.

Bar length can be determined by visual inspection of galaxy images \citep{kormendy1979, martin1995, hoyle2011}, by locating the maximum ellipticity of the galaxy isophotes \citep{wozniak1995, athanassoula2002, laine2002, marinova2007, aguerri2009}, by looking for variations of the isophotal position angle \citep{athanassoula2002, sheth2003, erwin2005}, or by structural decompositions of the galaxy surface brightness distribution \citep{prieto1997, prieto2001, aguerri2003, aguerri2005, laurikainen2005, laurikainen2009, gadotti2008, gadotti2011, weinzirl2009, kruk2018}. A typical bar radius is a few kpc \citep{marinova2007, aguerri2009}  and is correlated with other galaxy parameters, such as disc scale length, galaxy size, galaxy colour, and prominence of the bulge \citep[e.g.][]{aguerri2005, marinova2007, gadotti2011, hoyle2011}.

Bar strength is a parameter that measures the non-axisymmetric forces produced by the bar potential in the discs of galaxies \citep[][]{laurikainen2002}. It can be derived by measuring the bar torques \citep{combes1981, quillen1994, buta2001, laurikainen2007, salo2010}, bar ellipticity \citep{martinet1997, aguerri1999, whyte2002, marinova2007, aguerri2009} or Fourier decomposition of galaxy light \citep{ohta1990, marquez1996, aguerri2000, athanassoula2002, laurikainen2005}. Recently, \cite{kim2016} defined bar strength from the light deficit between the surface brightness profiles along the bar's major and minor axes.

The bar pattern speed $\Omega_{\rm p}$, defined as the rotational frequency of the bar, is an important dynamical parameter and its determination requires galaxy kinematics. Many indirect methods have been used to measure this parameter. Matching the modelled and observed surface gas distributions and/or gas velocity fields \citep[e.g.][]{sanders1980, hunter1988, england1990, garciaburillo1993, sempere1995a, lindblad1996a, lindblad1996b, laine1999, weiner2001, aguerri2001, perez2004, rautiainen2008, treuthardt2008}, with $\Omega_{\rm p}$ as one of the free parameters of the models, has been applied to 38 barred galaxies by \cite{rautiainen2008} to determine their pattern speeds. Some galaxy morphology features correlated with Lindblad resonances have also been extensively used to derive bar pattern speed, for example, position of galaxy rings \citep[e.g.][] {buta1986, buta1995, vegabeltran1997, munoztunon2004, perez2012}, changes in the morphology or phase of spiral arms near the corotation radius \citep[e.g.][]{canzian1993, canzian1997, puerari1997, aguerri1998, buta2009}, the offset and shape of dust lanes \citep{vanalbada1982, athanassoula1992}, or the morphology of the residual gas velocity field after rotation velocity subtraction \citep{sempere1995b, font2011, font2014}. These methods are based on the description of morphological features and are model-dependent. The most accurate method for measuring the bar pattern speed is the model independent method proposed by \citet[hereafter TW]{tremaine1984}, and we will employ their method in this work.

Usually, a bar is also parametrized by a distance--independent ratio ${\cal R}=R_{\rm CR}/a_{\rm b}$, where $R_{\rm CR}$ and $a_{\rm b}$ are the corotation radius and bar length. The corotation radius is the region of the galaxy where the gravitational and centrifugal forces cancel out in the rest frame of the bar. Thus the corotation radius can be derived from the bar pattern speed as ${\rm R_{CR}=V_{c}/\Omega_{p} }$, where $V_{\rm c}$ is the disc circular velocity. A self--consistent weak bar requires ${\cal R} > 1.0$, meaning that the bar cannot extend beyond the corotation radius \citep{contopoulos1980, athanassoula1980}. Studies of the dust lane shapes, using gas flow calculations in barred galaxy potentials predict ${\cal R} = 1.2 \pm 0.2$ \citep[see][]{athanassoula1992}. Bars are often classified into fast ($1.0 < {\cal R} < 1.4$) and slow (${\cal R} > 1.4$) bars. Most observed bars have $\cal{R}$ ratios smaller than 1.4 \citep[e.g.][]{elmegreen1996, rautiainen2008} and they have been interpreted as evidence for maximum discs (\citealt{debattista1998}, but see \citealt{athanassoula2014}). However there are also some bars compatible with being slow bars \citep{bureau1999, rautiainen2008, chemin2009}. Thus having an accurate measurement of the $\cal{R}$ values for barred galaxies is of great interest.

The TW method was applied to long--slit spectroscopy initially \citep[e.g.][]{debattista2002, corsini2003}. Nevertheless, difficulties in tracing different stellar populations between photometric and spectroscopic data (caused by different wavelength coverage and problems with the positioning of the pseudo slits in the photometric images) can affect the determinations. These problems can be solved by using integral field unit (IFU) spectroscopic data. Many galaxies now have IFU data obtained from different surveys, such as CALIFA \citep{sanchez2012}, SAMI \citep{bryant2015} and MaNGA \citep{bundy2015}. Here we will apply the TW method to a sample of MaNGA barred galaxies to derive their bar pattern speeds and study the dependence of $\Omega_{\rm p}$ on galaxy properties such as the central dark matter fraction.

The structure of this paper is as follows. Section \ref{sec_methods} briefly introduces the methods used in this work, including the TW method used to measure the bar pattern speeds, the stellar population synthesis (SPS) used to derive galaxy stellar ages and metallicities, and the JAM used to estimate the dark matter fractions. Section \ref{sec_data} describes the sample selection and stellar velocity maps used in this work. Section \ref{sec_gparas} presents the measurements of geometric parameters and bar strength of our sample galaxies. Bar pattern speeds and the dependences of the $\cal{R}$ parameter on galaxy properties such as the dark matter fraction are shown in Section \ref{sec_results}. Discussions and conclusions are shown in Sections \ref{sec_discussions} and \ref{sec_conclusions}, respectively. The appendices show several related tests we have performed using a simulated barred galaxy. The WMAP9 cosmological parameters ($\Omega_{m}= 0.286,\ \Omega_{L}= 0.714$, and $h= 0.693$) are used in this paper \citep{hinshaw2013}.

\section{Methods}
In this section, we introduce the methods used in this work. Section \ref{ssec_tw} introduces the TW method used to measure the bar pattern speeds. Section \ref{ssec_sps} briefly introduces the stellar population synthesis (SPS), from which we can obtain the mass weights used in the TW method and the stellar age and metallicity used in Section \ref{ssec_R}. Section \ref{ssec_jam} introduces the mass models we used in the Jeans Anisotropic modelling (JAM), which can give us estimations of dark matter fraction and circular velocity of our barred galaxies.

\label{sec_methods}
\subsection{The Tremaine \& Weinberg method}
\label{ssec_tw}
The TW method is a model--independent method for measuring the bar pattern speed based on the main assumptions that the galaxy has a single, well--defined pattern speed $\Omega_{\rm p}$ and the surface brightness of a tracer population satisfies the continuity equation. It can be expressed as a weighted mean velocity over a similarly weighted mean position:
\begin{equation}
\Omega_{\rm p} \sin i= \frac{ \int_{-\infty}^{+\infty}h(Y) \int_{-\infty}^{+\infty} \Sigma(X,Y)V_{\rm LOS}(X,Y) dXdY }{\int_{-\infty}^{+\infty}h(Y) \int_{-\infty}^{+\infty} \Sigma(X,Y) X dXdY}
\equiv \frac{ \langle V \rangle }{ \langle X \rangle } ,
\label{eq_tw}
\end{equation}
where (X,Y) are the Cartesian coordinates in the sky plane, with the origin at the center of the galaxy and the X-axis aligned with the line of nodes, i.e. the intersection of the sky plane and the disc plane. The disc inclination \emph{i} can be obtained from the ellipticity of the galaxy's outermost isophotes. $V_{\rm LOS}(X,Y)$ is the line of sight velocity measured from either long-slit or integral-field spectroscopy, while $\Sigma(X,Y)$ is usually the galaxy's surface brightness, and it can also represent the surface mass density obtained by SPS methods. The integrations of X and Y are formally over $-\infty < X,Y < +\infty$. Nevertheless, the X integration can be limited to $-X_{0} < X < X_{0}$ if the disc is axisymmetric at larger X, and the Y axis integration can be changed to an arbitrary range by the performance of weight function $h(Y)$. A weight function of $h(Y)=\delta(Y-Y_{0})$ corresponds to a slit or a pseudo slit parallel to the line of nodes with an offset by a distance $Y_{0}$, for the case in the long-slit spectroscopy and in IFU observations, respectively.

In practice, several slits or pseudo--slits parallel to the line of nodes are chosen for calculating the weighted average velocity and position. In each slit, the weighted average velocity and position for the axisymmetric disc is zero, so the non-zero integrations are the contribution of the bar that is not aligned to or perpendicular to the line of nodes. The centering errors in identifying the galaxy position centre $(X_{\rm c} , Y_{\rm c})$ and in measuring the systematic velocity $V_{\rm sys}$ can affect the measurement of $\Omega_{\rm p}$ significantly in long-slit spectroscopy. \cite{merrifield1995} refined the TW method and rewrote Eq. \ref{eq_tw} as:
\begin{equation}
    \Omega_{\rm p} \sin i= \frac{ \langle V \rangle - V_{\rm sys}}{ \langle X \rangle - X_{\rm c}} .
    \label{eq_mk}
\end{equation}
Thus, plotting $\langle V \rangle$ versus $\langle X \rangle$ for all slits produces a straight line with the slope representing $\Omega_{\rm p} \sin i$. Though in integral-field spectroscopy the centering errors are minimized by using a common reference frame, we still prefer to use the slopes from linearly fitting of $\langle V \rangle$ vs. $\langle X \rangle$.

There are two kinds of tracers typically used when applying the TW method: gas and stars. Although usually gas does not follow the continuity equation due to the presence of shocks, conversion between different gas phases, and star formation, the TW method has been successfully applied to gas \citep{zimmer2004, rand2004, hernandez2005, emsellem2006, fathi2007, chemin2009, gabbasov2009, fathi2009}. Some bar pattern speeds derived from the TW method using gas are consistent with values obtained from indirect methods and numerical simulations.

The stellar--based TW method has been applied largely to early--type barred galaxies \citep[e.g.][]{kent1987, merrifield1995, gerssen1999, debattista2002, aguerri2003, corsini2003, debattista2004, corsini2007}. In contrast, due to star formation and dust obscuration, the observed surface brightness in late--type galaxies does not always trace the mass distribution, and few pattern speeds have been obtained using this method for these galaxies \citep[e.g.][]{gerssen2003, treuthardt2007}. Nevertheless, experiments for investigating the effects of dust obscuration and star formation on the TW method using numerical simulations \citep{gerssen2007} suggest that it is possible to extend the application of the TW method to late-type barred galaxies.

The main sources of uncertainty in the TW method application are centering errors, low signal--to--noise ratio of the spectral data, uncertainties on the disc position angle (hereafter PA) and the inclination angle, dust obscuration and star formation and the number of slits (see \cite{corsini2011} for a detailed review). In integral--field spectroscopy, the centering error can be minimized by an accurate sample reference frame, and the signal--to--noise ratio can be increased by re--binning the pixels. For the PA error, \cite{debattista2003} demonstrates that an error of a few degrees in the disc PA can result in a large error in the estimation of $\Omega_{\rm p}$, because the misalignment between the PA of slits and the PA of the disc changes both the velocity and position integration. The maximum misalignment enabling reliable pattern speed measurements depends on the disc inclination and the bar orientation with respect to the line of nodes. For this reason, it is crucial to select samples with small PA and/or inclination errors when applying the TW method. We carefully select such a sample in Section \ref{ssec_sample}.

\subsection{Stellar population synthesis}
\label{ssec_sps}
The spectral energy distributions of galaxies encode many fundamental properties of unresolved stellar populations. These properties include star formation history, stellar metallicity and abundance patterns, stellar initial mass function, total mass in stars, and the physical state and quantity of dust and gas. Much effort has been devoted by the astronomical community in extracting such information from the spectral energy distributions of galaxies to study galaxy formation and evolution. The SPS method has been developed relying on stellar evolution theory, spectral library and initial mass function etc. to constrain the stellar age and metallicity distribution of a galaxy. For reviews of SPS see, e.g., \cite{walcher2011} and \cite{conroy2013}.

From the SPS of MaNGA IFU spectra, we obtain the stellar ages, metallicities and stellar mass-to-light ratios, i.e. stellar mass, of the galaxies in our sample. The stellar ages and metallicities are used to study correlations with the bar pattern speeds (see Section \ref{sssec_Rage}), and the stellar masses are used as the mass weights in calculating the integrals of the TW method (see Section \ref{ssec_op}) and in estimating the dark matter fractions. The spectra are Voronoi binned \citep{cappellari2003} to S/N=30 before fitting. We use the pPXF software \citep{cappellari2004, cappellari2017} and the MILES-based \citep{sanchez2006} SPS models of \cite{vazdekis2010}. A \cite{calzetti2000} reddening curve and a \cite{salpeter1955} initial mass function (IMF) are assumed in the modelling. In the model, we consider the IMF variation by correcting the M*/L values according to the Table 1 of \cite{Li2017}. In \cite{Li2017}, they compared the M*/Ls from different software packages and templates, and found that the uncertainties in M*/L are $\sim 0.1$ dex for young galaxies, and smaller for old galaxies.

\subsection{Mass models of barred galaxies}
\label{ssec_jam}
One main purpose of this paper is to study the dependence of the dimensionless parameter ${\cal R}$ on the dark matter fraction. We use the Jeans Anisotropic modelling (JAM) \citep[][]{cappellari2008} to estimate the circular velocities and the dark matter fractions. The mass model used in JAM has two components, stellar mass and dark halo. For the stellar mass distribution, we first calculate the deprojected SDSS r-band luminosity density using the Multi-Gaussian Expansion method \citep[][]{emsellem1994}. We then assume a constant stellar mass-to-light ratio to convert the luminosity density to a stellar mass distribution. In the deprojection, we use the inclination estimated from the apparent axis ratio (See Section \ref{ssec_ipa}). The constant stellar mass-to-light ratio is taken from the SPS described in Section \ref{ssec_sps}. We first calculate the averaging value within the effective radius, and then correct the Salpeter IMF based stellar mass-to-light ratio to a ratio based on a variable IMF, according to the relation in Table 1 of \cite{Li2017}. For the dark matter halo, we use a generalized NFW model \citep[see equation 2 of][]{cappellari2013}. The other details of the modelling process can be found in \cite{Li2016, Li2017}.

In \cite{Li2016}, the JAM method has been tested using cosmologically simulated galaxies. They found that the total mass of a galaxy is well constrained (1$\sigma$ error $\sim$ 10-18\%). This total mass can thus be used to derive the circular velocities, which are used to measure the dimensionless ratio ${\cal R}$. Due to the 0.1 dex uncertainty of M*/L for young galaxies, we finally take a systematic 12\% error for circular velocities of our sample galaxies. We have also checked the effectiveness of JAM in modelling a strongly barred simulation galaxy in Appendix \ref{ap: vc}. In this test, JAM recovers the circular velocity to about a 10\% error for different galaxy inclinations and bar orientations. We discuss this later in Section \ref{sssec_vcerr}.

\section{MaNGA Data on barred galaxies}
\label{sec_data}
\subsection{Sample selection}
\label{ssec_sample}
Mapping Nearby Galaxies at Apache Point Observatory (MaNGA) is aimed to investigate the internal kinematic structure and composition of gas and stars in an unprecedented sample of 10,000 nearby galaxies over the six--year lifetime of the survey (2014-2020). An overview of the project is presented in \cite{bundy2015}. Galaxies are selected from the NASA Sloan Atlas catalog of the SDSS Main Galaxy Legacy Area, with selection cuts applied to only redshift ($z \sim 0.02- 0.1$) and a color-based stellar mass estimate ($M_{*}> 10^{9}M_{\odot}$). The MaNGA sample is roughly separated into the Primary sample (60\%) with a spatial coverage to $1.5 R_{\rm e}$ and an average redshift $<z> =0.03$, and the Secondary sample (30\%) with larger spatial coverage ($2.5 R_{\rm e}$) and higher redshift ($<z> =0.045$). More details about the survey design, sample selection and optimization can be seen in \cite{wake2017} and \cite{yan2016}.

The MaNGA IFUs are taken by the IFU system mounted on the 2.5 m Sloan Telescope \citep{gunn2006}, which has 1423 fibres with $2''$ core diameters over a 3$^{\circ}$ diameter field of view. 17 IFUs are obtained simultaneously using 19-127 tightly-packed arrays of optical fibres, varying in size from 12.5$''$ to 32.5$''$ in diameter, with a distribution that is matched to the apparent size of galaxy targets on the sky. See \cite{drory2015} for more instrumental information. A defined three-point dither pattern \citep[see][]{law2015} is adopted to achieve uniform spatial sampling for all targets, for the regular hexagonal packing of MaNGA IFUs. The reconstructed PSF in combined datacubes after dithering and fibre sampling is 2.5$''$ (FWHM). The final pixel size for maps is 0.5$''$. Both the spatial size and resolution of galaxy targets are sufficient for studying the barred galaxy properties.

The MaNGA fibres feed light into two dual-channel BOSS spectrographs \citep{smee2013}, each with a red and blue channel that provide simultaneous wavelength coverage from 3600 to 10,300 \AA\ with a mid-range resolution of $R \sim 2000$. After roughly 3--hour dithered exposures, the S/N per fibre per angstrom at the outskirts of targets is between 4 to 8. For more details about observation strategy and the data reduction process, see \cite{law2015} and \cite{law2016}.

Our sample is from the first MaNGA public data release as part of SDSS Data Release 13 \citep[DR13;][]{albareti2017}, which contains 1390 IFU galaxies. Besides the bar vote fraction from Galaxy Zoo2 \citep{willett2013}, three of us selected the candidate barred sample separately, which has 234 galaxies in total. A more detailed examination identified a bar sample containing 168 galaxies, in which galaxies have at least two identifications of strong bars in the SDSS g-r-i three bands combined images. The Hubble types are also given for these galaxies through visual inspection. Of these 168 barred galaxies, the ELLIPSE fitting routine \citep{jedrzejewski1987} can only be applied to 137 galaxies to determine their geometrical parameters, such as the PAs of the bar and the disc, disc inclination, and bar length. The TW method can not be applied to galaxies with bars too parallel or too perpendicular to the disc major axis, because they will have nearly zero weighted mean velocities and positions in pseudo slits. Also the TW method is difficult to apply to lowly or highly inclined galaxies, because the former have small line-of-sight velocities, large velocity errors and bar PA errors, while for the latter bars are hard to identify and to choose pseudo slits. Therefore, we apply constraints on the disc axis ratio and the PA difference between the disc ($PA_{\rm d}$) and the bar ($PA_{\rm b}$). These are $0.3 < b/a <0.8$ and $10^{\circ}< |PA_{\rm d} - PA_{\rm b}| <80^{\circ}$ (in this criterion, we use the disc PA derived from the galaxy image). These two constraints reduce the sample to 74 galaxies. By excluding galaxies with bad velocity dispersion maps or low quality velocity maps (typically from galaxies with low IFU coverage and with disturbed structures), our final bar sample contains 53 galaxies. While 53 galaxies is a modest number, this is the largest sample so far used to study bar pattern speeds and their dependence on galaxy properties.

\begin{figure}
  \centering
  \includegraphics[width=\columnwidth]{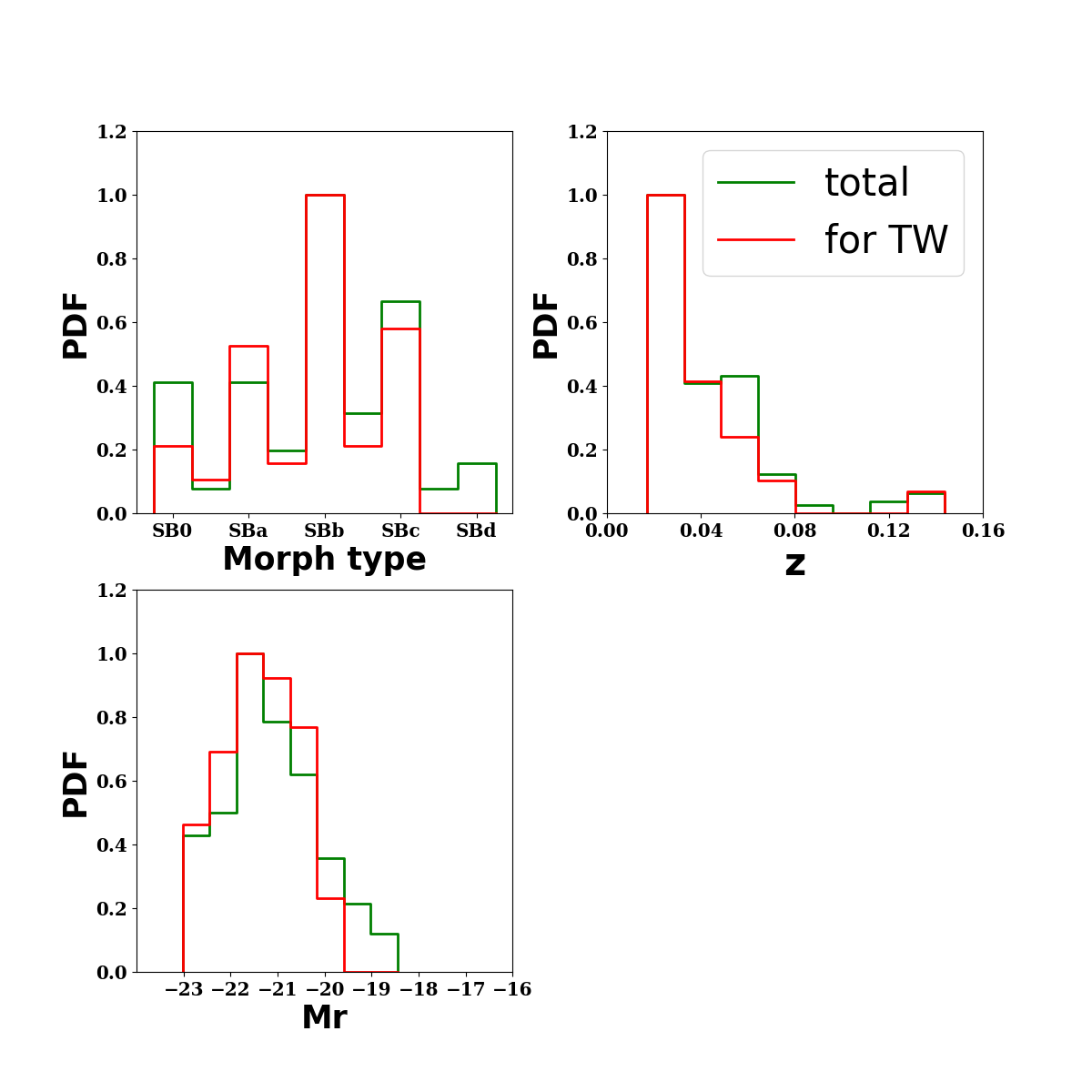}
  \caption{Normalized distribution of morphological types (upper left), redshift (upper right) and r-band absolute magnitudes (lower left) of the DR13 MaNGA barred galaxies (green line) and the final sample selected in this paper (red line).}
  \label{sdis}%
\end{figure}

\subsection{Stellar velocity maps}
\label{ssec_vmap}
Though our sample is chosen from the DR13, the stellar kinematics we use in this work are from the second public MaNGA release as part of SDSS DR14 \citep{abolfathi2018}. We use the newer data set since it contains several improvements in the data reduction pipeline (DRP) and the data analysis pipeline (DAP) over the DR13. See the MaNGA website for more details.

Stellar kinematics are extracted from the spectral datacubes using the MaNGA data analysis pipeline (K. Westfall et al. 2017, in preparation). Firstly, the spaxels of the datacube are Voronoi-binned \citep{cappellari2003} to S/N=10. The stellar velocity and velocity dispersion are obtained by fitting the spectra using the Penalised Pixel-Fitting (pPXF) method \citep{cappellari2004, cappellari2017}. The absorption lines are fitted using a subset of the MILES \citep{sanchez2006, falcon2011} stellar library, MILES-THIN. In the stellar velocity maps, the systemic velocity is subtracted by using the average velocity of the stars in the central 3$''$ aperture. The stellar velocity is used for computing the weighted mean velocity of the TW method and for measuring the kinematic PAs. It is also used in the JAM \citep[e.g.][]{cappellari2008, Li2016} method for obtaining the circular velocity and dark matter fraction.

\section{Measurements of parameters of barred galaxies}
\label{sec_gparas}
\subsection{Inclination and position angles}
\label{ssec_ipa}
To measure the pattern speed of a barred galaxy by the TW method, several pseudo slits are placed along the major axis of the disc, i.e. the line of nodes. The position angle of bar is also needed to infer the de--projected bar lengths. The position angles of the disc and of the bar, and the inclination of the galaxy can be estimated by analysing the galaxy's isophotes \citep{wozniak1995, aguerri2000b}. In this paper, r--band galaxy isophotes are fitted with ellipses using the {\tt ELLIPSE} routine from the IRAF package \citep{jedrzejewski1987} to obtain the ellipticity and PA radial profiles of a galaxy.

\begin{figure}
   \centering
   \includegraphics[width=\columnwidth]{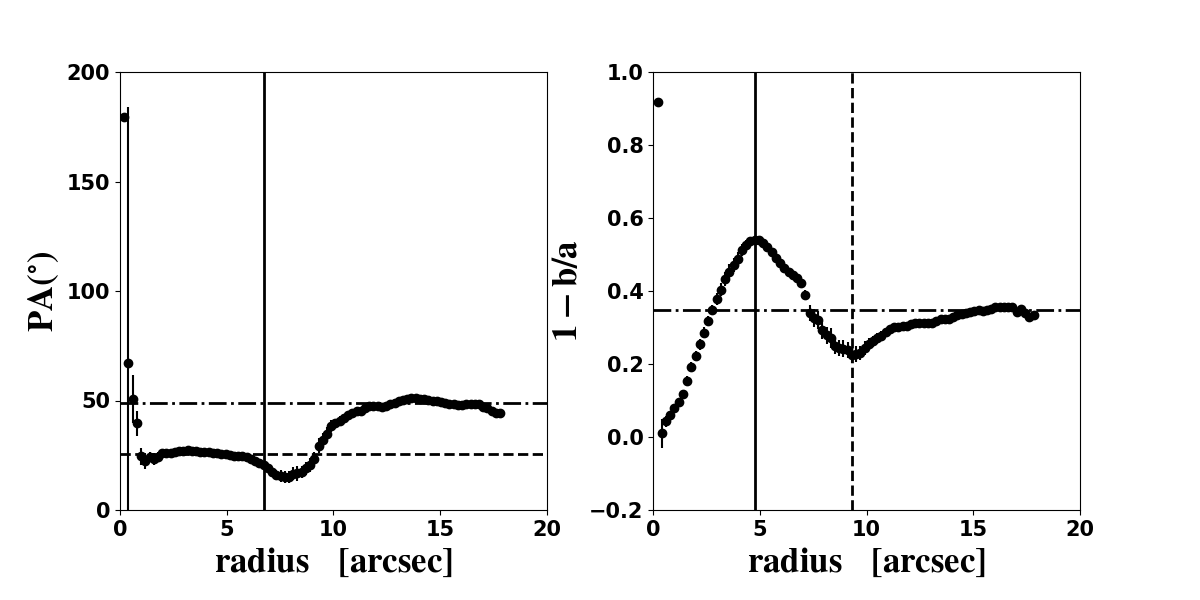}
   \caption{Results from the ellipse fitting of the r-band isophotes of an example galaxy (manga-8439-6102). The image of this galaxy is shown in the upper left panel of Fig. \ref{pattern}. The left and right panels are the ellipticity and PA radial profiles, respectively. The horizontal dotted dash lines show the measured PA (photometric PA, i.e. $PA_{\rm d,k}$) and ellipticity of the disc in each panel, and the horizontal dashed line in the PA radial profile indicates the PA of the bar. Moreover the vertical solid line in the left panel is the bar length ($a_{\rm b,pa}$) inferred from the PA radial profile, at which the PA value changes by 5 degrees relative to the bar PA, i.e. the dashed line. The vertical solid and the dotted dash lines in the right panel are the radii of the local maximum (i.e. $a_{\rm b,e}$) and minimum of the ellipticity profile.}
   \label{ellipse}%
\end{figure}

The ellipticity radial profile of a barred spiral commonly increases from almost zero to a local maximum and then decreases towards a local minimum. These two extremes result from the transition from the domination of a central bar to that of the disc. Assuming the spiral has a rounder outer disc, the ellipticity reaches a constant value b/a at large radii, and the inclination satisfies $\cos i = b/a$ under the thin disc approximation. At the same time, the PA radial profile also comes to a constant value, corresponding to the disc PA. The inclination and PA of the disc are computed by averaging the outer isophotes. For the central bar, we always choose the values of ellipticity and PA when the ellipticity profile reaches the local maximum. Fig. \ref{ellipse} shows the ellipticity and PA radial profiles from {\tt ELLIPSE} fitting of an example galaxy manga-8439-6102.

\begin{figure}
   \centering
   \includegraphics[width=\columnwidth]{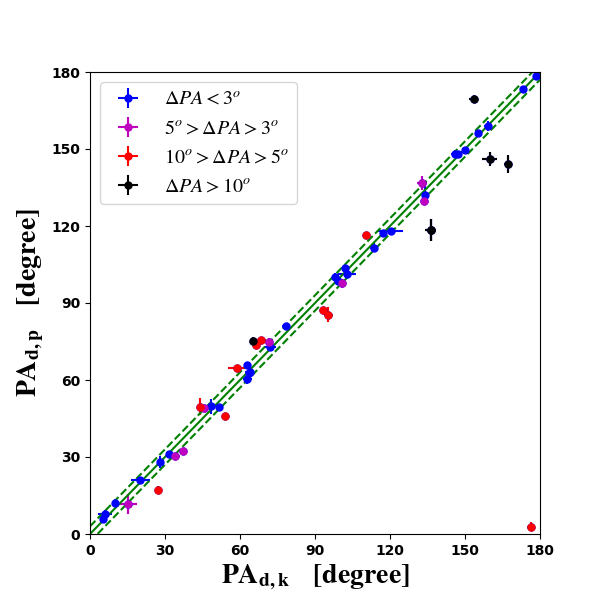}
   \caption{Comparison of the photometric disc PA, $PA_{\rm d,p}$, and the kinematic disc PA, $PA_{\rm d,k}$. The sample is separated into four subsamples according to the PA difference between the kinematic PA and the photometric PA, i.e. $\Delta PA= |PA_{\rm d,p} - PA_{\rm d,k}|$: $\Delta PA <3^{\circ}$ (blue, 29 galaxies), $3^{\circ} < \Delta PA < 5^{\circ}$ (magenta, 8 galaxies), $5^{\circ} < \Delta PA < 10^{\circ}$ (red, 10 galaxies) and $\Delta PA >10^{\circ}$ (black, 6 galaxies). The blue dashed lines label the 3 degrees PA range.}
   \label{fig: pak vs pap}%
\end{figure}

The disc PA is an important parameter for accurately measuring the pattern speed, but the determination of the photometric disc PA ($PA_{\rm d,p}$) described above may be problematic due to faint outer isophotes or disturbances from strong spiral arms or galaxy companions. There is another type of disc PA called the kinematic PA ($PA_{\rm d,k}$), which is derived from the velocity map using a Python program {\tt fit\_kinematic\_pa.py}\footnote{ \url{http://www-astro.physics.ox.ac.uk/\~mxc/software/} } written by Michele Cappellari. It implements the method presented in Appendix C of \cite{krajnovic2006} to measure the global kinematic PA from integral field observations of galaxy stellar or gas kinematics. The method finds the best angle that gives the lowest difference between the observed velocity map and its symmetrized map. This software has been used to study the stellar kinematical misalignment of early-type galaxies in \cite{cappellari2007} and \cite{krajnovic2011}. For our sample, a comparison of the photometric PAs and the kinematic PAs is shown in Fig. \ref{fig: pak vs pap}. For 29 galaxies out of 53 in the sample the kinematic and photometric PAs are the same within 3 degrees.

\subsection{Bar length}
\label{ssec_ab}
The bar length ($a_{\rm b}$) is difficult to estimate, because of the shape of the surface brightness profile of the bar, or the transition from the bar to the spiral arms. Visual determination from r--band images is a direct method, but it is difficult when the bar end is ill-defined because of the spiral arms. Several methods have been developed during the past few decades. Here we use three of the most popular methods to estimate the bar length: the ellipticity radial profile, the PA radial profile and the Fourier decomposition.

The first two estimations of bar length, ellipticity ($a_{\rm b,e}$) and PA ($a_{\rm b,pa}$) radial profiles, take advantage of fitting ellipses to photometric isophotes. These profiles record the transition of domination of the radial surface brightness profile from the round central bulge (if prominent), to the elongated bar, and then to the disc and spiral arms. Generally, the central isophotes are almost circular, either because of the centre spherical bulge  and/or the seeing effects. As one goes further out, the bar dominates the isophotes, thus the isophotes appear as concentric ellipses with nearly identical position angles and higher ellipticities relative to the disc. The disc eventually takes over the radial surface brightness distribution, and isophotes become concentric ellipses with the major axis aligned to the disc PA reaching a constant axis ratio $b/a = \cos i$. From this radial transform of isophotes, different methods have been proposed to estimate the bar length \citep[e.g.][]{marquez1999, athanassoula2002, micheldansac2006, aguerri2009}. Note, however, that this method supposes that bar isophotes can be well approximated by ellipses, while it is well known that generalized ellipses \citep{athanassoula1990, gadotti2008} are necessary at least for many strongly barred, often early type galaxies. This will be discussed further in Section \ref{sssec_aberr}.

As stated previously, for the radial ellipticity profile of a barred galaxy, the ellipticity increases from a central value (zero) to a local maximum, where the bar significantly dominates the isophotes. It then decreases to a local minimum, which corresponds to the end of the bar and transition to the disc-dominated isophotes. Thus the local maximum and minimum represent two extreme cases \citep[][]{micheldansac2006}, and can be understood as the lower and upper limits of the bar length. We adopt the radius reaching the local maximum ellipticity as the first measurement of the bar length (see the solid line in the right panel of Fig. \ref{ellipse}), and as a lower limit of the bar length. The column $a_{\rm b,e}$ in Table \ref{tab: geom} shows the bar lengths estimated using this method.

The second bar length estimation is from the PA radial profile. When the bar dominates the surface brightness, isophotes show nearly constant position angle, and then change to the orientation of the outer disc at large radii \citep[e.g.][]{wozniak1995,aguerri2000b}. The bar length is determined at the radius ($a_{\rm b,pa}$) where the position angle changes by $\Delta PA$ with respect to the value when the ellipticity reaches the local maximum. As adopted in previous literature \citep[e.g.][]{aguerri2015}, we take the value $\Delta PA =5^{\circ}$. There may be some correlation between the estimation of $a_{\rm b,e}$ and $a_{\rm b,pa}$, because the radial profiles of ellipticity and PA both result from the transition from the bar-dominated isophotes to disc-dominated ones. The values of $a_{\rm b,pa}$ determined by this method are listed as column 7 in Table \ref{tab: geom}.

\begin{figure}
   \centering
   \includegraphics[width=\columnwidth]{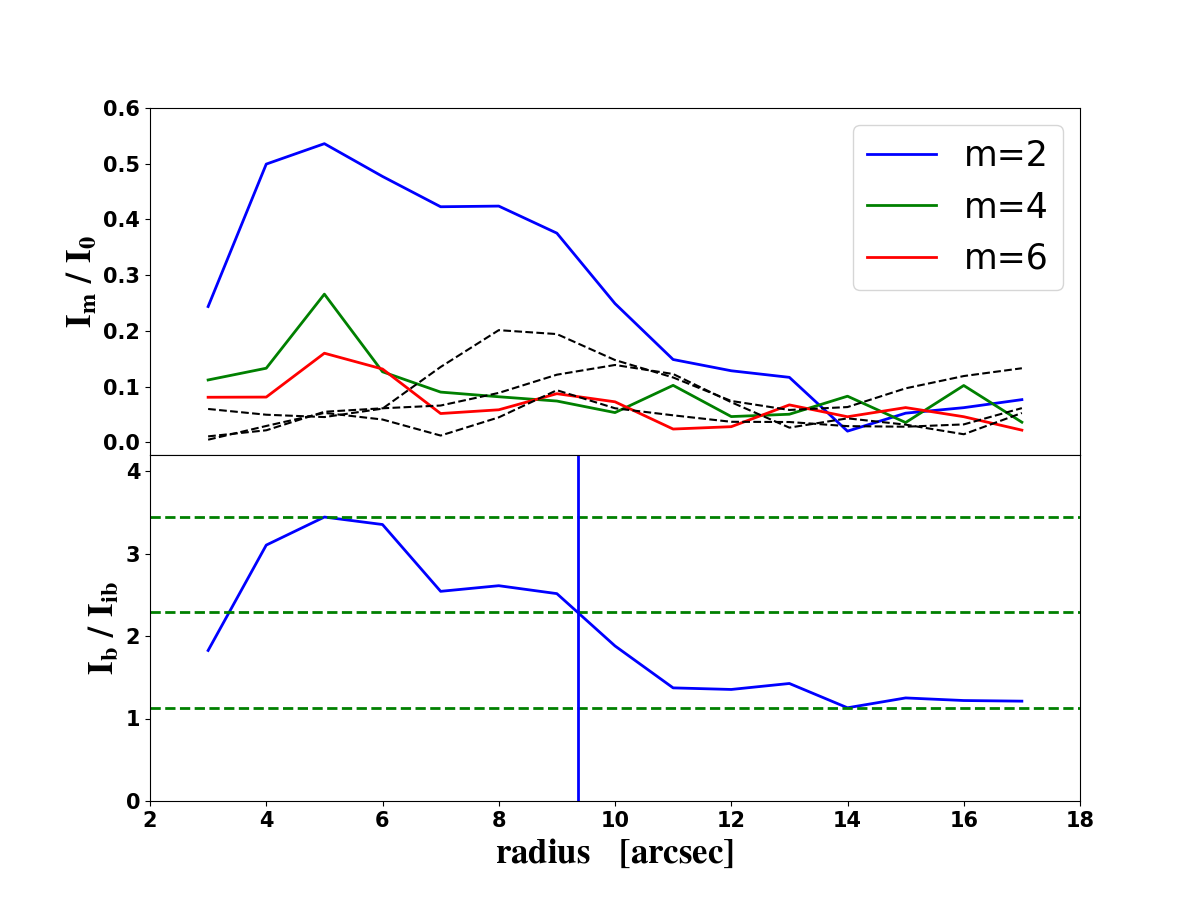}
   \caption{Components from the Fourier decomposition of de-projected azimuthal brightness profile (upper panel) and contrast of bar and the inter--bar intensities $I_{\rm b}/I_{\rm ib}$ (bottom panel) for the galaxy manga-8439-6102. The coloured full lines and the grey dashed lines in upper panel are the even and the odd modes of fourier components, respectively. The horizontal dashed lines in the lower panel are the maximum, median and minimum of the contrasts $I_{\rm b}/I_{\rm ib}$. The outer radius after the peak, i.e. the vertical solid line in the lower panel, is the bar length $a_{\rm b,f}$.}
     \label{f3}%
\end{figure}

The third method we use to estimate the bar length ($a_{\rm b,f}$) is the Fourier decomposition of the de-projected surface brightness profile \citep[e.g.][]{ohta1990, aguerri2000}. Fourier decomposition has been extensively used to characterise the strength of the bar relative to the disc. In this method, the pixels are de--projected to the face--on case according to the inclination angle. Then the de--projected pixels are assigned to mesh grids with radius bins of 0.5$''$ and azimuthal angle bins of $3^{\circ}$. These azimuthal profiles vary with radius in both amplitude and shape. There are prominent humps with a period of $180^{\circ}$ in the bar-dominated regions, which can also show us information on the bar length. After de--projection, the azimuthal profiles are decomposed into Fourier series. Throughout the bar region, the relative amplitudes of even components ($I_{0}, I_{2}, I_{4}, I_{6}$, i.e. the m = 0, 2, 4, and 6 terms of the Fourier decomposition) are much larger than those of odd components. The bar intensity $I_{\rm b}$ and inter--bar intensity $I_{\rm ib}$ are the intensities at the peak and at the bottom of the an azimuthal profile, respectively. To reduce noise fluctuations, they are defined as $I_{\rm b} = I_{0} + I_{2} + I_{4} + I_{6}$ and $I_{\rm ib} = I_{0} - I_{2} + I_{4} - I_{6}$. \cite{ohta1990} defined the bar region as the region with $I_{\rm b}/I_{\rm ib} > 2$. We use a modified criterion by \cite{aguerri2000}, in which the bar region is determined as the region where $I_{\rm b}/I_{\rm ib} > 0.5\times [(I_{\rm b}/I_{\rm ib})_{\rm max} - (I_{\rm b}/I_{\rm ib})_{\rm min}] + (I_{\rm b}/I_{\rm ib})_{\rm min}$. The bar length ($a_{\rm b,f}$) is identified as the outer radius at which $I_{\rm b}/I_{\rm ib} = 0.5\times [(I_{\rm b}/I_{\rm ib})_{\rm max} - (I_{\rm b}/I_{\rm ib})_{\rm min}] + (I_{\rm b}/I_{\rm ib})_{\rm min}$. This method has been checked with numerical simulations and the accuracy is within 8\% except for very thin homogeneous bars with quite large axis ratios \citep[][]{athanassoula2002}. The values of $a_{\rm b,f}$ for our sample are shown in Table \ref{tab: geom}. Fig. \ref{f3} shows an example of the Fourier decomposition procedure.

The effectiveness of these three methods depends on the shapes of the surface brightness profiles of the bars \citep{aguerri2009}, and requires accurate multi--component surface brightness decomposition of photometrical images. Here we simply take the average of $a_{\rm b,e}$, $a_{\rm b,pa}$ and $a_{\rm b,f}$ as the final bar length $a_{\rm b}$, and the lowest and the biggest differences between $a_{\rm b}$ and $a_{\rm b,e}$, $a_{\rm b,pa}$ and $a_{\rm b,f}$ as the lower and upper uncertainties.

\begin{figure}
    \centering
    \includegraphics[width=\columnwidth]{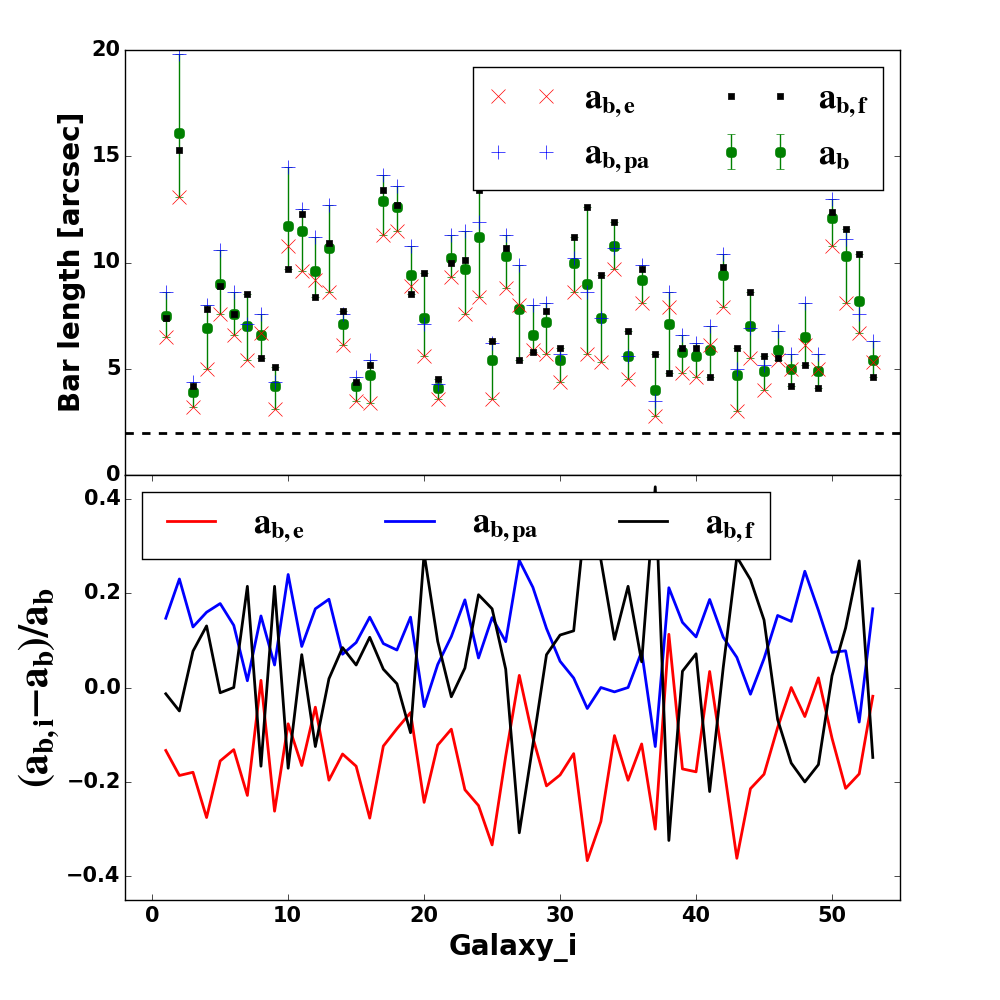}
    \caption{Upper panel: three bar length measurements. The blue crosses and black filled squares are $a_{\rm b,pa}$ from the PA radial profiles and $a_{\rm b,f}$ from the Fourier decomposition. The red crosses are the bar length $a_{\rm b,e}$ from ellipticity profiles, corresponding to the local maximum ellipticities. The bar length $a_{\rm b,pa}$ and $a_{\rm b,e}$ are de-projected using the kinematic PAs. The green filled circles are the average ($a_{\rm b}$) of these three bar lengths, with error bars corresponding to the smallest and the largest measurements. Lower panel: the relative error of three bar lengths to the average bar length. The red, blue, and black lines are the bar length from the ellipticity radial profile, the PA radial profile and the Fourier decomposition, respectively. Galaxy indices of both panels are in the same sequence as galaxies in the Table \ref{tab: galaxies}.}
    \label{fig: ab}
\end{figure}

\subsection{Bar strength}
\label{ssec_sb}
\begin{figure}
\centering
\includegraphics[width=\columnwidth]{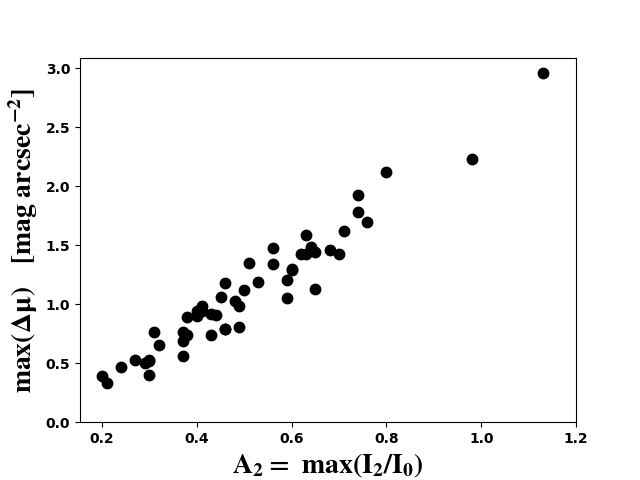}
\caption{Comparison of bar strength from the Fourier decomposition $\rm A_{2}= max( I_{2}/I_{0} )$ and that from the light deficit $\rm max( \Delta\mu )$.}
\label{sb_comp}
\end{figure}

Besides the above three measurements used to derive the scaled pattern speed ${\cal R}$, we have also measured the bar strength for our samples using two different methods. The first utilises the maximum of the ratio between the m=2 and m=0 terms of the Fourier decomposition of the surface brightness profile, i.e. $\rm A_{2}= max( I_{2}/I_{0} )$ \citep[e.g.][]{aguerri2000, athanassoula2002}. The second is the largest surface brightness difference between the de-projected surface brightness profiles along the major and minor axes of the bar, i.e. $\rm max( \Delta\mu )$ \citep[e.g.][]{kim2016}. Both measurements of bar strength are shown in Fig. \ref{sb_comp}. As one can see, they show a good correlation. From now on we only use $A_{2}$ when discussing bar strength.

\section{Results}
\label{sec_results}
\subsection{Pattern speed}
\label{ssec_op}
The pattern speed of our barred galaxies is measured using the non--parametric TW method described in Eq. (1). This method uses the stellar weighted average velocity $\langle V \rangle$ and position $\langle X \rangle$ for several slits parallel to the line of nodes of galaxies with some offsets. For an axisymmetric disc, the mean velocity and position in each slit should be zero. Thus only the non-axisymmetric features, such as a bar or spirals, which should not be aligned to the major or minor axes of the galaxy, contribute to a non--zero mean velocity and position.

The first step in estimating the pattern speed is to choose several pseudo slits using the disc PA, which can be the photometric $PA_{\rm d,p}$ or the kinematic $PA_{\rm d,k}$. We have used three to five pseudo slits with width of 0.5$''$ and a minimum interval of 1$''$ to avoid overlapped pixels. We have also tried a larger slit width of 1$''$, and found it makes no difference for most galaxies. Slit offsets are chosen to avoid the central bulge and bad pixels. For the slit length, we usually take a length of 1.0-1.2 times the effective radius $R_{\rm e}$. Though the TW integrals in Eq. \ref{eq_tw} are over $-\infty < X < \infty$, an integral range of $-X_{\rm max} < X < X_{\rm max}$ is enough if this region reaches the axisymmetric part of the disc. A larger integration region will introduce errors from outer low S/N pixels. The influence of the slit length on the performance of the TW method was tested with the simulation in Appendix \ref{ap: slit}, and is discussed in Section \ref{sssec_operr}.

\begin{figure*}
   \centering
   \includegraphics[width=0.6\textwidth]{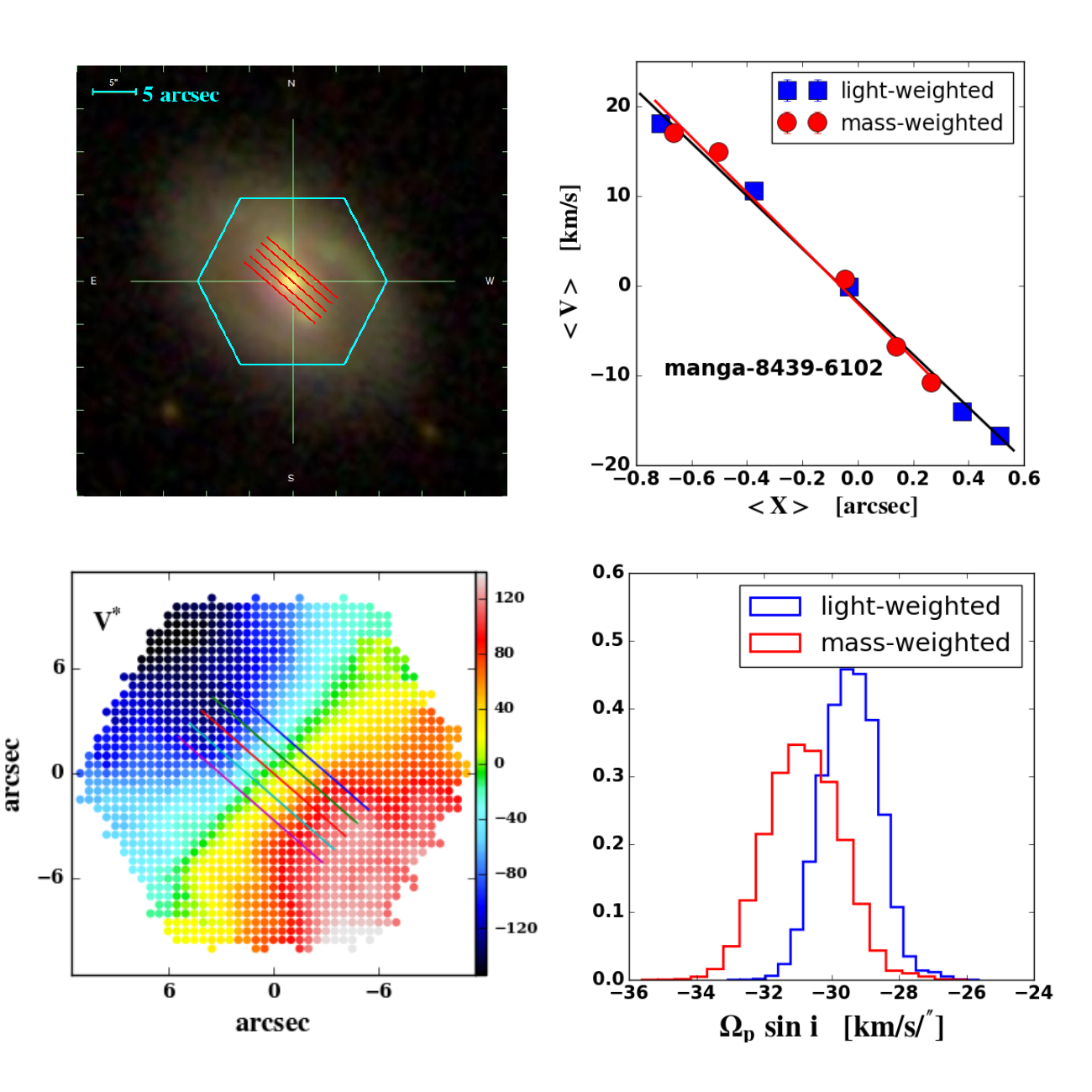}
   \caption{Our example galaxy (manga-8439-6102) for measuring the bar pattern speed. The upper left is the g-r-i three bands combined image of this galaxy, with a hexagon indicating the MaNGA fibre bundle and the red lines indicating the pseudo-slits we choose. The lower left is the velocity map of this example and the five solid lines indicate the pseudo slits. The upper right is the $\langle V \rangle$ vs. $\langle X \rangle$ plot, in which the blue and red lines represent the light--weighted average and mass--weighted average, respectively. The lower right is histograms of light--weighted (blue) and mass--weighted (red) linear fitting slopes $\Omega_{\rm p} \sin i$ of the $\langle V \rangle$ vs. $\langle X \rangle$ plots, taking the PA and velocity errors into account.}
     \label{pattern}%
\end{figure*}

After the pseudo slits have been chosen, two weights can be further used to calculate $\langle V \rangle$ and $\langle X \rangle$, i.e. luminosity weight from spectroscopic data and mass weight from the SPS. The luminosity weights are computed by summing up all the flux for each spectrum of the datacube in wavelength range from 4500 \AA~ to 4650 \AA~, chosen to avoid prominent emission lines. The mass weights are taken from stellar population modelling performed using the pPXF method. Besides the weights used for computing photometric integrals, there are two methods for computing the average stellar velocity $\langle V \rangle$ in each pseudo slit. One is by computing the velocity integrals in the numerator of Eq. (1) by just summing up all the weighted velocities with pixels located inside the pseudo slits. Another method sums all the raw spectra (with weights) inside each pseudo slit into one new, single spectrum. Then it is analysed using the pPXF method as explained previously and $\langle V \rangle$ is the radial velocity obtained from the fit to this single spectrum. But as shown in Fig. 7 of \cite{aguerri2015}, there is no significant difference in the values of $\Omega_{\rm p} \sin i$ for these two methods, so we just use the former, which is simpler.

In principle, each ratio of $\langle V \rangle$ and $\langle X \rangle$ can give us an estimate of the pattern speed. However it is better to fit all the $\langle V \rangle$ vs. $\langle X \rangle$ points with a straight line to avoid the centering error and systemic velocity, and thus the slope of the straight line in Eq. \ref{eq_mk} is the projected pattern speed $\Omega_{\rm p} \sin i$. The upper right panel of Fig. \ref{pattern} shows the  $\langle V \rangle$ vs. $\langle X \rangle$ data points and their linear fits for the example galaxy manga-8439-6102. Position errors are not taken into consideration because they are quite small. The final pattern speed and its uncertainty are taken as the median and 16 and 84 percentiles of all the slopes with 1000 Gaussian randomly distributed PAs and 1000 velocity maps. The histograms of both light--weighted ($\Omega_{\rm p,l} \sin i$) and mass--weighted ($\Omega_{\rm p,m} \sin i$) of these pattern speeds for the same example galaxy are shown in the lower right panel of Fig. \ref{pattern}. The values of  $\Omega_{\rm p,l} \sin i$ and  $\Omega_{\rm p,m} \sin i$ are listed in Table \ref{tab: op_pap}.

\begin{figure*}
   \centering
   \includegraphics[width=0.8\textwidth]{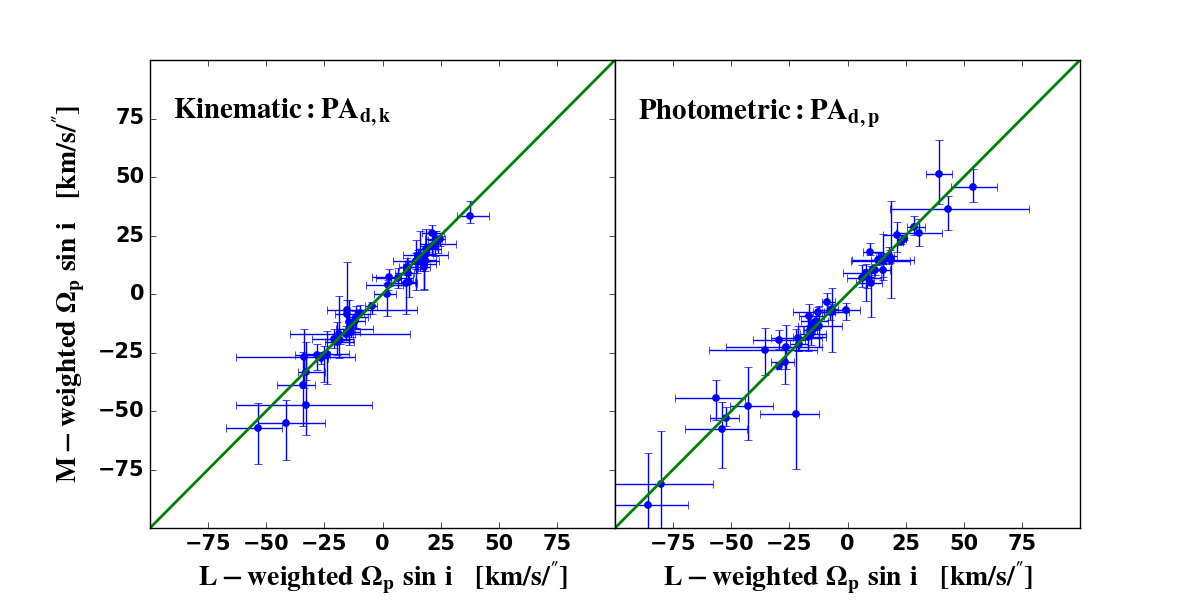}
   \caption{Comparison of the light-weighted and the mass-weighted pattern speeds $\Omega_{\rm p}\sin i$ measured using the kinematic PAs, $PA_{\rm d,k}$ (left panel) or the photometric PAs, $PA_{\rm d,p}$ (right panel).}
    \label{compare}%
\end{figure*}

\begin{figure}
   \centering
   \includegraphics[width=\columnwidth]{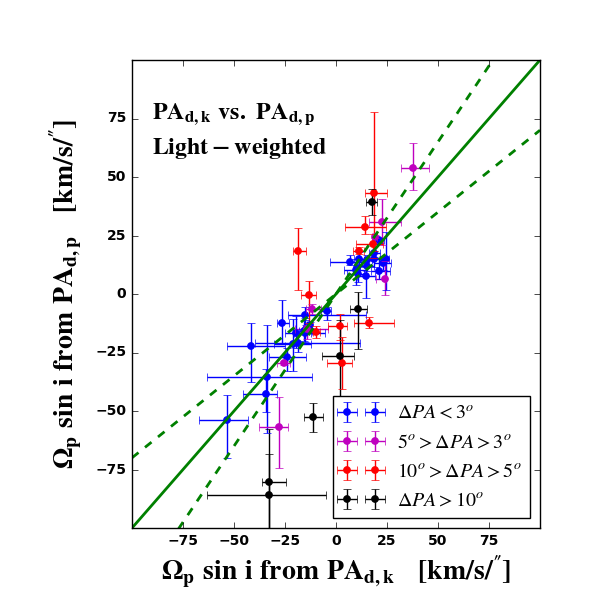}
   \caption{Comparison of the light-weighted pattern speeds $\Omega_{\rm p,l}\sin i$ measured using the kinematic PAs, $PA_{\rm d,k}$ (X axis) and the photometric PAs, $PA_{\rm d,p}$ (Y axis). Galaxies are separated into four groups according to the difference between the kinematic PA and the photometric PA, i.e. $\Delta PA= |PA_{\rm d,p} - PA_{\rm d,k}|$: $\Delta PA <3^{\circ}$ (blue), $3^{\circ} < \Delta PA < 5^{\circ}$ (magenta), $5^{\circ} < \Delta PA < 10^{\circ}$ (red) and $\Delta PA >10^{\circ}$ (black). The dashed lines are 30\% differences relative to the X axis.}
    \label{compare_pk}%
\end{figure}

Fig. \ref{compare} shows the comparison of light-weighted and mass-weighted pattern speeds obtained using the kinematic PAs or photometric PAs. Generally, the light-weighted pattern speeds agree with mass-weighted ones within $1\sigma$ for both types of PA. We need to emphasize that the mass weights are from spaxels Voronoi-binned to S/N=30, which is different from the kinematic data that is Voronoi-binned to S/N=10. We have also compared the light-weighted and mass-weighted pattern speeds in which the mass weights and the kinematics are calculated both from Voronoi-binned to S/N=20 data. These pattern speeds agree well with each other again. In our later studies of dependences of the pattern speeds on galaxy properties, we use only the light-weighted pattern speeds. As shown in Fig. \ref{compare_pk}, the light-weighted pattern speeds measured using the kinematic and photometric PAs show some differences. Some galaxies even show different signs in pattern speed values. We will return to this in Section \ref{sssec_operr}.

\subsection{Dependence on Galaxy Properties}
\label{ssec_R}
With circular velocities, bar lengths and bar pattern speeds estimated above, we can derive the scaled pattern speeds ${\cal R}= R_{\rm CR}/a_{\rm b} = ( V_{\rm c}/ \Omega_{\rm p} ) / a_{\rm b}$, and study its dependence on galaxy properties such as the dark matter fraction, the stellar age and metallicity.

\subsubsection{Dependence on the dark matter fraction}
\label{sssec_Rfdm}
\begin{figure*}
\centering
\includegraphics[width=1.0\textwidth]{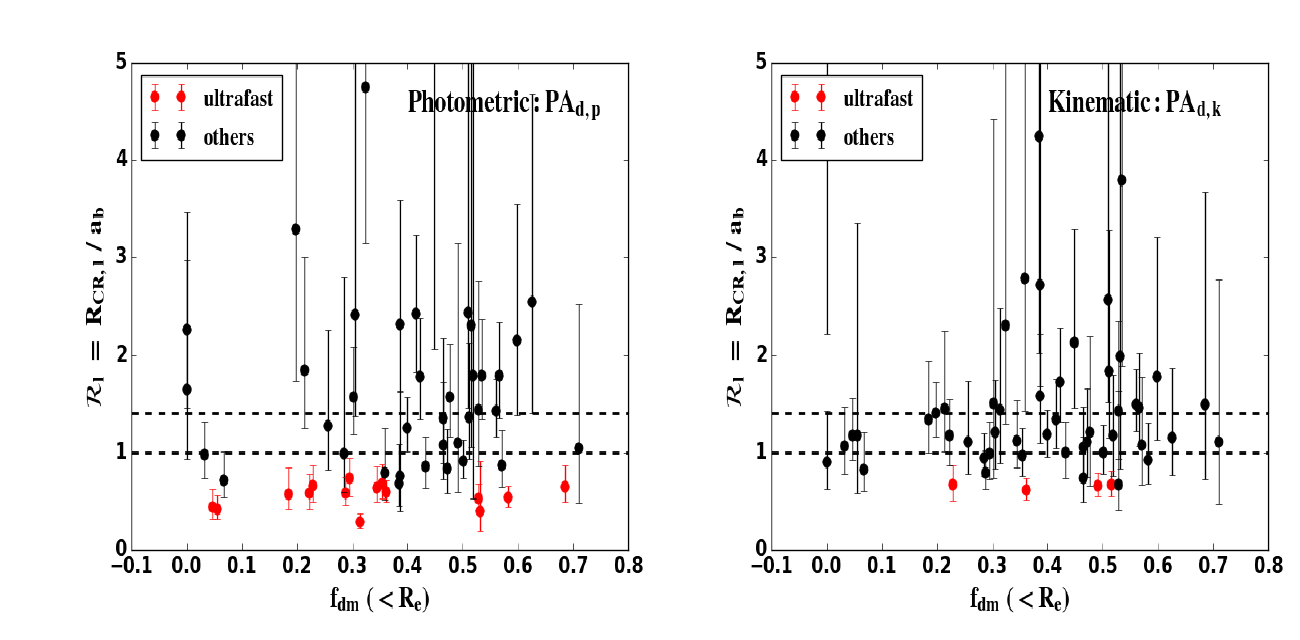}
\caption{Dependence of the light-weighted ${\cal R}$ value on the dark matter fraction inside the effective radius. The left column is for pattern speeds measured using the photometric PAs, $PA_{\rm d,p}$, while the right column uses the kinematic PAs, $PA_{\rm d,k}$. The red points are ultrafast bars with the $1\sigma$ upper limit of ${\cal R}$ smaller than one. Two horizontal dashed lines label the ${\cal R}= 1$ and ${\cal R}= 1.4$.}
\label{dm_frac}
\end{figure*}

The dark matter fraction is estimated by the JAM method. The light-weighted ${\cal R}_{\rm l}$ value as a function of dark matter fraction inside one effective radius for our sample is shown in Fig. \ref{dm_frac}. For the pattern speeds measured using the photometric and kinematic PAs, there is no significant trend. A prominent difference between the photometric and kinematic results is the number of ultrafast bars, defined as having $1 \sigma$ upper limit of ${\cal R}$ smaller than 1. There are 15 and 4 ultrafast bars for photometric and kinematic PAs, respectively. These ultrafast bar galaxies apparently have bars extending beyond the corotation radii of the galaxies. Such bars are supposedly unphysical because the main orbit family constituting bars (the $x_{1}$ orbits) stops at corotation while its extension has orbits elongated perpendicular to the bar \citep{contopoulos1980}. This problem will be further discussed in Section \ref{ssec_Rerr}.

\begin{figure*}
\centering
\includegraphics[width=1.0\textwidth]{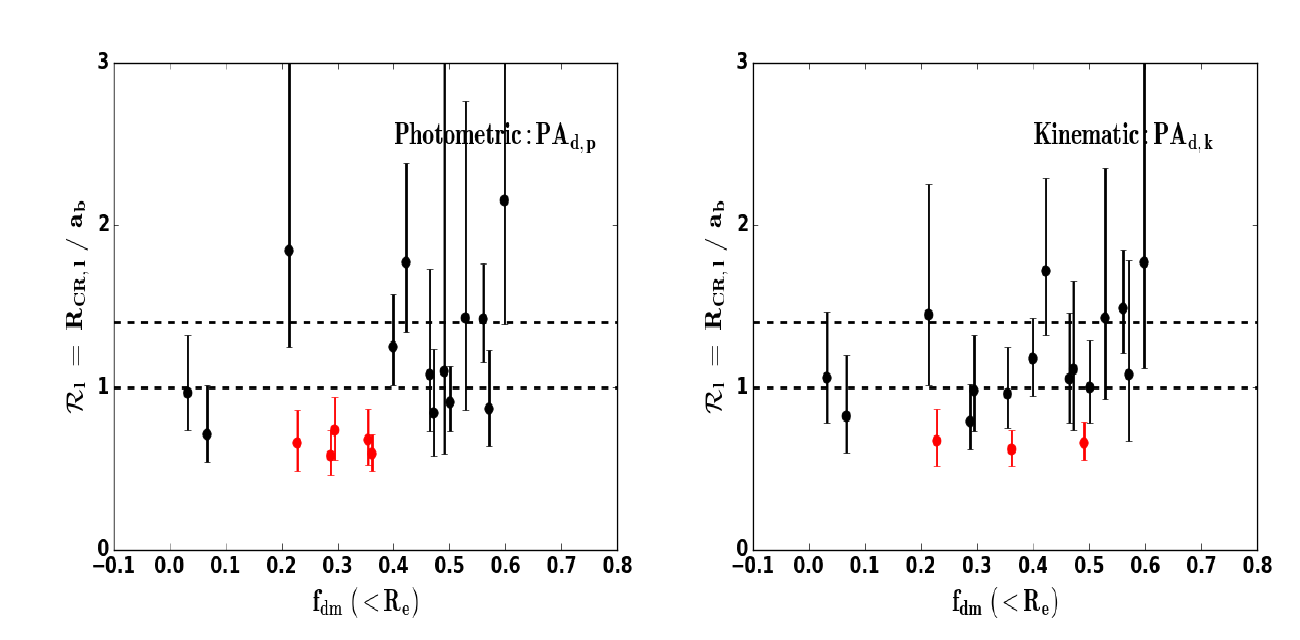}
\caption{Same as Fig. \ref{dm_frac} but for galaxies with difference between the kinematic PA and the photometric PA smaller than 5 degrees and linear fitting error of the $\langle V \rangle$ vs. $\langle X \rangle$ plot smaller than 20\%.}
\label{dm_frac_ref}
\end{figure*}

We examine a smaller sample of 19 barred galaxies, for which the difference between the kinematic and photometric PAs is less than 5$^{\circ}$ and the linear fitting errors of $\langle V \rangle$ vs. $\langle X \rangle$ are smaller than 20\%. The dependence of the ${\cal R}$ values on the dark matter fraction for this refined sample is shown in Fig. \ref{dm_frac_ref}. For this sample, no trends were found either and so the lack of correlation is not due to large measurement errors.

\subsubsection{Dependence on stellar age and metallicity}
\label{sssec_Rage}
\begin{figure*}
\centering
\includegraphics[width=0.8\textwidth]{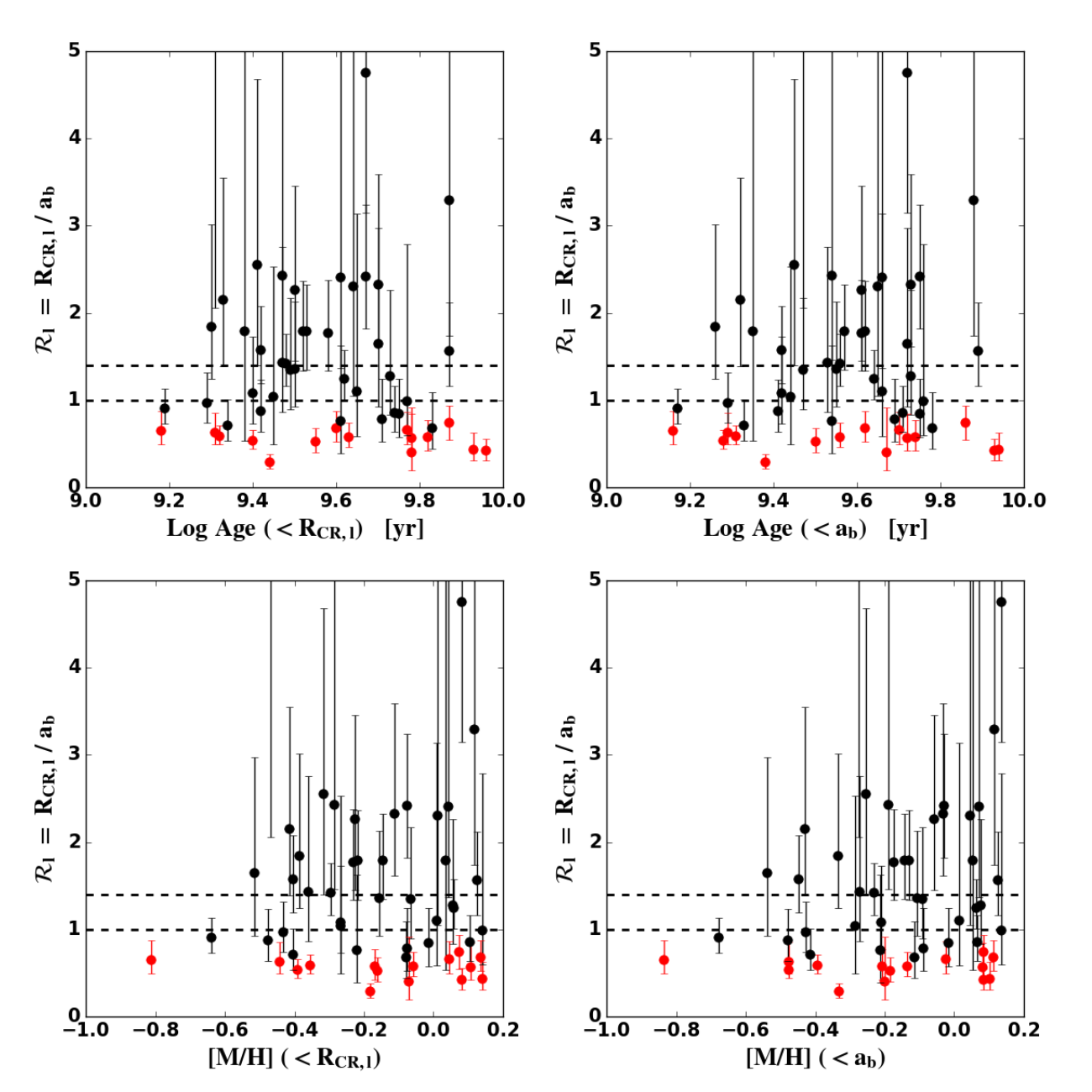}
\caption{Dependence of the light-weighted ${\cal R}$ value on the stellar age (upper row) and the stellar metallicity (lower row) for the average within the corotation radius (left column) and the bar region (right column). Two horizontal dashed lines in each figure label ${\cal R}= 1$ and ${\cal R}= 1.4$. Colour coding is the same as in Fig. \ref{dm_frac}.}
\label{fig_agez}
\end{figure*}

Stellar age and metallicity are derived from stellar population synthesis of MaNGA IFU spectra. Fig. \ref{fig_agez} shows the ${\cal R}$ value as a function of stellar age and metallicity for the average inside the corotation radius or inside the bar region. The bar region is defined as an ellipse with the bar length as the major axis, and the minor axis is determined by the local maximum ellipticity. The ${\cal R}$ values are derived using photometric PAs. There are no significant trends between the ${\cal R}$ values and the stellar age and metallicity. The ${\cal R}$ values derived from the kinematic PAs do not show significant trends either.

\subsubsection{Dependence on bar strength}
\label{sssec_Rsb}
\begin{figure*}
\centering
\includegraphics[width=0.8\textwidth]{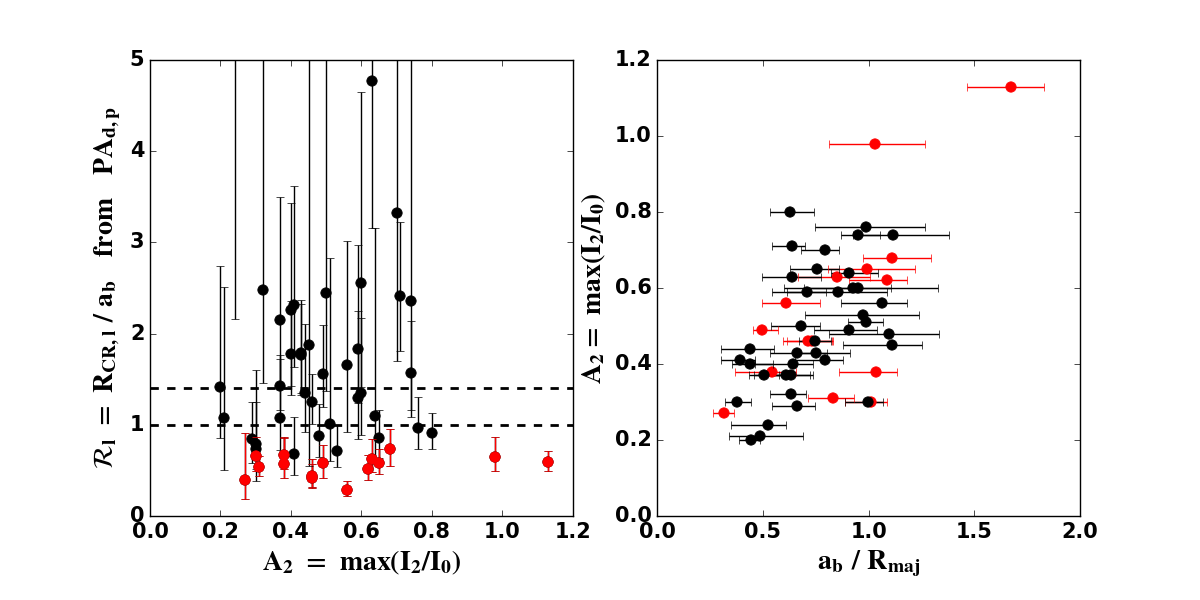}
\caption{Bar strength as a function of the light-weighted ${\cal R}$ value measured using the photometric PAs (left) and bar length normalized by the radius $R_{\rm maj}$ (right). $R_{\rm maj}$ is the length of the major axis at 1.35 times $R_{\rm e}$; this factor is used to make up for the offset mainly resulting from the cutoff of the luminosity profile in the MGE $R_{\rm e}$ calculation \protect \citep[see Fig. 7 of][]{cappellari2013}. Colors are the same as in Fig. \ref{dm_frac}.}
\label{sb_ab}
\end{figure*}

Bar strength is an important parameter that measures the non-axisymmetric forces produced by the bar potential in the disc of galaxies. It is an indicator of the bar slowdown rate and correlates well with the angular momentum absorbed by the spheroidal components of a galaxy \citep{athanassoula2003}. As the angular momentum is transformed from the bar to the spheroid and particularly to the regions around its resonances, the bar becomes longer and slender, and the bar pattern speed decreases. This implies that the values of both the corotation radius and the bar length decrease, but does not give any indication of what their ratio will do \citep[see][section 4.7.2 for a discussion]{athanassoula2013}. \cite{athanassoula2014} showed that this ratio can stay nearly constant with time in simulations with initial conditions including a triaxial halo and a gas fraction higher than 20\%, as one would expect for galaxies at higher redshift. Besides, we have a spread of galaxy masses in our sample. Thus the lack of a trend in the left panel of Fig. \ref{sb_ab} does not disagree with any known theoretical result. A similar result was found for ${\cal R}$ values measured using kinematic PAs. We do however see a trend that larger bar strength galaxies have longer bar lengths, as shown in the right panel of Fig. \ref{sb_ab}. In the right panel, the bar length is scaled by $R_{\rm maj}$, which is the length of the major axis at 1.35 times $R_{\rm e}$. This factor is used to make up for the offset mainly due to the cutoff in the luminosity profile in the MGE $R_{\rm e}$ calculation \citep[see Fig. 7 of][]{cappellari2013}. Note that, given the definition of the bar strength used and of the results of Fig. \ref{sb_comp}, this is not a trivial result. It shows that as the bar becomes stronger it also becomes longer, in good agreement with simulations.

\section{Discussion}
\label{sec_discussions}
\subsection{Uncertainties of ${\cal R}$ }
\label{ssec_Rerr}
\subsubsection{Uncertainties of pattern speed}
\label{sssec_operr}
The measurement of pattern speed using the TW method depends on the measurement of galaxy's PA. According to the simulation work of \cite{debattista2003}, a PA error of 5$^{\circ}$ can result in an error of about 44 percent in ${\cal R}$. In our work, we have used two kinds of PA to measure the pattern speed. One is the photometric PA derived from ellipse fitting of isophotes of SDSS r-band images. The outer isophotes can be influenced by strong structures (such as spiral arms, rings, etc.) in the outer regions of the disc, and these subsequently influence the estimation of the disc PA. A faint disc, a non-circular disc or a disc contaminated by nearby bright sources will also suffer from the PA measurement problem. The photometric PA is obtained by averaging the PAs of the relatively flat parts of the outer PA radial profile, but the choice of the flat region is somewhat arbitrary. The PA error is taken as the maximum of the standard deviation of the outer PAs and the statistical error given by the {\tt ELLIPSE} program. Most PA errors for our galaxies are between 1 and 2 degrees, and are usually smaller than the errors of the kinematic PAs.

Kinematic PAs are derived from stellar velocity maps, according to the difference between the observed and the symmetrized velocity maps. Thus this method depends on the IFU coverage and the symmetry degree of the velocity map. Using the DR13 or DR14 MaNGA data, or using the velocity map Voronoi binned to $S/N = 10$ or binned to $S/N = 20$ only yields consistent PAs to within an error of 3 degrees. For about 60\% of our sample, shown as blue dots in Fig. \ref{fig: pak vs pap}, the photometric and kinematic PAs have differences smaller than 3 degrees. The symmetry of the disc velocity field is possibly influenced by the existence of the bar. As shown in Fig. 4 of \cite{sancisi1979} or Fig. 14 of \cite{duval1983}, the inner parts of the disc velocity field are twisted by the bar. This effect may influence the measurement of the kinematic PAs, and this is examined with the simulation in Appendix \ref{ap: pak}. As shown in Fig. \ref{sim_pak}, galaxy inclination, IFU coverage and PA difference between the disc and the bar influence the kinematic PA measurement. More specifically, galaxies with lower inclination angles, smaller IFU coverages and the PA differences between the disc and the bar closer to 45$^{\circ}$ are more seriously influenced by the bar twisting effect, and have larger measurement errors. The bar twisting effect may contribute most to the difference in the ${\cal R}$ ratios measured using kinematic PAs and photometric PAs. We have checked a subsample containing 25 galaxies with $20^{\circ}< |PA_{\rm d,k} - PA_{\rm b}| <35^{\circ}$ or $55^{\circ}< |PA_{\rm d,k} - PA_{\rm b}| <70^{\circ}$. We find no trends between ${\cal R}$ and other galaxy parameters, especially the dark matter fraction. Thus this effect will not change our main results.

The pseudo slits chosen to measure the pattern speed may also influence the behaviour of the TW method. We test the influence of slit position and length on the TW method using the simulation in Appendix \ref{ap: slit}. Slit interval and width only make the measured pattern speed a little more dispersed. Nevertheless, slit length always shows the same behaviour, namely that the value of the pattern speed first increases with slit length and then, for slits longer than 1.2 times the bar length, it stays nearly flat, as shown in Fig. \ref{sim_sl_i} and Fig. \ref{sim_sl_dpa}. In our measurements for MaNGA galaxies, same slits with the largest slit lengths allowed by the IFU data are chosen for randomly sampled PA and velocity map. This length is usually about $1.2 R_{\rm e}$. From the distribution of ratio between the bar length and the radius $R_{\rm maj}$ correlated to $R_{\rm e}$ shown in Fig. \ref{sb_ab}, some galaxies with bar length larger than the effective radius may underestimate bar pattern speeds, and thus overestimate ${\cal R}$ values. This may make some contribution to the large number of galaxies with ${\cal R} >1.4$.

\begin{figure*}
\centering
\includegraphics[width=0.8\textwidth]{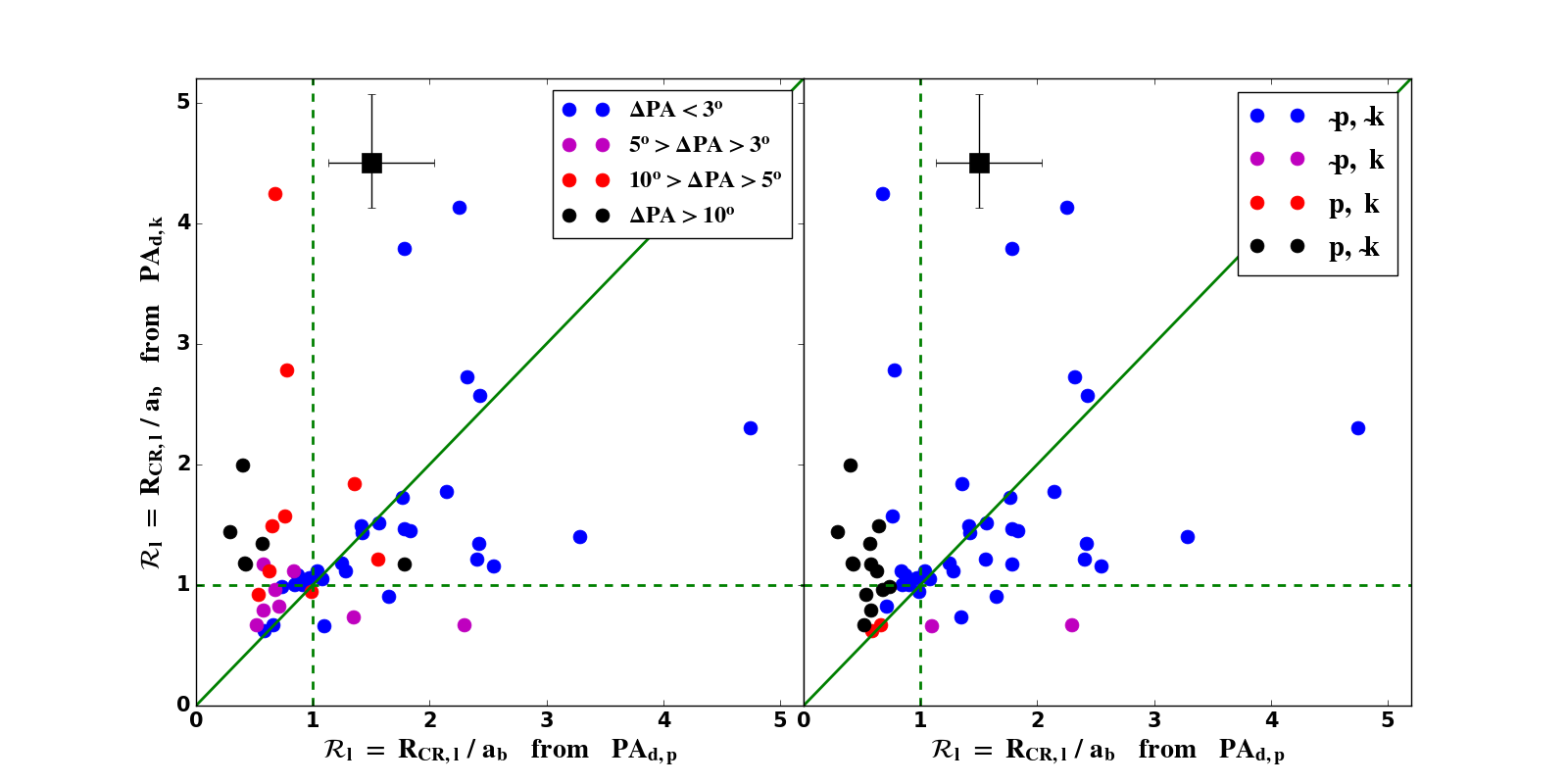}
\caption{Comparison of the light-weighted ${\cal R}_{\rm l}$ parameters using the photometric and the kinematic PAs. Left: Galaxies are separated into four groups according to the PA difference between the photometric PA and the kinematic PA: $\Delta PA <3^{\circ}$ (blue), $3^{\circ} < \Delta PA < 5^{\circ}$ (magenta), $5^{\circ} < \Delta PA < 10^{\circ}$ (red) and $\Delta PA >10^{\circ}$ (black). Right: Galaxies are separated into four groups according to if it is an ultrafast bar: $\mathbf{ \sim p,\sim k }$ (blue) means that they are not ultrafast bars in both; $\mathbf{ \sim p,\ k }$ (magenta) means that they are ultrafast bars only for the kinematic PAs; $\mathbf{ p,\ k }$ (red) means that they are ultrafast in both; $\mathbf{ p,\sim k }$ (black) means that they are ultrafast only for the photometric PAs. For clarity, the error bars are omitted but typical errors are shown on the square on the upper left.}
\label{fig: R_dpa}
\end{figure*}

After comparing the pattern speeds measured using different data (DR13 and DR14) or different slit positions, we find that the PA difference is the main reason for the difference in pattern speeds measured by using these two types of PA. As shown in the left panel of Fig. \ref{fig: R_dpa}, galaxies with PA difference $\Delta PA <5^{\circ}$ distribute diagonally in the panel. Galaxies with $\Delta PA >5^{\circ}$ show larger ${\cal R}$ values measured using kinematic PAs than those measured using photometric PAs. This could be due to the method used for estimating the kinematic PA \citep{krajnovic2006}, which seeks to minimize any asymmetry in the velocity field. This will automatically reduce the size of the $\langle V \rangle$ integrals in Eq. \ref{eq_tw}, and then systematically underestimate  $\Omega_{p}$. We use the mock IFUs in our Appendix \ref{ap: pak} to test this possibility. We measure the pattern speeds for the true disc PA and the measured kinematic PA, and find that the latter will usually lead to lower pattern speed, except for some highly inclined mock IFUs. The estimation method for kinematic PAs can indeed lead to underestimating the pattern speed. Besides, the influences of the PA difference between the kinematic PA and the photometric PA on the pattern speed measurement vary for different galaxies. Some galaxies with $\Delta PA <5^{\circ}$ have larger ${\cal R}$ value differences than ones with $\Delta PA >5^{\circ}$, as shown in the left panel of Fig. \ref{fig: R_dpa}.

MaNGA IFUs have different bundles, from 19 fibres to 127 fibres, varying in size from 12.5$''$ to 32.5$''$ in diameter. Thus different galaxies are covered differently observationally. For some galaxies, the outer parts of their discs are not observed, which influences the pseudo slit lengths, and so the application of the TW method. Another problem for some galaxies is that they have quite a faint disc relative to the bar, and their velocity maps have no reliable disc velocities to be used in estimating pattern speeds. We have excluded some galaxies in the loosely constrained sample for these two reasons, for example manga-8484-12703 and manga-8132-6101 respectively. (The latter seems to have quite good linear fitting of $\langle V \rangle$ vs. $\langle X \rangle$ and quite large bar length, about 25 kpc as our estimation.)

Besides the uncertainties inherent to observations, there are some other factors that may influence the application of the TW method. This method is based on a well-defined pattern speed and the continuity equation of tracers. If the galaxy has two bars, or a spiral rotating with a different pattern speed than the bar \citep{tagger1987, sellwood1988, sygnet1988}, applying the TW method is not straightforward. In addition, the continuity equation of stellar tracers can be influenced by significant star formation and dust obscuration. Nevertheless, \cite{gerssen2007} show that it is possible to extend the application of the TW method to the stellar component of late-type barred galaxies. We have also checked the correlation between pattern speed and dust extinction derived from SPS and find that the measured pattern speeds do not correlate with dust extinction. As for star formation, we find no evidence of correlation either.

Another potential problem that may influence the application of the TW method is the time evolution of the bar pattern speed. In our tests with simulation data in the Appendices, TW performs well but it is only a snapshot after the bar has evolved for about 8 Gyr. Checking the performance of the TW method for different stages of evolution of the bar, and for bars of different strengths would be very interesting but is beyond the scope of this paper. We will discuss them elsewhere.

\subsubsection{Bar length uncertainties}
\label{sssec_aberr}
The three methods we use to measure the bar length have been widely used in previous works \citep[][etc.]{athanassoula2002, micheldansac2006, aguerri2009} and have been tested with simulations \citep{athanassoula2002, aguerri2009}. Indeed, a given method could be better suited to one galaxy, while a different method could be better to another. The suitability of a method can depend on the orientation, and/or the strength of the bar, and/or whether it also has spirals or an inner ring etc.  Our three bar lengths are consistent with their average to about 20 percent, as shown in the lower panel of Fig. \ref{fig: ab}.

Another concern in measuring the bar length with {\tt ELLIPSE} fitting is the assumption that bar isophotes can be well approximated by ellipses, while it is well known that generalized ellipses \citep{athanassoula1990, gadotti2008} are necessary at least for many strongly barred, often early type galaxies. We have tried to fit generalized ellipses with different shape parameters, b/a ratios and different data distributions with simple ellipses. We found an overestimation of about 5\% of the bar length for rectangular-like isophotoes. However, this fraction will be influenced by the galaxy inclination, the angle between bar and disc major axes, and the intrinsic bar shape (box-like or triaxial). This effect needs further examination but is beyond the scope of this paper.

\subsubsection{Circular velocity uncertainties}
\label{sssec_vcerr}
The third uncertainty in determining ${\cal R}$ might be the estimation of the circular velocity $V_{\rm c}$. Typically, the circular velocity is derived from the observed stellar streaming velocity. However the mean rotation velocity of a population of stars will fall below the circular velocity due to asymmetric drift, which is difficult to correct \citep[see Chapter 4.8.2 of ][]{binney2008}. In this work, we calculate the circular velocity from the galaxy's total mass derived by the JAM method. This method can constrain the total mass to a 10-18\% 1$\sigma$ error according to the tests in \cite{Li2016}. However, due to the 0.1 dex uncertainty in M*/L for young galaxies, we finally take a systematic 12\% error for circular velocities. This error is systematic and can vary from galaxy to galaxy. In Appendix \ref{sim_vc}, we have checked the performance of the JAM method in a simulated strongly barred galaxy. As shown in Fig. \ref{sim_vc}, the circular velocities are recovered within an error range of about 10\% for different galaxy inclinations and different bar orientations. This test reinforces our decision to use JAM circular velocities in this work.

\begin{figure}
\centering
\includegraphics[width=\columnwidth]{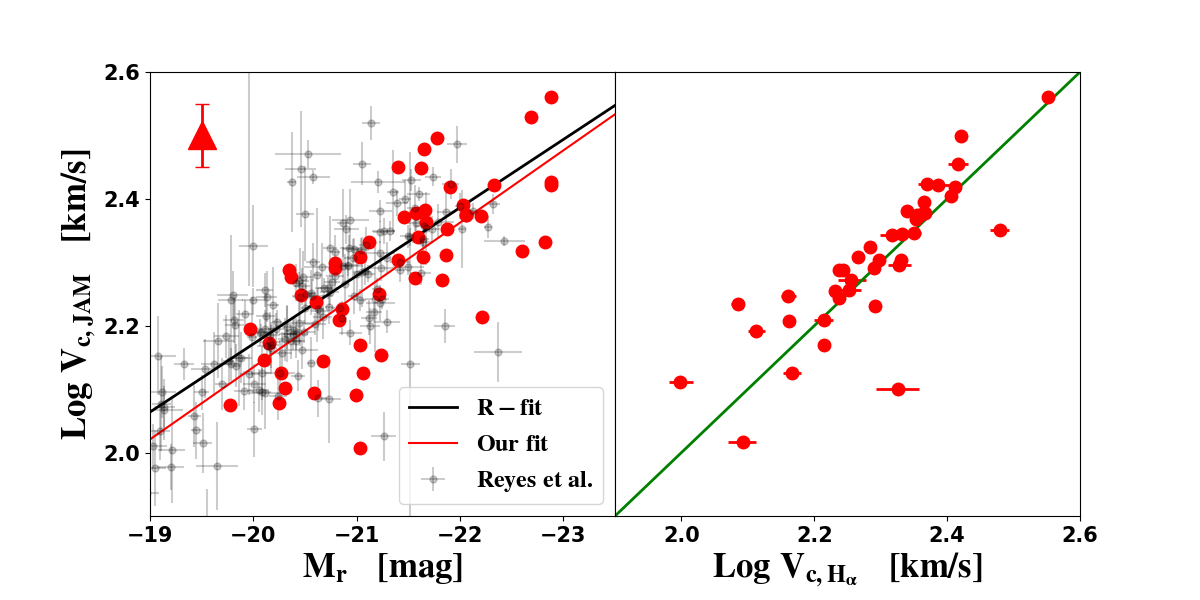}
\caption{Left panel: The Tully-Fisher relation for circular velocities from JAM modelling. The grey dots are the sample from \protect \cite{reyes2011}. The red points are for our sample. The solid lines are linear fits of the Tully-Fisher relation for the Reyes et al. sample (black) and our sample (red). A 0.05 dex systematic error of $V_{\rm c}$ from JAM is given as a triangle in the upper left corner. Right panel: comparison of circular velocities from JAM and from $H_{\rm \alpha}$ emission lines. Error bar of $V_{\rm c}$ from JAM is same as that in the left panel.}
\label{TF_relation}
\end{figure}

In addition, we have checked the Tully-Fisher relation \citep[TF;][]{tully1977} of our bar sample and compared the JAM circular velocities with $H_{\rm \alpha}$ emission line velocities. The Tully-Fisher relation is a correlation between the galaxy circular velocity at large radii and their absolute magnitudes. The left panel of Fig. \ref{TF_relation} shows the TF relation for our galaxies in comparison with other spiral galaxies obtained from the literature \citep[see][]{reyes2011}. Our sample has a similar TF relation to that of spiral galaxies in the literature. The right panel of Fig. \ref{TF_relation} is the comparison of our circular velocities with $H_{\rm \alpha}$ emission line velocities, which are the average of the outer flat regions. Our circular velocities are consistent with those gas velocities in a range about 0.08 dex. If we believe gas moves circularly, our estimations of circular velocities are not the main uncertainty in determining ${\cal R}$.

\subsubsection{A general assessment of uncertainties}
\label{sssec_Rerr}
For a general assessment of uncertainties in ${\cal R}$ measurement, we have checked several subsamples with more accurate estimation in one parameter at a time. These include a subsample with $|PA_{\rm d,k} - PA_{\rm d,p}| <3^{\circ}$ (29 galaxies), a subsample containing about 22 galaxies with larger ratios of slit length to bar length, a subsample with three bar length estimations consistent with their averages within 20\% (28 galaxies) and a subsample containing 24 galaxies which have circular velocities consistent with the Tully-Fisher relation within 0.05 dex. For all these subsamples, we still find no trend between ${\cal R}$ and the dark matter fraction and other galaxy parameters. 

In addition, we perform the following test to see whether large observational errors can obscure real correlations: To do this, we assume a correlation between ${\cal R}$ and the dark matter fraction lying on the diagonal of Fig. \ref{dm_frac}, i.e. ${\cal R}= 1.25f_{\rm dm} +1$ or ${\cal R}=  2.5f_{\rm dm} +1$. Then we take as a measure of the errors the lowest point of the ultrafast bars, i.e. the distance of the most ultrafast galaxy from the ${\cal R}=1$ line (0.71 and 0.39 for $PA_{\rm d,p}$ and $PA_{\rm d,k}$, respectively). Then we randomly sample ${\cal R}$ according to the mock correlation and the errors. In both mock observations using both type of disc PAs, we can still find a trend between ${\cal R}$ and the dark matter fraction despite of the scatters. This means that the lack of correlation in our results is not due to the errors in ${\cal R}$.

To summarise, in this and the previous section, and throughout this paper we have discussed a number of uncertainties that can influence our results. Their nature and amplitude, however, are such that our main results will not be affected, as already discussed. Furthermore, in all cases we have tried the trends not only using our complete sample, but also subsamples which, although containing fewer galaxies, are of higher quality in that they contain only galaxies which have the most accurate estimations of the respective parameters. In all cases we found good agreement. Specifically for the lack of correlation between the ${\cal R}$ measurement and the dark matter fraction in the inner parts, we tried the effect of the uncertainties on mock correlations and found that these uncertainties were not sufficient to produce this lack of correlation.

\subsection{Ultrafast bars}
\label{ssec_ultra}
An important concern in our work is that we have too many ultrafast bars, which place an upper limit on ${\cal R}$ smaller than 1. Theoretically, it is unphysical because the bar cannot end beyond the corotation radius of the galaxies \citep[see][]{contopoulos1980, athanassoula1980}. The main family of orbits constituting the bar (the $x_{1}$ orbits) stops at corotation while its extension has orbits elongated perpendicular to the bar \citep{contopoulos1980}. Furthermore, the response to a bar forcing is bar-like only up to corotation and becomes spiral beyond it \citep{athanassoula1980}. There are 15 and 15 ultrafast bars measured using the photometric PAs by the light-weighted and mass-weighted, respectively. These numbers decline to 4 and 5 when the kinematic PAs are used. More accurately, as shown in the right panel of Fig. \ref{fig: R_dpa}, for the light-weighted results, there are 13 bars which are ultrafast when we only use the photometric PA to measure pattern speed, and there are 2 galaxies which are ultrafast bars when only using the kinematic PA to measure pattern speed. There are only 2 galaxies which are ultrafast bars when using both kinds of PA to measure the pattern speed with light-weighting. Basically, the kinematic PAs lead to smaller pattern speeds and less ultrafast bars. The remaining 2 galaxies which are ultrafast bars in both results can also be physical for some reasons unknown.

\begin{figure*}
\centering
\includegraphics[width=0.8\textwidth]{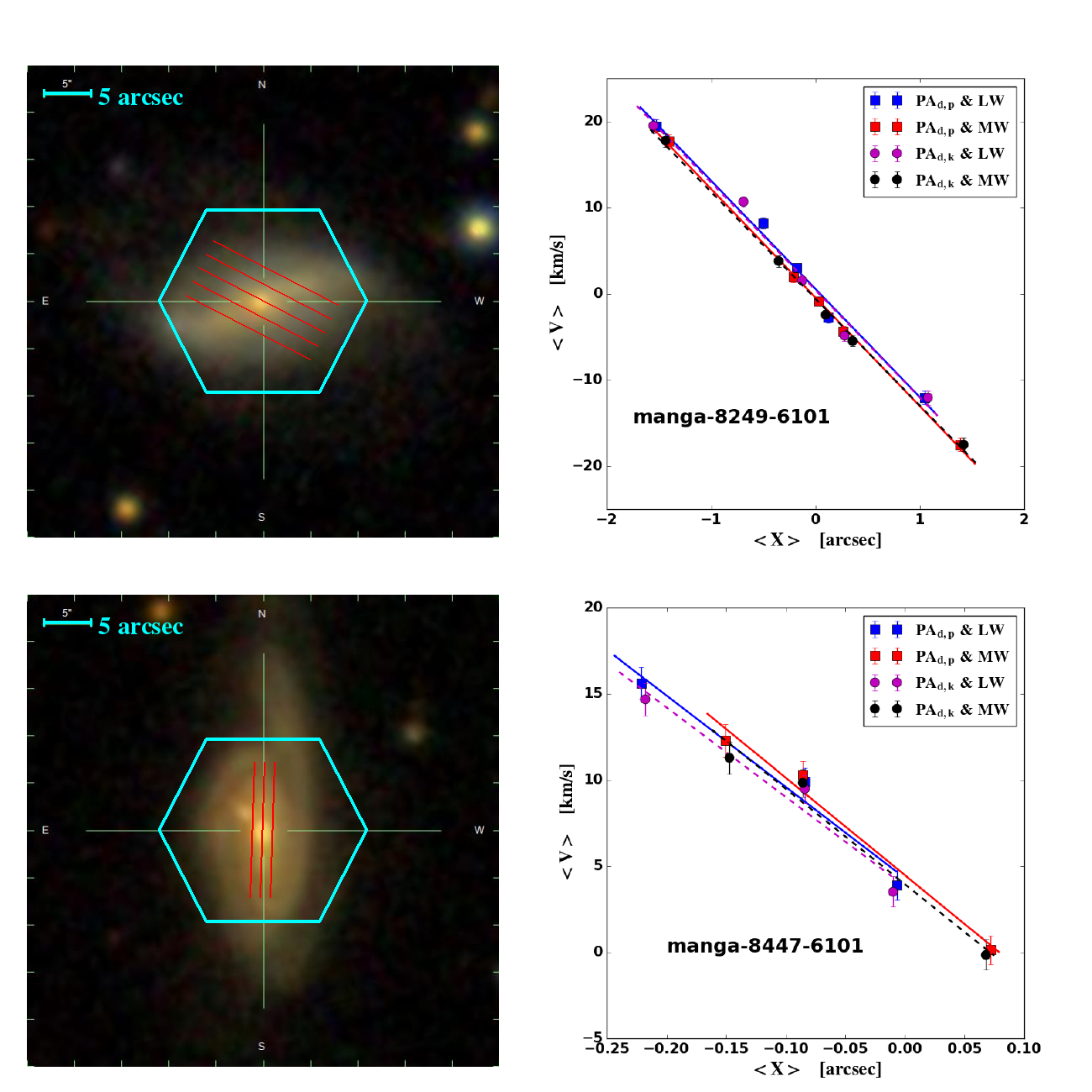}
\caption{Images (left column) and $\langle V \rangle$ vs. $\langle X \rangle$ plots (right column) for two galaxies which are ultrafast in both kinematic and photometric results: manga-8249-6101 (top row), manga-8447-6101 (bottom row). The red lines in left column are the pseudo slits aligned with photometric PAs. The blue and the red squares are the light-weighted and the mass-weighted mean position and mean velocity obtained using photometric PAs, respectively. While the magenta and black dots are those measured using kinematic PAs.}
\label{fig: ultra2}
\end{figure*}

The two galaxies that are both ultrafast measured using the kinematic PA and the photometric PA are manga-8249-6101 and manga-8447-6101. They are both in our refined sample, i.e. the two red dots with lower dark matter fraction in the kinematic results of Fig. \ref{dm_frac_ref}. Their PA differences between the kinematic PA and the photometric PA are 0.6 and 0.2 degree, respectively. Their images and $\langle V \rangle$ vs. $\langle X \rangle$ plots are shown in Fig. \ref{fig: ultra2}. The galaxy manga-8447-6101 has an 11.8 degrees PA difference between the disc and the bar and a companion close to the bar. These may influence the performance of the TW method though it has good linear fitting in the $\langle V \rangle$ vs. $\langle X \rangle$ plot. The galaxy manga-8249-6101 is a typical barred galaxy, and the slit length has covered the bright disc. The pattern speed measurement for this galaxy is reliable. Thus the probability that this galaxy has an ultrafast bar is high.

Ultrafast bars have also been found in several studies on barred galaxies. \cite[e.g.][]{aguerri2015} apply the TW method to 15 barred galaxies and find that about three galaxies have high probability to have an ultrafast bar. They discuss that their ultrafast bars are not likely due to the unknown errors in the TW method or due to the presence of dust lanes. Other measurements of the bar pattern speeds using the potential-density phase-shift have also found several ultrafast bars with ${\cal R} < 1$ \citep{buta2009}. They argue that some of them could be true ultrafast bars and not artefacts due to wrong measurements. Further research in this direction is needed.

As mentioned above, several previous investigations found ultrafast bars. In our sample we find one bar with high probability to be ultrafast (manga-8249-6101), but we need to emphasize that it could result from errors introduced by a number of approximations and hypotheses, especially those in measuring the bar pattern speed and the bar length. As shown in Fig. \ref{fig: R_dpa} and Fig. \ref{fig: ultra2}, there are only two ultrafast bars when the kinematic PA and photometric PA are consistent. This means the estimation accuracy of the disc PA influences rather strongly the estimation of the pattern speed. Nevertheless, in observations, both the kinematic PA and the photometric PA estimates rely on several assumptions and suffer from some problems. Therefore, more accurate PA estimation for deriving the pattern speed and more reliable bar length estimation are necessary for more accurate $\cal{R}$ ratios to ascertain the reality of ultrafast bars.

\subsection{Bar slowdown}
\label{ssec_slowdown}
Here we study the dependence of pattern speed on galaxy properties such as the dark matter fraction. We want to study the influence of these galaxy properties on the bar slowdown. There are no trends between ${\cal R}$ and the stellar age, metallicity and bar strength. These trends may have been buried by the combined action of many factors that can influence the evolution of bars.

The common factors that influence bar slowdown, i.e. the angular momentum loss from bars, include dark matter fraction, dark matter halo shape, initial gas fraction, bulge mass and size, disc-to-halo mass ratio and velocity dispersions of all components. Some of these have been studied in \cite{athanassoula2014} in detail. In outline, more highly triaxial halos and higher initial gas fractions will have higher final pattern speeds according to the comparison of simulations in Fig. 3 of \cite{athanassoula2014}. More extended bulges and colder spheroidal components will absorb more angular momentum from the bar, and thus will lead to slower pattern speeds. For the disc-to-halo mass ratio, a 20 percent difference can bring about 5.7 times difference in the amount of angular momentum transferred \citep[see Fig. 10 of][]{athanassoula2003}. It is possible that all these factors, even including the difference in galaxy total mass, can hide correlations between the pattern speed and other properties.

Another concern is that in the observations we do not know the evolution time and the initial pattern speed of the bar. Even though we can link the angular momentum loss of the bar to its strength \citep{athanassoula2003, athanassoula2013}, we do not know for how long it has been evolving and how fast it evolves. Therefore we do not know if our galaxies with the same pattern speeds are in the early stage of a slow evolution, or in the late stage of a fast evolution. The stellar ages and metallicities we calculate in previous sections can not characterise the bar evolution time, because stars can be trapped in the bar well after their formation, in which case the bar would be younger than its stars, or stars may be born in the bar region well after bar formation, in which case the bar would be much older than the stars. More studies need be done to understand the bar evolution time.

\section{Conclusions}
\label{sec_conclusions}
We have successfully applied the model--independent TW method to estimate the bar pattern speeds for 53 barred galaxies from their MaNGA IFU data. The sample of galaxies was selected from the MaNGA data sets of SDSS DR13 according to the PA difference between the bar and the disc, and the disc axis ratio. We have used the updated MaNGA DR14 IFU kinematic data for improvements resulting from the upgraded DAP and DRP packages. This sample is the largest so far to use the TW method, and spans a wide range of morphological types from SB0 to SBc with a peak at SBb-SBbc.

We have measured both photometric PAs and kinematic PAs for our sample galaxies. About 60\% of them are consistent within 3 degrees. For each type of PA, we have used light from the spectrum and mass from the SPS as weights in the integrals of the TW method. These two weightings give consistent pattern speed measurements, while two types of PAs can lead to different pattern speeds. For galaxies with larger PA differences ($\Delta PA > 3^{\circ}$), the kinematic PA usually gives lower pattern speeds than that given by the photometric PA, which leads to slower bars and fewer ultrafast bars. This may be due to the kinematic PA estimation, which by seeking to minimize any asymmetry in the velocity field, underestimates the true pattern speed. Thus a robust determination of the galaxy PA is essential for the estimation of bar pattern speed.

With three independent bar length estimations and circular velocities derived using the total mass profile from the JAM method, we can determine the dimensionless parameter ${\cal R}= R_{\rm CR}/a_{\rm b} = ( V_{\rm c}/ \Omega_{\rm p} ) / a_{\rm b}$. We have studied the dependence of the ${\cal R}$ parameter on galaxy properties such as the dark matter fraction inside the effective radius from the JAM method, the stellar age and metallicity from the SPS and the bar strength from the photometric image. We find a positive correlation that galaxies with larger bar lengths have larger bar strengths. No clear trends between the parameter ${\cal R}$ and these galaxy properties can be found in the results from both types of PAs. This could be due to the fact that, as suggested by simulations, the bar slowdown process and angular momentum exchange involve many factors, which prevent us from seeing any correlations.

In the future, MaNGA will obtain IFU data for about 10, 000 galaxies. A better sample can be defined using the difference between the photometric PA and the kinematic PA. This will help reduce the uncertainties in measuring the bar pattern speed and aid establishing more convincing correlations between the bar pattern speed and galaxy properties.

\section*{Acknowledgements}
We thank Cheng Cheng for help in using the IRAF {\tt ELLIPSE} fitting program to measure galaxy geometric parameters and photometric PAs. We also thank Yanfei Zou, Juntai Shen and Martin Bureau for discussions about this work, through which we have greatly improved our understanding of the TW method. We would also thank the reviewer for his insightful report. This work was supported by the National Key Basic Research and Development Program of China (No. 2018YFA0404501 to SM), and by the National Science Foundation of China (Grant No. 11333003, 11390372 and 11761131004 to SM). E. Athanassoula thanks the CNES for financial support. This work was granted access to the HPC resources of CINES under the allocation 2017-A0020407665 attributed by GENCI (Grand Equipement National de Calcul Intensif).

Funding for the Sloan Digital Sky Survey IV has been provided by the Alfred P. Sloan Foundation, the U.S. Department of Energy Office of Science, and the Participating Institutions. SDSS acknowledges support and resources from the Center for High-Performance Computing at the University of Utah. The SDSS web site is www.sdss.org.

SDSS is managed by the Astrophysical Research Consortium for the Participating Institutions of the SDSS Collaboration including the Brazilian Participation Group, the Carnegie Institution for Science, Carnegie Mellon University, the Chilean Participation Group, the French Participation Group, Harvard-Smithsonian Center for Astrophysics, Instituto de Astrof{\'i}sica de Canarias, The Johns Hopkins University, Kavli Institute for the Physics and Mathematics of the Universe (IPMU) / University of Tokyo, Lawrence Berkeley National Laboratory, Leibniz Institut f{\"u}r Astrophysik Potsdam (AIP), Max-Planck-Institut f{\"u}r Astronomie (MPIA Heidelberg), Max-Planck-Institut f{\"u}r Astrophysik (MPA Garching), Max-Planck-Institut f{\"u}r Extraterrestrische Physik (MPE), National Astronomical Observatories of China, New Mexico State University, New York University, University of Notre Dame, Observat{\'o}rio Nacional / MCTI, The Ohio State University, Pennsylvania State University, Shanghai Astronomical Observatory, United Kingdom Participation Group, Universidad Nacional Aut{\'o}noma de M{\'e}xico, University of Arizona, University of Colorado Boulder, University of Oxford, University of Portsmouth, University of Utah, University of Virginia, University of Washington, University of Wisconsin, Vanderbilt University, and Yale University.





\begin{thebibliography}{99}
\bibitem[Abolfathi et al.(2018)]{abolfathi2018} Abolfathi, B., Aguado, D.~S., Aguilar, G., et al.\ 2018, \apjs, 235, 42
\bibitem[Aguerri et al.(1998)]{aguerri1998} Aguerri, J.~A.~L., Beckman, J.~E., \& Prieto, M.\ 1998, \aj, 116, 2136
\bibitem[Aguerri(1999)]{aguerri1999} Aguerri, J.~A.~L.\ 1999, \aap, 351, 43 
\bibitem[Aguerri et al.(2000a)]{aguerri2000} Aguerri, J.~A.~L., Mu{\~n}oz-Tu{\~n}{\'o}n, C., Varela, A.~M., \& Prieto, M.\ 2000, \aap, 361, 841 
\bibitem[Aguerri et al.(2000b)]{aguerri2000b} Aguerri, J.~A.~L., Varela, A.~M., Prieto, M., \& Mu{\~n}oz-Tu{\~n}{\'o}n, C.\ 2000, \aj, 119, 1638
\bibitem[Aguerri et al.(2001)]{aguerri2001} Aguerri, J.~A.~L., Hunter, J.~H., Prieto, M., et al.\ 2001, \aap, 373, 786
\bibitem[Aguerri et al.(2003)]{aguerri2003} Aguerri, J.~A.~L., Debattista, V.~P., \& Corsini, E.~M.\ 2003, \mnras, 338, 465 
\bibitem[Aguerri et al.(2005)]{aguerri2005} Aguerri, J.~A.~L., Elias-Rosa, N., Corsini, E.~M., \& Mu{\~n}oz-Tu{\~n}{\'o}n, C.\ 2005, \aap, 434, 109 
\bibitem[Aguerri et al.(2009)]{aguerri2009} Aguerri, J.~A.~L., M{\'e}ndez-Abreu, J., \& Corsini, E.~M.\ 2009, \aap, 495, 491
\bibitem[Aguerri et al.(2015)]{aguerri2015} Aguerri, J.~A.~L., M{\'e}ndez-Abreu, J., Falc{\'o}n-Barroso, J., et al.\ 2015, \aap, 576, A102 
\bibitem[Albareti et al.(2017)]{albareti2017} Albareti, F.~D., Allende Prieto, C., Almeida, A., et al.\ 2017, \apjs, 233, 25
\bibitem[Antoja et al.(2014)]{antoja2014} Antoja, T., Helmi, A., Dehnen, W., et al.\ 2014, \aap, 563, A60
\bibitem[Athanassoula(1980)]{athanassoula1980} Athanassoula, E.\ 1980, \aap, 88, 184
\bibitem[Athanassoula et al.(1990)]{athanassoula1990} Athanassoula, E., Morin, S., Wozniak, H., et al.\ 1990, \mnras, 245, 130
\bibitem[Athanassoula(1992)]{athanassoula1992} Athanassoula, E.\ 1992, \mnras, 259, 345 
\bibitem[Athanassoula \& Misiriotis(2002)]{athanassoula2002} Athanassoula, E., \& Misiriotis, A.\ 2002, \mnras, 330, 35
\bibitem[Athanassoula(2003)]{athanassoula2003} Athanassoula, E.\ 2003, \mnras, 341, 1179 
\bibitem[Athanassoula et al.(2013a)]{athanassoula2013} Athanassoula, E., Machado, R.~E.~G., \& Rodionov, S.~A.\ 2013, \mnras, 429, 1949
\bibitem[Athanassoula(2013b)]{athanassoula2013b} Athanassoula, E.\ 2013, Secular Evolution of Galaxies, by Jes{\'u}s Falc{\'o}n-Barroso, and Johan H. Knapen, Cambridge, UK: Cambridge University Press, 2013, p.305
\bibitem[Athanassoula(2014)]{athanassoula2014} Athanassoula, E.\ 2014, \mnras, 438, L81 
\bibitem[Barazza et al.(2008)]{barazza2008} Barazza, F.~D., Jogee, S., \& Marinova, I.\ 2008, \apj, 675, 1194
\bibitem[Binney \& Tremaine(2008)]{binney2008} Binney, J., \& Tremaine, S.\ 2008, Galactic Dynamics: Second Edition, by James Binney and Scott Tremaine.~ISBN 978-0-691-13026-2 (HB).~Published by Princeton University Press, Princeton, NJ USA, 2008.
\bibitem[Blitz \& Spergel(1991)]{blitz1991} Blitz, L., \& Spergel, D.~N.\ 1991, \apj, 379, 631 
\bibitem[Bryant et al.(2015)]{bryant2015} Bryant, J.~J., Owers, M.~S., Robotham, A.~S.~G., et al.\ 2015, \mnras, 447, 2857
\bibitem[Bundy et al.(2015)]{bundy2015} Bundy, K., Bershady, M.~A., Law, D.~R., et al.\ 2015, \apj, 798, 7 
\bibitem[Bureau et al.(1999)]{bureau1999} Bureau, M., Freeman, K.~C., Pfitzner, D.~W., \& Meurer, G.~R.\ 1999, \aj, 118, 2158 
\bibitem[Buta(1986)]{buta1986} Buta, R.\ 1986, \apjs, 61, 609 
\bibitem[Buta et al.(1995)]{buta1995} Buta, R., van Driel, W., Braine, J., et al.\ 1995, \apj, 450, 593 
\bibitem[Buta \& Block(2001)]{buta2001} Buta, R., \& Block, D.~L.\ 2001, \apj, 550, 243 
\bibitem[Buta \& Zhang(2009)]{buta2009} Buta, R.~J., \& Zhang, X.\ 2009, \apjs, 182, 559 
\bibitem[Buta et al.(2015)]{buta2015} Buta, R.~J., Sheth, K., Athanassoula, E., et al.\ 2015, \apjs, 217, 32
\bibitem[Calzetti et al.(2000)]{calzetti2000} Calzetti, D., Armus, L., Bohlin, R.~C., et al.\ 2000, \apj, 533, 682 
\bibitem[Canzian(1993)]{canzian1993} Canzian, B.\ 1993, \apj, 414, 487
\bibitem[Canzian \& Allen(1997)]{canzian1997} Canzian, B., \& Allen, R.~J.\ 1997, \apj, 479, 723 
\bibitem[Cappellari \& Copin(2003)]{cappellari2003} Cappellari, M., \& Copin, Y.\ 2003, \mnras, 342, 345 
\bibitem[Cappellari \& Emsellem(2004)]{cappellari2004} Cappellari, M., \& Emsellem, E.\ 2004, \pasp, 116, 138 
\bibitem[Cappellari et al.(2007)]{cappellari2007} Cappellari, M., Emsellem, E., Bacon, R., et al.\ 2007, \mnras, 379, 418
\bibitem[Cappellari(2008)]{cappellari2008} Cappellari, M.\ 2008, \mnras, 390, 71
\bibitem[Cappellari et al.(2013)]{cappellari2013} Cappellari, M., Scott, N., Alatalo, K., et al.\ 2013, \mnras, 432, 1709
\bibitem[Cappellari(2017)]{cappellari2017} Cappellari, M.\ 2017, \mnras, 466, 798 
\bibitem[Chemin \& Hernandez(2009)]{chemin2009} Chemin, L., \& Hernandez, O.\ 2009, \aap, 499, L25 
\bibitem[Combes \& Sanders(1981)]{combes1981} Combes, F., \& Sanders, R.~H.\ 1981, \aap, 96, 164 
\bibitem[Conroy(2013)]{conroy2013} Conroy, C.\ 2013, \araa, 51, 393
\bibitem[Contopoulos(1980)]{contopoulos1980} Contopoulos, G.\ 1980, \aap, 81, 198 
\bibitem[Corsini et al.(2003)]{corsini2003} Corsini, E.~M., Debattista, V.~P., \& Aguerri, J.~A.~L.\ 2003, \apjl, 599, L29 
\bibitem[Corsini et al.(2007)]{corsini2007} Corsini, E.~M., Aguerri, J.~A.~L., Debattista, V.~P., et al.\ 2007, \apjl, 659, L121 
\bibitem[Corsini(2011)]{corsini2011} Corsini, E.~M.\ 2011, Memorie della Societa Astronomica Italiana Supplementi, 18, 23
\bibitem[Debattista \& Sellwood(1998)]{debattista1998} Debattista, V.~P., \& Sellwood, J.~A.\ 1998, \apjl, 493, L5
\bibitem[Debattista \& Sellwood(2000)]{debattista2000} Debattista, V.~P., \& Sellwood, J.~A.\ 2000, \apj, 543, 704 
\bibitem[Debattista et al.(2002)]{debattista2002} Debattista, V.~P., Corsini, E.~M., \& Aguerri, J.~A.~L.\ 2002, \mnras, 332, 65
\bibitem[Debattista(2003)]{debattista2003} Debattista, V.~P.\ 2003, \mnras, 342, 1194 
\bibitem[Debattista \& Williams(2004)]{debattista2004} Debattista, V.~P., \& Williams, T.~B.\ 2004, \apj, 605, 714
\bibitem[de Vaucouleurs(1964)]{devaucouleurs1964} de Vaucouleurs, G.\ 1964, The Galaxy and the Magellanic Clouds, 20, 195 
\bibitem[Drory et al.(2015)]{drory2015} Drory, N., MacDonald, N., Bershady, M.~A., et al.\ 2015, \aj, 149, 77
\bibitem[Duval \& Athanassoula(1983)]{duval1983} Duval, M.~F., \& Athanassoula, E.\ 1983, \aap, 121, 297
\bibitem[Elmegreen(1996)]{elmegreen1996} Elmegreen, B.\ 1996, IAU Colloq.~157: Barred Galaxies, 91, 197
\bibitem[Emsellem et al.(1994)]{emsellem1994} Emsellem, E., Monnet, G., \& Bacon, R.\ 1994, \aap, 285, 723
\bibitem[Emsellem et al.(2006)]{emsellem2006} Emsellem, E., Fathi, K., Wozniak, H., et al.\ 2006, \mnras, 365, 367 
\bibitem[England et al.(1990)]{england1990} England, M.~N., Gottesman, S.~T., \& Hunter, J.~H., Jr.\ 1990, \apj, 348, 456
\bibitem[Erwin(2005)]{erwin2005} Erwin, P.\ 2005, \mnras, 364, 283 
\bibitem[Eskridge et al.(2000)]{eskridge2000} Eskridge, P.~B., Frogel, J.~A., Pogge, R.~W., et al.\ 2000, \aj, 119, 536
\bibitem[Falc{\'o}n-Barroso et al.(2011)]{falcon2011} Falc{\'o}n-Barroso, J., S{\'a}nchez-Bl{\'a}zquez, P., Vazdekis, A., et al.\ 2011, \aap, 532, A95 
\bibitem[Fathi et al.(2007)]{fathi2007} Fathi, K., Toonen, S., Falc{\'o}n-Barroso, J., et al.\ 2007, \apjl, 667, L137 
\bibitem[Fathi et al.(2009)]{fathi2009} Fathi, K., Beckman, J.~E., Pi{\~n}ol-Ferrer, N., et al.\ 2009, \apj, 704, 1657
\bibitem[Font et al.(2011)]{font2011} Font, J., Beckman, J.~E., Epinat, B., et al.\ 2011, \apjl, 741, L14 
\bibitem[Font et al.(2014)]{font2014} Font, J., Beckman, J.~E., Querejeta, M., et al.\ 2014, \apjs, 210, 2
\bibitem[Gabbasov et al.(2009)]{gabbasov2009} Gabbasov, R.~F., Repetto, P., \& Rosado, M.\ 2009, \apj, 702, 392 
\bibitem[Gadotti(2008)]{gadotti2008} Gadotti, D.~A.\ 2008, \mnras, 384, 420 
\bibitem[Gadotti(2011)]{gadotti2011} Gadotti, D.~A.\ 2011, \mnras, 415, 3308 
\bibitem[Garcia-Burillo et al.(1993)]{garciaburillo1993} Garcia-Burillo, S., Combes, F., \& Gerin, M.\ 1993, \aap, 274, 148
\bibitem[Gerssen et al.(1999)]{gerssen1999} Gerssen, J., Kuijken, K., \& Merrifield, M.~R.\ 1999, \mnras, 306, 926 
\bibitem[Gerssen et al.(2003)]{gerssen2003} Gerssen, J., Kuijken, K., \& Merrifield, M.~R.\ 2003, \mnras, 345, 261
\bibitem[Gerssen \& Debattista(2007)]{gerssen2007} Gerssen, J., \& Debattista, V.~P.\ 2007, \mnras, 378, 189 
\bibitem[Gunn et al.(2006)]{gunn2006} Gunn, J.~E., Siegmund, W.~A., Mannery, E.~J., et al.\ 2006, \aj, 131, 2332 
\bibitem[Hernandez et al.(2005)]{hernandez2005} Hernandez, O., Wozniak, H., Carignan, C., et al.\ 2005, \apj, 632, 253 
\bibitem[Hinshaw et al.(2013)]{hinshaw2013} Hinshaw, G., Larson, D., Komatsu, E., et al.\ 2013, \apjs, 208, 19 
\bibitem[Hoyle et al.(2011)]{hoyle2011} Hoyle, B., Masters, K.~L., Nichol, R.~C., et al.\ 2011, \mnras, 415, 3627
\bibitem[Hunter et al.(1988)]{hunter1988} Hunter, J.~H., Jr., England, M.~N., Gottesman, S.~T., Ball, R., \& Huntley, J.~M.\ 1988, \apj, 324, 721 
\bibitem[Jedrzejewski(1987)]{jedrzejewski1987} Jedrzejewski, R.~I.\ 1987, \mnras, 226, 747 
\bibitem[Krajnovi{\'c} et al.(2006)]{krajnovic2006} Krajnovi{\'c}, D., Cappellari, M., de Zeeuw, P.~T., \& Copin, Y.\ 2006, \mnras, 366, 787 
\bibitem[Krajnovi{\'c} et al.(2011)]{krajnovic2011} Krajnovi{\'c}, D., Emsellem, E., Cappellari, M., et al.\ 2011, \mnras, 414, 2923 
\bibitem[Kent(1987)]{kent1987} Kent, S.~M.\ 1987, \aj, 93, 1062 
\bibitem[Kim et al.(2016)]{kim2016} Kim, T., Gadotti, D.~A., Athanassoula, E., et al.\ 2016, \mnras, 462, 3430 
\bibitem[Knapen et al.(2000)]{knapen2000} Knapen, J.~H., Shlosman, I., \& Peletier, R.~F.\ 2000, \apj, 529, 93
\bibitem[Kormendy(1979)]{kormendy1979} Kormendy, J.\ 1979, \apj, 227, 714
\bibitem[Kruk et al.(2018)]{kruk2018} Kruk, S.~J., Lintott, C.~J., Bamford, S.~P., et al.\ 2018, \mnras, 473, 4731
\bibitem[Laine \& Heller(1999)]{laine1999} Laine, S., \& Heller, C.~H.\ 1999, \mnras, 308, 557 
\bibitem[Laine et al.(2002)]{laine2002} Laine, S., Shlosman, I., Knapen, J.~H., \& Peletier, R.~F.\ 2002, \apj, 567, 97 
\bibitem[Laurikainen \& Salo(2002)]{laurikainen2002} Laurikainen, E., \& Salo, H.\ 2002, \mnras, 337, 1118
\bibitem[Laurikainen et al.(2005)]{laurikainen2005} Laurikainen, E., Salo, H., \& Buta, R.\ 2005, \mnras, 362, 1319 
\bibitem[Laurikainen et al.(2007)]{laurikainen2007} Laurikainen, E., Salo, H., Buta, R., \& Knapen, J.~H.\ 2007, \mnras, 381, 401
\bibitem[Laurikainen et al.(2009)]{laurikainen2009} Laurikainen, E., Salo, H., Buta, R., \& Knapen, J.~H.\ 2009, \apjl, 692, L34 
\bibitem[Law et al.(2015)]{law2015} Law, D.~R., Yan, R., Bershady, M.~A., et al.\ 2015, \aj, 150, 19 
\bibitem[Law et al.(2016)]{law2016} Law, D.~R., Cherinka, B., Yan, R., et al.\ 2016, \aj, 152, 83
\bibitem[Li et al.(2016)]{Li2016} Li, H., Li, R., Mao, S., et al.\ 2016, \mnras, 455, 3680
\bibitem[Li et al.(2017)]{Li2017} Li, H., Ge, J., Mao, S., et al.\ 2017, \apj, 838, 77 
\bibitem[Lindblad et al.(1996a)]{lindblad1996a} Lindblad, P.~A.~B., Lindblad, P.~O., \& Athanassoula, E.\ 1996, \aap, 313, 65 
\bibitem[Lindblad \& Kristen(1996b)]{lindblad1996b} Lindblad, P.~A.~B., \& Kristen, H.\ 1996, \aap, 313, 733 
\bibitem[Long et al.(2013)]{long2013} Long, R.~J., Mao, S., Shen, J., \& Wang, Y.\ 2013, \mnras, 428, 3478 
\bibitem[Marinova \& Jogee(2007)]{marinova2007} Marinova, I., \& Jogee, S.\ 2007, \apj, 659, 1176
\bibitem[Marquez et al.(1996)]{marquez1996} Marquez, I., Moles, M., \& Masegosa, J.\ 1996, \aap, 310, 401
\bibitem[M{\'a}rquez et al.(1999)]{marquez1999} M{\'a}rquez, I., Durret, F., Gonz{\'a}lez Delgado, R.~M., et al.\ 1999, \aaps, 140, 1
\bibitem[Martin(1995)]{martin1995} Martin, P.\ 1995, \aj, 109, 2428
\bibitem[Martinet \& Friedli(1997)]{martinet1997} Martinet, L., \& Friedli, D.\ 1997, \aap, 323, 363 
\bibitem[Martinez-Valpuesta et al.(2006)]{martinezvalpuesta2006} Martinez-Valpuesta, I., Shlosman, I., \& Heller, C.\ 2006, \apj, 637, 214 
\bibitem[Masters et al.(2011)]{masters2011} Masters, K.~L., Nichol, R.~C., Hoyle, B., et al.\ 2011, \mnras, 411, 2026
\bibitem[Men{\'e}ndez-Delmestre et al.(2007)]{menendezdelmestre2007} Men{\'e}ndez-Delmestre, K., Sheth, K., Schinnerer, E., Jarrett, T.~H., \& Scoville, N.~Z.\ 2007, \apj, 657, 790
\bibitem[Merrifield \& Kuijken(1995)]{merrifield1995} Merrifield, M.~R., \& Kuijken, K.\ 1995, \mnras, 274, 933
\bibitem[Michel-Dansac \& Wozniak(2006)]{micheldansac2006} Michel-Dansac, L., \& Wozniak, H.\ 2006, \aap, 452, 97
\bibitem[Mu{\~n}oz-Tu{\~n}{\'o}n et al.(2004)]{munoztunon2004} Mu{\~n}oz-Tu{\~n}{\'o}n, C., Caon, N., \& Aguerri, J.~A.~L.\ 2004, \aj, 127, 58 
\bibitem[Nair \& Abraham(2010)]{nair2010} Nair, P.~B., \& Abraham, R.~G.\ 2010, \apjl, 714, L260
\bibitem[Ohta et al.(1990)]{ohta1990} Ohta, K., Hamabe, M., \& Wakamatsu, K.-I.\ 1990, \apj, 357, 71
\bibitem[P{\'e}rez et al.(2004)]{perez2004} P{\'e}rez, I., Fux, R., \& Freeman, K.\ 2004, \aap, 424, 799 
\bibitem[P{\'e}rez et al.(2012)]{perez2012} P{\'e}rez, I., Aguerri, J.~A.~L., \& M{\'e}ndez-Abreu, J.\ 2012, \aap, 540, A103 
\bibitem[Portail et al.(2015)]{portail2015} Portail, M., Wegg, C., Gerhard, O., \& Martinez-Valpuesta, I.\ 2015, \mnras, 448, 713
\bibitem[Prieto et al.(1997)]{prieto1997} Prieto, M., Gottesman, S.~T., Aguerri, J.-A.~L., \& Varela, A.-M.\ 1997, \aj, 114, 1413 
\bibitem[Prieto et al.(2001)]{prieto2001} Prieto, M., Aguerri, J.~A.~L., Varela, A.~M., \& Mu{\~n}oz-Tu{\~n}{\'o}n, C.\ 2001, \aap, 367, 405 
\bibitem[Puerari \& Dottori(1997)]{puerari1997} Puerari, I., \& Dottori, H.\ 1997, \apjl, 476, L73
\bibitem[Quillen et al.(1994)]{quillen1994} Quillen, A.~C., Frogel, J.~A., \& Gonzalez, R.~A.\ 1994, \apj, 437, 162
\bibitem[Rand \& Wallin(2004)]{rand2004} Rand, R.~J., \& Wallin, J.~F.\ 2004, \apj, 614, 142 
\bibitem[Rautiainen et al.(2008)]{rautiainen2008} Rautiainen, P., Salo, H., \& Laurikainen, E.\ 2008, \mnras, 388, 1803 
\bibitem[Reyes et al.(2011)]{reyes2011} Reyes, R., Mandelbaum, R., Gunn, J.~E., Pizagno, J., \& Lackner, C.~N.\ 2011, \mnras, 417, 2347 
\bibitem[Salo et al.(2010)]{salo2010} Salo, H., Laurikainen, E., Buta, R., \& Knapen, J.~H.\ 2010, \apjl, 715, L56
\bibitem[Salpeter(1955)]{salpeter1955} Salpeter, E.~E.\ 1955, \apj, 121, 161 
\bibitem[S{\'a}nchez-Bl{\'a}zquez et al.(2006)]{sanchez2006} S{\'a}nchez-Bl{\'a}zquez, P., Peletier, R.~F., Jim{\'e}nez-Vicente, J., et al.\ 2006, \mnras, 371, 703 
\bibitem[S{\'a}nchez et al.(2012)]{sanchez2012} S{\'a}nchez, S.~F., Kennicutt, R.~C., Gil de Paz, A., et al.\ 2012, \aap, 538, A8 
\bibitem[Sancisi et al.(1979)]{sancisi1979} Sancisi, R., Allen, R.~J., \& Sullivan, W.~T., III 1979, \aap, 78, 217
\bibitem[Sanders \& Tubbs(1980)]{sanders1980} Sanders, R.~H., \& Tubbs, A.~D.\ 1980, \apj, 235, 803 
\bibitem[Sellwood \& Sparke(1988)]{sellwood1988} Sellwood, J.~A., \& Sparke, L.~S.\ 1988, \mnras, 231, 25P
\bibitem[Sellwood(2006)]{sellwood2006a} Sellwood, J.~A.\ 2006, \apj, 637, 567 
\bibitem[Sellwood \& Debattista(2006)]{sellwood2006b} Sellwood, J.~A., \& Debattista, V.~P.\ 2006, \apj, 639, 868
\bibitem[Sempere et al.(1995a)]{sempere1995a} Sempere, M.~J., Combes, F., \& Casoli, F.\ 1995, \aap, 299, 371 
\bibitem[Sempere et al.(1995b)]{sempere1995b} Sempere, M.~J., Garcia-Burillo, S., Combes, F., \& Knapen, J.~H.\ 1995, \aap, 296, 45 
\bibitem[Sheth et al.(2003)]{sheth2003} Sheth, K., Regan, M.~W., Scoville, N.~Z., \& Strubbe, L.~E.\ 2003, \apjl, 592, L13 
\bibitem[Smee et al.(2013)]{smee2013} Smee, S.~A., Gunn, J.~E., Uomoto, A., et al.\ 2013, \aj, 146, 32 
\bibitem[Sygnet et al.(1988)]{sygnet1988} Sygnet, J.~F., Tagger, M., Athanassoula, E., \& Pellat, R.\ 1988, \mnras, 232, 733
\bibitem[Tagger et al.(1987)]{tagger1987} Tagger, M., Sygnet, J.~F., Athanassoula, E., \& Pellat, R.\ 1987, \apjl, 318, L43
\bibitem[Tremaine \& Weinberg(1984)]{tremaine1984} Tremaine, S., \& Weinberg, M.~D.\ 1984, \apjl, 282, L5 
\bibitem[Treuthardt et al.(2007)]{treuthardt2007} Treuthardt, P., Buta, R., Salo, H., \& Laurikainen, E.\ 2007, \aj, 134, 1195 
\bibitem[Treuthardt et al.(2008)]{treuthardt2008} Treuthardt, P., Salo, H., Rautiainen, P., \& Buta, R.\ 2008, \aj, 136, 300
\bibitem[Tully \& Fisher(1977)]{tully1977} Tully, R.~B., \& Fisher, J.~R.\ 1977, \aap, 54, 661 
\bibitem[van Albada \& Sanders(1982)]{vanalbada1982} van Albada, T.~S., \& Sanders, R.~H.\ 1982, \mnras, 201, 303
\bibitem[Vazdekis et al.(2010)]{vazdekis2010} Vazdekis, A., S{\'a}nchez-Bl{\'a}zquez, P., Falc{\'o}n-Barroso, J., et al.\ 2010, \mnras, 404, 1639 
\bibitem[Vega Beltran et al.(1997)]{vegabeltran1997} Vega Beltran, J.~C., Corsini, E.~M., Pizzella, A., \& Bertola, F.\ 1997, \aap, 324, 485 
\bibitem[Villa-Vargas et al.(2009)]{villavargas2009} Villa-Vargas, J., Shlosman, I., \& Heller, C.\ 2009, \apj, 707, 218
\bibitem[Wake et al.(2017)]{wake2017} Wake, D.~A., Bundy, K., Diamond-Stanic, A.~M., et al.\ 2017, \aj, 154, 86
\bibitem[Walcher et al.(2011)]{walcher2011} Walcher, J., Groves, B., Budav{\'a}ri, T., \& Dale, D.\ 2011, \apss, 331, 1 
\bibitem[Weinberg(1985)]{weinberg1985} Weinberg, M.~D.\ 1985, \mnras, 213, 451 
\bibitem[Weiner et al.(2001)]{weiner2001} Weiner, B.~J., Sellwood, J.~A., \& Williams, T.~B.\ 2001, \apj, 546, 931 
\bibitem[Weinzirl et al.(2009)]{weinzirl2009} Weinzirl, T., Jogee, S., Khochfar, S., Burkert, A., \& Kormendy, J.\ 2009, \apj, 696, 411
\bibitem[Whyte et al.(2002)]{whyte2002} Whyte, L.~F., Abraham, R.~G., Merrifield, M.~R., et al.\ 2002, \mnras, 336, 1281
\bibitem[Willett et al.(2013)]{willett2013} Willett, K.~W., Lintott, C.~J., Bamford, S.~P., et al.\ 2013, \mnras, 435, 2835 
\bibitem[Wozniak et al.(1995)]{wozniak1995} Wozniak, H., Friedli, D., Martinet, L., Martin, P., \& Bratschi, P.\ 1995, \aaps, 111, 115
\bibitem[Yan et al.(2016)]{yan2016} Yan, R., Bundy, K., Law, D.~R., et al.\ 2016, \aj, 152, 197
\bibitem[Zimmer et al.(2004)]{zimmer2004} Zimmer, P., Rand, R.~J., \& McGraw, J.~T.\ 2004, \apj, 607, 285 
\end{thebibliography}


\clearpage
\begin{table*}
\caption{Main parameters of 53 selected MaNGA barred galaxies.}
\large
\label{tab: galaxies}
\begin{tabular}{l c c c c c c c c}
\hline
\hline
Plate-ifu & RA  & DEC & Morph. Type & $R_{\rm e}$ & $M_{\rm r}$ & $z$ & $V_{\rm c,jam}$ & $f_{\rm dm}(<R_{\rm e})$ \\
          & ($^{\circ}$) & ($^{\circ}$) &    & ($''$) & (mag) & & (${\rm km/s}$) & \\
   (1) & (2) & (3) & (4) & (5) & (6) & (7) & (8) & (9) \\
\hline 
7495-12704    &    205.4384 &    27.0048 & SBbc   &  8.65 &   -21.40 & 0.0289 &  202 & 0.56 \\
 7962-12703    &    261.2173 &    28.0783 & SBab   &  8.34 &   -22.33 & 0.0477 &  264 & 0.29 \\
 7990-3704    &    262.0749 &    56.7748 & SB0   &  3.83 &   -20.15 & 0.0291 &  149 & 0.47 \\
 7990-9101    &    259.7555 &    57.1735 & SBc   &  4.51 &   -19.77 & 0.0280 &  119 & 0.60 \\
 7992-6104    &    255.2795 &    64.6769 & SBc   &  8.78 &   -20.31 & 0.0271 &  126 & 0.50 \\
 8082-6102    &    49.9459 &     0.5846 & SB0   &  6.91 &   -21.46 & 0.0242 &  235 & 0.26 \\
 8083-6102    &    51.1150 &    -0.0863 & SBa   &  4.70 &   -21.62 & 0.0365 &  281 & 0.32 \\
 8083-12704    &    50.6968 &     0.1494 & SBbc   &  13.32 &   -21.03 & 0.0228 &  102 & 0.53 \\
 8133-3701    &    112.0793 &    43.3021 & SBb   &  2.39 &   -20.10 & 0.0437 &  140 & 0.57 \\
 8134-6102    &    114.9245 &    45.9126 & SB0a   &  5.98 &   -21.40 & 0.0320 &  282 & 0.48 \\
 8137-9102    &    117.0386 &    43.5907 & SBb   &  6.68 &   -21.07 & 0.0311 &  133 & 0.53 \\
 8140-12701    &    116.9303 &    41.3864 & SBa   &  5.69 &   -20.61 & 0.0286 &  173 & 0.29 \\
 8140-12703    &    117.8985 &    42.8801 & SBb   &  9.85 &   -21.87 & 0.0320 &  205 & 0.46 \\
 8243-6103    &    129.1749 &    53.7272 & SB0   &  4.75 &   -21.65 & 0.0315 &  302 & 0.20 \\
 8244-3703    &    131.9928 &    51.6010 & SB0   &  2.50 &   -21.03 & 0.0483 &  204 & 0.35 \\
 8247-3701    &    136.6714 &    41.3651 & SB0a   &  4.83 &   -20.59 & 0.0250 &  124 & 0.00 \\
 8249-6101    &    137.5625 &    46.2933 & SBc   &  4.64 &   -20.27 & 0.0267 &  134 & 0.36 \\
 8254-9101    &    161.2617 &    43.7048 & SBa   &  8.00 &   -21.78 & 0.0253 &  313 & 0.28 \\
 8256-6101    &    163.7348 &    41.4985 & SBa   &  6.06 &   -20.79 & 0.0246 &  199 & 0.49 \\
 8257-3703    &    166.6557 &    46.0388 & SBb   &  4.03 &   -20.34 & 0.0250 &  194 & 0.03 \\
 8257-6101    &    165.2613 &    44.8882 & SBc   &  5.77 &   -20.86 & 0.0294 &  169 & 0.53 \\
 8274-6101    &    163.7348 &    41.4985 & SBa   &  6.09 &   -20.79 & 0.0246 &  195 & 0.52 \\
 8312-12702    &    245.2709 &    39.9174 & SBc   &  7.23 &   -21.24 & 0.0320 &  143 & 0.34 \\
 8312-12704    &    247.3041 &    41.1509 & SBb   &  7.47 &   -21.00 & 0.0296 &  123 & 0.47 \\
 8313-9101    &    239.6975 &    41.9381 & SBb   &  6.76 &   -21.87 & 0.0387 &  226 & 0.45 \\
 8317-12704    &    193.7040 &    44.1556 & SBa   &  7.14 &   -22.68 & 0.0543 &  338 & 0.42 \\
 8318-12703    &    196.2324 &    47.5036 & SBb   &  9.09 &   -22.21 & 0.0393 &  236 & 0.51 \\
 8320-6101    &    206.6275 &    22.7060 & SBb   &  5.22 &   -20.37 & 0.0266 &  189 & 0.57 \\
 8326-3704    &    214.8502 &    45.9008 & SBa   &  3.83 &   -20.25 & 0.0265 &  120 & 0.52 \\
 8326-6102    &    215.0179 &    47.1213 & SBb   &  2.95 &   -22.06 & 0.0704 &  237 & 0.00 \\
 8330-12703    &    203.3746 &    40.5297 & SBbc   &  7.51 &   -20.67 & 0.0269 &  140 & 0.58 \\
 8335-12701    &    215.3953 &    40.3581 & SBb   &  4.39 &   -21.66 & 0.0633 &  241 & 0.63 \\
 8439-6102    &    142.7782 &    49.0797 & SBab   &  4.54 &   -21.64 & 0.0339 &  203 & 0.07 \\
 8439-12702    &    141.5393 &    49.3102 & SBa   &  8.10 &   -21.57 & 0.0269 &  239 & 0.40 \\
 8440-12704    &    136.1423 &    41.3978 & SBb   &  4.56 &   -21.12 & 0.0270 &  215 & 0.42 \\
 8447-6101    &    206.1333 &    40.2400 & SBb   &  4.48 &   -22.89 & 0.0753 &  364 & 0.23 \\
 8452-3704    &    157.5390 &    47.2784 & SBc   &  4.34 &   -19.97 & 0.0251 &  157 & 0.71 \\
 8452-12703    &    156.8057 &    48.2448 & SBb   &  8.13 &   -22.83 & 0.0610 &  214 & 0.18 \\
 8481-12701    &    236.7613 &    54.3409 & SBa   &  4.59 &   -21.91 & 0.0669 &  262 & 0.43 \\
 8482-9102    &    242.9559 &    49.2287 & SBb   &  3.54 &   -21.59 & 0.0580 &  219 & 0.39 \\
 8482-12703    &    245.5031 &    49.5208 & SBbc   &  9.65 &   -22.21 & 0.0496 &  164 & 0.39 \\
 8482-12705    &    244.2167 &    50.2822 & SBb   &  7.39 &   -22.06 & 0.0417 &  237 & 0.51 \\
 8486-6101    &    238.0396 &    46.3198 & SBc   &  3.56 &   -21.57 & 0.0589 &  188 & 0.21 \\
 8548-6102    &    245.5224 &    46.6242 & SBc   &  3.85 &   -20.83 & 0.0478 &  162 & 0.69 \\
 8548-6104    &    245.7474 &    46.6753 & SBc   &  2.73 &   -20.47 & 0.0480 &  177 & 0.30 \\
 8549-12702    &    241.2714 &    45.4430 & SBb   &  6.72 &   -22.03 & 0.0433 &  246 & 0.22 \\
 8588-3701    &    248.1406 &    39.1310 & SBb   &  4.43 &   -22.88 & 0.1303 &  267 & 0.05 \\
 8601-12705    &    250.1231 &    39.2351 & SBc   &  6.66 &   -21.21 & 0.0297 &  178 & 0.53 \\
 8603-12701    &    248.1406 &    39.1310 & SBb   &  4.42 &   -22.88 & 0.1303 &  265 & 0.06 \\
 8603-12703    &    247.2826 &    40.6650 & SBa   &  6.55 &   -21.04 & 0.0300 &  148 & 0.36 \\
 8604-12703    &    247.7642 &    39.8385 & SBab   &  9.08 &   -21.67 & 0.0305 &  231 & 0.30 \\
 8612-6104    &    255.0069 &    38.8160 & SBb   &  8.60 &   -21.83 & 0.0356 &  187 & 0.31 \\
 8612-12702    &    253.9464 &    39.3105 & SBc   &  8.26 &   -22.60 & 0.0631 &  208 & 0.39 \\ 
\hline
\end{tabular}
\begin{tablenotes}
\small
\item Note: Columns are: (1) MaNGA ID of galaxy; (2) galaxy right ascension; (3) galaxy declination; (4) morphological type; (5) effective radius in the r-band from SDSS-DR9; (6)  absolute r-band magnitude from SDSS-DR9; (7) redshift of the galaxy; (8) circular velocity $V_{\rm c,jam}$ from the total mass density profile of JAM modelling; (9) dark matter fraction inside one effective radius from JAM modelling.
\end{tablenotes}
\end{table*}

\clearpage
\begin{table*}
\caption{Geometric parameters of 53 selected MaNGA barred galaxies.}
\label{tab: geom}
\begin{tabular}{l c c c c c c c c c c}
\hline
\hline
Plate-ifu & $i$  & $PA_{\rm d,p}$ & $PA_{\rm d,k}$ & $PA_{\rm b}$ & $a_{\rm b,e}$ & $a_{\rm b,pa}$ & $a_{\rm b,f}$ & $a_{\rm b}$ & $A_{\rm 2}$ & $\Delta\mu$ \\
       & ($^{\circ}$) & ($^{\circ}$) & ($^{\circ}$) & ($^{\circ}$) &  ($''$) & ($''$) & ($''$) & ($''$) &  & \\
   (1) & (2) & (3) & (4) & (5) & (6) & (7) & (8) & (9) & (10) & (11)\\
\hline
 7495-12704    & 52.2$\pm$0.6 &    173.4$\pm$0.8 &    173.0 $\pm$0.6 &    144.9$\pm$0.7 &   6.5 &   8.6 &   7.4 &   7.5$^{+1.1}_{-1.0}$  &   0.37 &   0.69 \\ 
 7962-12703    & 61.2$\pm$1.2 &    32.4$\pm$0.4 &    37.0 $\pm$0.9 &    49.8$\pm$0.4 &   13.1 &   19.8 &   15.3 &   16.1$^{+3.7}_{-3.0}$  &   0.65 &   1.44 \\ 
 7990-3704    & 39.4$\pm$1.4 &    11.6$\pm$3.8 &    15.2 $\pm$3.4 &    36.6$\pm$0.8 &   3.2 &   4.4 &   4.2 &   3.9$^{+0.5}_{-0.7}$  &   0.29 &   0.50 \\ 
 7990-9101    & 71.8$\pm$0.2 &    21.0$\pm$0.6 &    20.0 $\pm$3.8 &    33.6$\pm$0.8 &   5.0 &   8.0 &   7.8 &   6.9$^{+1.1}_{-1.9}$  &   0.37 &   0.56 \\ 
 7992-6104    & 46.7$\pm$1.8 &    7.9$\pm$1.4 &     6.0 $\pm$2.8 &    153.2$\pm$1.1 &   7.6 &   10.6 &   8.9 &   9.0$^{+1.6}_{-1.4}$  &   0.80 &   2.12 \\ 
 8082-6102    & 41.3$\pm$0.5 &    98.7$\pm$0.9 &    99.0 $\pm$0.9 &    19.1$\pm$0.5 &   6.6 &   8.6 &   7.6 &   7.6$^{+1.0}_{-1.0}$  &   0.59 &   1.21 \\ 
 8083-6102    & 70.4$\pm$0.2 &    65.7$\pm$0.3 &    62.8 $\pm$0.9 &    76.6$\pm$0.5 &   5.4 &   7.1 &   8.5 &   7.0$^{+1.5}_{-1.6}$  &   0.63 &   1.43 \\ 
 8083-12704    & 41.7$\pm$0.9 &    144.1$\pm$3.6 &    167.0 $\pm$1.4 &    119.6$\pm$1.5 &   6.7 &   7.6 &   5.5 &   6.6$^{+1.0}_{-1.1}$  &   0.27 &   0.53 \\ 
 8133-3701    & 44.6$\pm$1.1 &    101.2$\pm$1.8 &    102.8 $\pm$3.5 &    145.1$\pm$1.0 &   3.1 &   4.4 &   5.1 &   4.2$^{+0.9}_{-1.1}$  &   0.48 &   1.03 \\ 
 8134-6102    & 53.8$\pm$0.9 &    87.4$\pm$0.4 &    93.0 $\pm$0.8 &    11.0$\pm$1.2 &   10.8 &   14.5 &   9.7 &   11.7$^{+2.8}_{-2.0}$  &   0.74 &   1.93 \\ 
 8137-9102    & 43.3$\pm$2.2 &    136.8$\pm$2.7 &    132.8 $\pm$1.9 &    126.6$\pm$0.6 &   9.6 &   12.5 &   12.3 &   11.5$^{+1.0}_{-1.9}$  &   0.62 &   1.43 \\ 
 8140-12701    & 37.8$\pm$1.3 &    60.2$\pm$1.6 &    62.8 $\pm$1.8 &    128.0$\pm$0.8 &   9.2 &   11.2 &   8.4 &   9.6$^{+1.6}_{-1.2}$  &   0.68 &   1.46 \\ 
 8140-12703    & 55.0$\pm$0.6 &    28.0$\pm$2.3 &    28.0 $\pm$1.1 &    17.0$\pm$0.9 &   8.6 &   12.7 &   10.9 &   10.7$^{+2.0}_{-2.1}$  &   0.37 &   0.77 \\ 
 8243-6103    & 59.1$\pm$0.6 &    12.1$\pm$1.4 &     9.8 $\pm$0.6 &    55.5$\pm$1.3 &   6.1 &   7.6 &   7.7 &   7.1$^{+0.6}_{-1.0}$  &   0.70 &   1.43 \\ 
 8244-3703    & 46.1$\pm$1.1 &    74.8$\pm$1.6 &    71.5 $\pm$1.6 &    31.3$\pm$0.8 &   3.5 &   4.6 &   4.4 &   4.2$^{+0.4}_{-0.7}$  &   0.38 &   0.74 \\ 
 8247-3701    & 37.9$\pm$1.5 &    49.7$\pm$2.9 &    48.5 $\pm$4.4 &    162.3$\pm$0.9 &   3.4 &   5.4 &   5.2 &   4.7$^{+0.7}_{-1.3}$  &   0.40 &   0.90 \\ 
 8249-6101    & 48.7$\pm$1.4 &    62.9$\pm$1.9 &    63.5 $\pm$1.6 &    106.3$\pm$0.7 &   11.3 &   14.1 &   13.4 &   12.9$^{+1.2}_{-1.6}$  &   1.13 &   2.96 \\ 
 8254-9101    & 44.1$\pm$1.6 &    17.3$\pm$1.5 &    27.2 $\pm$0.8 &    134.6$\pm$1.2 &   11.5 &   13.6 &   12.7 &   12.6$^{+1.0}_{-1.1}$  &   0.51 &   1.35 \\ 
 8256-6101    & 51.4$\pm$2.6 &    132.2$\pm$3.3 &    134.0 $\pm$0.9 &    59.2$\pm$0.8 &   8.9 &   10.8 &   8.5 &   9.4$^{+1.4}_{-0.9}$  &   0.64 &   1.49 \\ 
 8257-3703    & 58.3$\pm$0.6 &    156.1$\pm$0.6 &    155.2 $\pm$1.2 &    133.9$\pm$0.7 &   5.6 &   7.1 &   9.5 &   7.4$^{+2.1}_{-1.8}$  &   0.76 &   1.70 \\ 
 8257-6101    & 45.0$\pm$2.2 &    159.0$\pm$2.1 &    159.2 $\pm$1.5 &    119.7$\pm$1.0 &   3.6 &   4.3 &   4.5 &   4.1$^{+0.4}_{-0.5}$  &   0.20 &   0.39 \\ 
 8274-6101    & 54.0$\pm$1.1 &    129.6$\pm$1.2 &    133.5 $\pm$1.0 &    59.2$\pm$0.8 &   9.3 &   11.3 &   10.0 &   10.2$^{+1.1}_{-0.9}$  &   0.74 &   1.78 \\ 
 8312-12702    & 42.9$\pm$1.1 &    85.5$\pm$3.0 &    95.2 $\pm$1.8 &    120.9$\pm$1.1 &   7.6 &   11.5 &   10.1 &   9.7$^{+1.8}_{-2.1}$  &   0.63 &   1.59 \\ 
 8312-12704    & 46.1$\pm$0.7 &    30.3$\pm$1.2 &    34.0 $\pm$1.8 &    151.3$\pm$1.0 &   8.4 &   11.9 &   13.4 &   11.2$^{+2.2}_{-2.8}$  &   0.60 &   1.30 \\ 
 8313-9101    & 38.6$\pm$0.7 &    116.3$\pm$0.8 &    110.5 $\pm$1.0 &    156.5$\pm$1.6 &   3.6 &   6.2 &   6.3 &   5.4$^{+0.9}_{-1.8}$  &   0.24 &   0.47 \\ 
 8317-12704    & 69.2$\pm$0.3 &    103.7$\pm$0.5 &    101.8 $\pm$0.9 &    126.7$\pm$0.7 &   8.8 &   11.3 &   10.7 &   10.3$^{+1.0}_{-1.5}$  &   0.71 &   1.62 \\ 
 8318-12703    & 61.8$\pm$0.9 &    46.0$\pm$0.7 &    53.8 $\pm$0.9 &    86.4$\pm$0.9 &   8.0 &   9.9 &   5.4 &   7.8$^{+2.1}_{-2.4}$  &   0.44 &   0.91 \\ 
 8320-6101    & 50.0$\pm$0.6 &    5.9$\pm$0.4 &     5.0 $\pm$1.1 &    67.8$\pm$1.0 &   5.9 &   8.0 &   5.8 &   6.6$^{+1.4}_{-0.8}$  &   0.43 &   0.92 \\ 
 8326-3704    & 50.4$\pm$1.1 &    146.1$\pm$2.6 &    159.8 $\pm$3.0 &    126.8$\pm$1.1 &   5.7 &   8.1 &   7.7 &   7.2$^{+0.9}_{-1.5}$  &   0.45 &   1.06 \\ 
 8326-6102    & 51.9$\pm$0.9 &    148.0$\pm$1.7 &    145.8 $\pm$1.6 &    43.1$\pm$1.9 &   4.4 &   5.7 &   6.0 &   5.4$^{+0.6}_{-1.0}$  &   0.56 &   1.34 \\ 
 8330-12703    & 45.0$\pm$0.5 &    75.4$\pm$1.1 &    68.5 $\pm$1.9 &    49.2$\pm$0.8 &   8.6 &   10.2 &   11.2 &   10.0$^{+1.2}_{-1.4}$  &   0.31 &   0.77 \\ 
 8335-12701    & 67.0$\pm$0.5 &    81.2$\pm$1.0 &    78.2 $\pm$1.4 &    104.3$\pm$0.8 &   5.7 &   8.6 &   12.6 &   9.0$^{+3.6}_{-3.3}$  &   0.60 &   1.29 \\ 
 8439-6102    & 49.3$\pm$0.5 &    48.9$\pm$0.7 &    45.5 $\pm$1.1 &    25.6$\pm$0.9 &   5.3 &   7.4 &   9.4 &   7.4$^{+2.0}_{-2.1}$  &   0.53 &   1.19 \\ 
 8439-12702    & 55.1$\pm$0.4 &    31.3$\pm$0.6 &    31.5 $\pm$0.5 &    145.3$\pm$0.9 &   9.7 &   10.7 &   11.9 &   10.8$^{+1.1}_{-1.1}$  &   0.46 &   1.18 \\ 
 8440-12704    & 57.9$\pm$0.4 &    149.7$\pm$1.1 &    150.0 $\pm$0.8 &    116.3$\pm$1.4 &   4.5 &   5.6 &   6.8 &   5.6$^{+1.2}_{-1.1}$  &   0.43 &   0.74 \\ 
 8447-6101    & 63.9$\pm$0.8 &    178.4$\pm$1.6 &    178.2 $\pm$1.2 &    10.2$\pm$1.0 &   8.1 &   9.9 &   9.7 &   9.2$^{+0.7}_{-1.1}$  &   0.30 &   0.40 \\ 
 8452-3704    & 59.7$\pm$0.3 &    72.7$\pm$0.8 &    72.0 $\pm$2.5 &    52.7$\pm$1.1 &   2.8 &   3.5 &   5.7 &   4.0$^{+1.7}_{-1.2}$  &   0.21 &   0.33 \\ 
 8452-12703    & 45.7$\pm$2.4 &    75.1$\pm$1.7 &    65.0 $\pm$1.2 &    32.7$\pm$1.1 &   7.9 &   8.6 &   4.8 &   7.1$^{+1.5}_{-2.3}$  &   0.38 &   0.89 \\ 
 8481-12701    & 49.2$\pm$0.8 &    148.0$\pm$1.0 &    147.0 $\pm$1.2 &    86.9$\pm$1.2 &   4.8 &   6.6 &   6.0 &   5.8$^{+0.8}_{-1.0}$  &   0.65 &   1.13 \\ 
 8482-9102    & 62.6$\pm$0.6 &    63.2$\pm$1.4 &    63.8 $\pm$1.9 &    86.4$\pm$1.0 &   4.6 &   6.2 &   6.0 &   5.6$^{+0.6}_{-1.0}$  &   0.41 &   0.99 \\
 8482-12703    & 42.4$\pm$0.9 &    2.9$\pm$1.7 &    176.2 $\pm$1.5 &    132.0$\pm$1.7 &   6.1 &   7.0 &   4.6 &   5.9$^{+1.1}_{-1.3}$  &   0.41 &   0.94 \\ 
 8482-12705    & 63.0$\pm$1.0 &    117.2$\pm$0.9 &    117.0 $\pm$1.0 &    100.9$\pm$0.6 &   7.9 &   10.4 &   9.8 &   9.4$^{+1.0}_{-1.5}$  &   0.32 &   0.66 \\ 
 8486-6101    & 40.4$\pm$1.2 &    111.5$\pm$1.7 &    113.5 $\pm$1.4 &    81.4$\pm$1.1 &   3.0 &   5.0 &   6.0 &   4.7$^{+1.3}_{-1.7}$  &   0.59 &   1.05 \\ 
 8548-6102    & 54.1$\pm$0.4 &    64.7$\pm$0.8 &    58.8 $\pm$3.6 &    179.2$\pm$1.3 &   5.5 &   6.9 &   8.6 &   7.0$^{+1.6}_{-1.5}$  &   0.98 &   2.23 \\ 
 8548-6104    & 62.2$\pm$1.6 &    118.1$\pm$0.4 &    120.2 $\pm$5.0 &    136.8$\pm$0.6 &   4.0 &   5.2 &   5.6 &   4.9$^{+0.7}_{-0.9}$  &   0.49 &   0.81 \\ 
 8549-12702    & 54.3$\pm$2.6 &    97.6$\pm$1.6 &    100.8 $\pm$1.0 &    149.7$\pm$0.9 &   5.4 &   6.8 &   5.5 &   5.9$^{+0.9}_{-0.5}$  &   0.49 &   0.99 \\ 
 8588-3701    & 40.4$\pm$1.7 &    118.6$\pm$4.3 &    136.2 $\pm$1.9 &    158.7$\pm$1.2 &   5.0 &   5.7 &   4.2 &   5.0$^{+0.7}_{-0.8}$  &   0.46 &   0.79 \\ 
 8601-12705    & 68.3$\pm$0.5 &    49.4$\pm$0.6 &    51.5 $\pm$0.9 &    64.8$\pm$0.8 &   6.1 &   8.1 &   5.2 &   6.5$^{+1.6}_{-1.3}$  &   0.40 &   0.94 \\ 
 8603-12701    & 41.1$\pm$1.4 &    118.6$\pm$4.3 &    136.2 $\pm$2.1 &    158.7$\pm$1.2 &   5.0 &   5.7 &   4.1 &   4.9$^{+0.8}_{-0.8}$  &   0.46 &   0.79 \\ 
 8603-12703    & 58.0$\pm$0.6 &    73.5$\pm$1.5 &    66.5 $\pm$1.5 &    93.0$\pm$0.5 &   10.8 &   13.0 &   12.4 &   12.1$^{+0.9}_{-1.3}$  &   0.30 &   0.52 \\ 
 8604-12703    & 48.8$\pm$1.0 &    100.1$\pm$1.6 &    97.8 $\pm$1.0 &    150.9$\pm$1.1 &   8.1 &   11.1 &   11.6 &   10.3$^{+1.3}_{-2.2}$  &   0.50 &   1.12 \\ 
 8612-6104    & 42.4$\pm$2.3 &    169.6$\pm$1.5 &    153.5 $\pm$1.8 &    92.7$\pm$2.2 &   6.7 &   7.6 &   10.4 &   8.2$^{+2.2}_{-1.5}$  &   0.56 &   1.48 \\ 
 8612-12702    & 52.3$\pm$1.0 &    49.6$\pm$3.3 &    44.0 $\pm$1.4 &    75.2$\pm$0.7 &   5.3 &   6.3 &   4.6 &   5.4$^{+0.9}_{-0.8}$  &   0.30 &   0.53 \\  
\hline
\end{tabular}
\begin{tablenotes}
\small
\item Note: Columns: (1) MaNGA plate-ifu of galaxy; (2) galaxy inclination measured from the ellipticity radial profile using ellipse fitting of r-band SDSS image; (3) galaxy photometric PA measured from the PA radial profile of ellipse fitting using r-band SDSS image; (4) galaxy kinematic PA measured from the velocity map using fit\_kinematic\_pa.py program; (5) bar PA defined as the PA with the local maximum ellipticity; (6) bar length defined as the radius with the local maximum ellipticity; (7) bar length measured when the PA changes by more than 5$^{\circ}$ relative to the bar PA; (8) bar length obtained from the ratio of bar and inter--bar intensities calculated by Fourier decomposition; (9) average of the former three bar lengths, with errors corresponding to the maximum differences between the mean and the three estimations; (10) bar strength estimated by the maximum of $m=2$ term of Fourier decomposition; (11) bar strength obtained from the surface brightness deficit between profiles along the major and the minor axes of the bar.
\end{tablenotes}
\end{table*}

\clearpage
\begin{table*}
\caption{Luminosity-weighted and mass-weighted pattern speeds and corotation radii of MaNGA barred galaxies measured using the photometric PAs.}
\label{tab: op_pap}
\tabcolsep=15.0pt
\begin{tabular}{@{\extracolsep{\fill}}l c c c c c c}
\hline
\hline
Plate-ifu  & $\Omega_{\rm p,l} \sin i$ & $\rm{R}_{\rm CR,l}$ & ${\cal{R}}_{\rm l}$ & $\Omega_{\rm p,m} \sin i$ & $\rm{R}_{\rm CR,m}$ & ${\cal{R}}_{\rm m}$ \\
       & (${\rm km/s/''}$) & ($''$) & & (${\rm km/s/''}$) & ($''$) &  \\
   (1) & (2) & (3) & (4) & (5) & (6) & (7) \\
\hline
  7495-12704    &    -15.0$^{+ 1.8}_{- 1.4}$  &    10.7$^{+ 1.8}_{- 1.6}$  & 1.43$^{+0.33}_{-0.28}$  &    -12.9$^{+ 2.2}_{- 2.2}$  &    12.4$^{+ 3.1}_{- 2.3}$  & 1.68$^{+0.48}_{-0.38}$  \\ 
 7962-12703    &    24.4$^{+ 0.8}_{- 0.6}$  &     9.4$^{+ 1.2}_{- 1.1}$  & 0.58$^{+0.16}_{-0.12}$  &    23.4$^{+ 0.7}_{- 0.4}$  &     9.8$^{+ 1.2}_{- 1.2}$  & 0.60$^{+0.17}_{-0.12}$  \\ 
 7990-3704    &    30.7$^{+ 9.8}_{- 9.7}$  &     3.1$^{+ 1.5}_{- 0.8}$  & 0.84$^{+0.42}_{-0.26}$  &    26.3$^{+ 6.0}_{- 6.0}$  &     3.6$^{+ 1.1}_{- 0.8}$  & 0.96$^{+0.35}_{-0.25}$  \\ 
 7990-9101    &     8.6$^{+ 2.8}_{- 3.3}$  &    13.2$^{+ 8.4}_{- 3.5}$  & 2.15$^{+1.39}_{-0.77}$  &     7.1$^{+ 3.7}_{- 4.6}$  &    16.1$^{+24.2}_{- 5.9}$  & 2.77$^{+3.66}_{-1.30}$  \\ 
 7992-6104    &    11.2$^{+ 0.8}_{- 0.7}$  &     8.2$^{+ 1.2}_{- 1.1}$  & 0.91$^{+0.22}_{-0.18}$  &    10.9$^{+ 0.7}_{- 0.7}$  &     8.4$^{+ 1.2}_{- 1.1}$  & 0.93$^{+0.22}_{-0.18}$  \\ 
 8082-6102    &    -16.8$^{+ 7.6}_{- 6.4}$  &     9.3$^{+ 7.6}_{- 2.8}$  & 1.28$^{+0.97}_{-0.44}$  &    -18.2$^{+ 9.2}_{- 6.1}$  &     8.7$^{+ 8.1}_{- 2.5}$  & 1.23$^{+1.05}_{-0.43}$  \\ 
 8083-6102    &    -8.8$^{+ 3.4}_{- 2.3}$  &    30.8$^{+19.2}_{- 7.6}$  & 4.73$^{+2.88}_{-1.61}$  &    -3.5$^{+ 3.8}_{- 3.4}$  &    72.1$^{+163.5}_{-33.7}$  & 11.85$^{+23.17}_{-6.48}$  \\ 
 8083-12704    &    -26.5$^{+15.6}_{-25.6}$  &     2.4$^{+ 3.4}_{- 1.1}$  & 0.39$^{+0.51}_{-0.19}$  &    -22.5$^{+ 9.2}_{- 9.6}$  &     3.0$^{+ 2.1}_{- 0.9}$  & 0.48$^{+0.32}_{-0.17}$  \\ 
 8133-3701    &    -27.1$^{+ 4.1}_{- 5.8}$  &     3.6$^{+ 0.8}_{- 0.7}$  & 0.88$^{+0.35}_{-0.24}$  &    -29.3$^{+ 6.0}_{- 9.3}$  &     3.3$^{+ 1.0}_{- 0.8}$  & 0.80$^{+0.36}_{-0.24}$  \\ 
 8134-6102    &    -12.6$^{+ 2.6}_{- 2.1}$  &    18.2$^{+ 5.2}_{- 3.4}$  & 1.56$^{+0.56}_{-0.41}$  &    -7.7$^{+ 2.0}_{- 1.9}$  &    29.4$^{+10.7}_{- 6.5}$  & 2.54$^{+1.01}_{-0.72}$  \\ 
 8137-9102    &    -15.1$^{+ 2.0}_{- 4.0}$  &     5.8$^{+ 1.4}_{- 1.2}$  & 0.53$^{+0.15}_{-0.13}$  &    -15.2$^{+ 1.9}_{- 3.6}$  &     5.8$^{+ 1.3}_{- 1.2}$  & 0.52$^{+0.15}_{-0.12}$  \\ 
 8140-12701    &    14.8$^{+ 3.1}_{- 2.3}$  &     7.1$^{+ 1.7}_{- 1.5}$  & 0.73$^{+0.22}_{-0.17}$  &    14.4$^{+ 2.2}_{- 1.8}$  &     7.3$^{+ 1.5}_{- 1.2}$  & 0.76$^{+0.19}_{-0.16}$  \\ 
 8140-12703    &    -15.8$^{+ 6.4}_{- 4.4}$  &    10.7$^{+ 7.0}_{- 2.6}$  & 1.07$^{+0.68}_{-0.35}$  &    -17.2$^{+ 6.2}_{- 4.8}$  &     9.8$^{+ 5.3}_{- 2.4}$  & 0.97$^{+0.54}_{-0.31}$  \\ 
 8243-6103    &    -12.3$^{+ 9.7}_{- 8.9}$  &    21.0$^{+41.5}_{- 9.0}$  & 3.31$^{+5.80}_{-1.58}$  &    -13.6$^{+ 8.4}_{- 9.2}$  &    18.9$^{+27.0}_{- 7.8}$  & 2.92$^{+3.83}_{-1.32}$  \\ 
 8244-3703    &    53.9$^{+10.5}_{- 9.5}$  &     2.7$^{+ 0.7}_{- 0.5}$  & 0.67$^{+0.20}_{-0.15}$  &    45.9$^{+ 7.4}_{- 6.6}$  &     3.2$^{+ 0.7}_{- 0.6}$  & 0.79$^{+0.21}_{-0.17}$  \\ 
 8247-3701    &    -7.3$^{+ 1.8}_{- 3.7}$  &    10.0$^{+ 3.9}_{- 3.2}$  & 2.27$^{+1.15}_{-0.81}$  &    -7.5$^{+ 1.6}_{- 3.5}$  &     9.9$^{+ 3.3}_{- 3.1}$  & 2.21$^{+1.03}_{-0.75}$  \\ 
 8249-6101    &    -13.2$^{+ 1.2}_{- 1.4}$  &     7.5$^{+ 1.2}_{- 1.1}$  & 0.59$^{+0.12}_{-0.10}$  &    -13.0$^{+ 1.0}_{- 1.4}$  &     7.7$^{+ 1.2}_{- 1.1}$  & 0.61$^{+0.12}_{-0.10}$  \\ 
 8254-9101    &    18.5$^{+ 9.9}_{-16.8}$  &    11.7$^{+23.3}_{- 4.2}$  & 0.96$^{+1.82}_{-0.36}$  &    14.6$^{+24.9}_{-16.2}$  &    13.1$^{+28.9}_{- 7.6}$  & 1.08$^{+2.23}_{-0.65}$  \\ 
 8256-6101    &    15.1$^{+11.6}_{-13.3}$  &     9.9$^{+20.7}_{- 4.2}$  & 1.10$^{+2.12}_{-0.51}$  &    14.1$^{+11.3}_{- 8.0}$  &    10.7$^{+13.7}_{- 4.7}$  & 1.16$^{+1.43}_{-0.54}$  \\ 
 8257-3703    &    22.9$^{+ 1.1}_{- 1.1}$  &     7.2$^{+ 0.9}_{- 0.9}$  & 0.97$^{+0.34}_{-0.24}$  &    22.3$^{+ 1.2}_{- 1.1}$  &     7.4$^{+ 1.0}_{- 1.0}$  & 1.00$^{+0.35}_{-0.25}$  \\ 
 8257-6101    &    -21.4$^{+10.6}_{-11.5}$  &     5.5$^{+ 5.5}_{- 2.0}$  & 1.42$^{+1.28}_{-0.56}$  &    -18.9$^{+ 5.2}_{- 5.3}$  &     6.3$^{+ 2.6}_{- 1.5}$  & 1.59$^{+0.64}_{-0.43}$  \\ 
 8274-6101    &     6.4$^{+ 7.9}_{- 6.8}$  &    22.6$^{+50.9}_{-11.7}$  & 2.33$^{+4.77}_{-1.26}$  &     6.8$^{+ 5.4}_{- 3.7}$  &    23.1$^{+25.8}_{-10.4}$  & 2.32$^{+2.40}_{-1.07}$  \\ 
 8312-12702    &    -16.1$^{+ 2.2}_{- 2.6}$  &     6.0$^{+ 1.3}_{- 1.0}$  & 0.63$^{+0.22}_{-0.14}$  &    -15.6$^{+ 2.4}_{- 2.6}$  &     6.2$^{+ 1.4}_{- 1.1}$  & 0.65$^{+0.23}_{-0.15}$  \\ 
 8312-12704    &    -6.5$^{+ 2.3}_{- 2.0}$  &    13.9$^{+ 7.9}_{- 3.6}$  & 1.33$^{+0.80}_{-0.44}$  &    -6.4$^{+ 2.1}_{- 1.9}$  &    13.8$^{+ 7.0}_{- 3.6}$  & 1.32$^{+0.74}_{-0.45}$  \\ 
 8313-9101    &    -0.4$^{+ 5.8}_{-11.9}$  &    26.6$^{+73.4}_{-15.7}$  & 6.31$^{+14.20}_{-4.12}$  &    -6.7$^{+ 2.7}_{- 4.5}$  &    20.4$^{+14.1}_{- 8.1}$  & 4.16$^{+3.38}_{-1.83}$  \\ 
 8317-12704    &    13.2$^{+ 3.1}_{- 3.0}$  &    24.0$^{+ 7.7}_{- 5.2}$  & 2.43$^{+0.83}_{-0.61}$  &    14.4$^{+ 3.0}_{- 2.1}$  &    21.6$^{+ 4.9}_{- 4.1}$  & 2.16$^{+0.58}_{-0.46}$  \\ 
 8318-12703    &    21.2$^{+ 4.3}_{- 5.8}$  &    10.0$^{+ 3.7}_{- 2.1}$  & 1.35$^{+0.74}_{-0.43}$  &    25.0$^{+ 5.6}_{- 7.0}$  &     8.4$^{+ 3.3}_{- 1.8}$  & 1.13$^{+0.65}_{-0.36}$  \\ 
 8320-6101    &    12.0$^{+ 2.4}_{- 2.1}$  &    12.0$^{+ 2.9}_{- 2.4}$  & 1.78$^{+0.54}_{-0.44}$  &    10.2$^{+ 2.8}_{- 2.2}$  &    14.0$^{+ 4.5}_{- 3.2}$  & 2.10$^{+0.75}_{-0.56}$  \\ 
 8326-3704    &    -6.6$^{+ 7.4}_{-17.0}$  &    11.7$^{+28.8}_{- 7.9}$  & 1.90$^{+4.12}_{-1.35}$  &    -7.6$^{+10.2}_{-16.9}$  &     9.4$^{+24.2}_{- 5.7}$  & 1.48$^{+3.40}_{-0.95}$  \\ 
 8326-6102    &    -22.1$^{+ 9.7}_{-15.5}$  &     8.2$^{+ 7.0}_{- 3.3}$  & 1.62$^{+1.35}_{-0.71}$  &    -51.5$^{+36.4}_{-23.3}$  &     3.7$^{+ 6.4}_{- 1.3}$  & 0.78$^{+1.19}_{-0.33}$  \\ 
 8330-12703    &    18.4$^{+ 1.7}_{- 1.5}$  &     5.3$^{+ 0.9}_{- 0.7}$  & 0.54$^{+0.12}_{-0.10}$  &    16.0$^{+ 2.7}_{- 2.6}$  &     6.2$^{+ 1.4}_{- 1.1}$  & 0.63$^{+0.17}_{-0.13}$  \\ 
 8335-12701    &     9.8$^{+ 5.5}_{- 3.3}$  &    22.0$^{+12.3}_{- 7.8}$  & 2.53$^{+2.13}_{-1.13}$  &    18.1$^{+ 3.8}_{- 1.6}$  &    11.9$^{+ 2.2}_{- 2.2}$  & 1.31$^{+0.78}_{-0.43}$  \\ 
 8439-6102    &    -29.4$^{+ 0.8}_{- 0.8}$  &     5.3$^{+ 0.6}_{- 0.7}$  & 0.71$^{+0.29}_{-0.17}$  &    -30.8$^{+ 1.1}_{- 1.1}$  &     5.0$^{+ 0.7}_{- 0.6}$  & 0.68$^{+0.28}_{-0.16}$  \\ 
 8439-12702    &    14.6$^{+ 2.0}_{- 2.4}$  &    13.4$^{+ 3.1}_{- 2.2}$  & 1.25$^{+0.31}_{-0.24}$  &    15.5$^{+ 2.2}_{- 1.8}$  &    12.6$^{+ 2.4}_{- 2.1}$  & 1.18$^{+0.25}_{-0.22}$  \\ 
 8440-12704    &    17.7$^{+ 3.7}_{- 2.1}$  &    10.0$^{+ 2.1}_{- 1.9}$  & 1.79$^{+0.59}_{-0.45}$  &    16.1$^{+ 3.9}_{- 3.3}$  &    11.2$^{+ 3.4}_{- 2.5}$  & 2.03$^{+0.77}_{-0.56}$  \\ 
 8447-6101    &    -53.8$^{+10.7}_{-16.1}$  &     5.9$^{+ 1.8}_{- 1.4}$  & 0.66$^{+0.21}_{-0.17}$  &    -57.6$^{+11.4}_{-16.8}$  &     5.6$^{+ 1.6}_{- 1.3}$  & 0.63$^{+0.18}_{-0.16}$  \\ 
 8452-3704    &    -35.6$^{+22.4}_{-23.9}$  &     3.8$^{+ 5.6}_{- 1.6}$  & 1.07$^{+1.39}_{-0.58}$  &    -24.0$^{+ 9.2}_{-10.7}$  &     5.6$^{+ 3.6}_{- 1.8}$  & 1.45$^{+1.15}_{-0.62}$  \\ 
 8452-12703    &    39.1$^{+ 5.6}_{- 5.2}$  &     3.9$^{+ 0.8}_{- 0.7}$  & 0.57$^{+0.28}_{-0.15}$  &    51.3$^{+14.4}_{-12.9}$  &     3.0$^{+ 1.1}_{- 0.8}$  & 0.45$^{+0.26}_{-0.15}$  \\ 
 8481-12701    &    -42.8$^{+10.8}_{- 7.6}$  &     4.7$^{+ 1.6}_{- 0.9}$  & 0.85$^{+0.31}_{-0.21}$  &    -47.8$^{+16.6}_{-14.5}$  &     4.2$^{+ 2.3}_{- 1.1}$  & 0.78$^{+0.40}_{-0.26}$  \\ 
 8482-9102    &    -16.6$^{+ 6.4}_{- 4.1}$  &    12.0$^{+ 7.1}_{- 2.9}$  & 2.34$^{+1.30}_{-0.70}$  &    -9.4$^{+ 5.3}_{- 6.8}$  &    20.3$^{+25.0}_{- 8.6}$  & 4.04$^{+4.37}_{-1.89}$  \\
 8482-12703    &    -29.3$^{+11.0}_{-11.2}$  &     3.8$^{+ 2.2}_{- 1.1}$  & 0.68$^{+0.42}_{-0.24}$  &    -19.8$^{+ 4.3}_{- 3.6}$  &     5.6$^{+ 1.7}_{- 1.1}$  & 0.99$^{+0.38}_{-0.27}$  \\ 
 8482-12705    &    10.1$^{+ 4.8}_{- 6.4}$  &    21.0$^{+29.1}_{- 7.2}$  & 2.47$^{+2.96}_{-0.98}$  &     4.6$^{+ 7.2}_{-14.3}$  &    26.1$^{+44.2}_{-12.1}$  & 3.09$^{+4.59}_{-1.58}$  \\ 
 8486-6101    &    15.1$^{+ 3.1}_{- 3.8}$  &     8.1$^{+ 2.8}_{- 1.7}$  & 1.84$^{+1.10}_{-0.61}$  &    10.4$^{+ 4.5}_{- 3.3}$  &    11.7$^{+ 5.6}_{- 3.7}$  & 2.66$^{+1.91}_{-1.05}$  \\ 
 8548-6102    &    28.7$^{+ 4.5}_{- 3.2}$  &     4.5$^{+ 0.9}_{- 0.7}$  & 0.65$^{+0.21}_{-0.16}$  &    28.4$^{+ 4.8}_{- 3.4}$  &     4.5$^{+ 0.9}_{- 0.7}$  & 0.65$^{+0.21}_{-0.16}$  \\ 
 8548-6104    &    -20.9$^{+ 3.8}_{- 4.0}$  &     7.5$^{+ 1.9}_{- 1.5}$  & 1.56$^{+0.51}_{-0.37}$  &    -21.6$^{+ 3.0}_{- 2.9}$  &     7.3$^{+ 1.5}_{- 1.2}$  & 1.52$^{+0.43}_{-0.33}$  \\ 
 8549-12702    &    -56.7$^{+12.7}_{-17.5}$  &     3.5$^{+ 1.1}_{- 0.9}$  & 0.58$^{+0.20}_{-0.16}$  &    -44.5$^{+ 7.9}_{- 9.4}$  &     4.4$^{+ 1.2}_{- 0.9}$  & 0.74$^{+0.22}_{-0.17}$  \\ 
 8588-3701    &    -80.3$^{+22.5}_{-22.3}$  &     2.1$^{+ 0.9}_{- 0.5}$  & 0.44$^{+0.19}_{-0.13}$  &    -81.1$^{+22.7}_{-26.2}$  &     2.1$^{+ 0.9}_{- 0.5}$  & 0.44$^{+0.18}_{-0.12}$  \\ 
 8601-12705    &    13.7$^{+ 2.8}_{- 1.2}$  &    11.7$^{+ 2.1}_{- 2.1}$  & 1.78$^{+0.60}_{-0.44}$  &    14.2$^{+ 1.8}_{- 1.0}$  &    11.4$^{+ 1.8}_{- 1.7}$  & 1.75$^{+0.55}_{-0.42}$  \\ 
 8603-12701    &    -85.8$^{+17.4}_{-17.1}$  &     2.0$^{+ 0.6}_{- 0.4}$  & 0.41$^{+0.15}_{-0.10}$  &    -90.0$^{+22.0}_{-26.0}$  &     1.9$^{+ 0.7}_{- 0.5}$  & 0.39$^{+0.16}_{-0.11}$  \\ 
 8603-12703    &    -13.5$^{+ 5.0}_{- 6.3}$  &     9.2$^{+ 5.6}_{- 3.0}$  & 0.79$^{+0.46}_{-0.28}$  &    -11.7$^{+ 4.1}_{- 3.8}$  &    10.8$^{+ 6.0}_{- 3.0}$  & 0.92$^{+0.49}_{-0.27}$  \\ 
 8604-12703    &     7.9$^{+ 3.8}_{- 9.7}$  &    21.2$^{+40.8}_{- 7.1}$  & 2.45$^{+3.94}_{-1.06}$  &     9.0$^{+ 3.7}_{-11.8}$  &    18.5$^{+38.3}_{- 5.7}$  & 2.13$^{+3.60}_{-0.89}$  \\ 
 8612-6104    &    -52.5$^{+ 6.0}_{- 6.5}$  &     2.4$^{+ 0.4}_{- 0.4}$  & 0.29$^{+0.09}_{-0.07}$  &    -53.0$^{+ 4.8}_{- 3.2}$  &     2.4$^{+ 0.4}_{- 0.3}$  & 0.29$^{+0.09}_{-0.07}$  \\ 
 8612-12702    &    43.1$^{+34.8}_{-24.7}$  &     3.8$^{+ 4.8}_{- 1.7}$  & 0.74$^{+0.87}_{-0.36}$  &    36.2$^{+ 5.5}_{- 8.7}$  &     4.6$^{+ 1.5}_{- 0.8}$  & 0.87$^{+0.31}_{-0.20}$  \\    
\hline
\end{tabular}
\begin{tablenotes}
\small
\item Note: Columns are: (1) MaNGA plate-ifu of galaxy; (2), (3) and (4) are the light--weighted pattern speed, corotation radius and dimensionless ratio ${\cal R}=R_{\rm CR}/a_{\rm b}$, respectively; while (5), (6), (7) are corresponding mass--weighted values.
\end{tablenotes}
\end{table*}

\begin{table*}
\caption{Luminosity-weighted and mass-weighted pattern speeds and corotation radii of MaNGA barred galaxies measured using the kinematic PAs.}
\label{tab: op_pak}
\tabcolsep=15.0pt
\begin{tabular}{@{\extracolsep{\fill}}l c c c c c c}
\hline
\hline
Plate-ifu  & $\Omega_{\rm p,l} \sin i$ & $\rm{R}_{\rm CR,l}$ & ${\cal{R}}_{\rm l}$ & $\Omega_{\rm p,m} \sin i$ & $\rm{R}_{\rm CR,m}$ & ${\cal{R}}_{\rm m}$ \\
       & (${\rm km/s/''}$) & ($''$) & & (${\rm km/s/''}$) & ($''$) &  \\
   (1) & (2) & (3) & (4) & (5) & (6) & (7) \\
\hline
 7495-12704    &    -14.5$^{+ 1.6}_{- 1.3}$  &    11.1$^{+ 1.8}_{- 1.7}$  & 1.50$^{+0.35}_{-0.29}$  &    -12.0$^{+ 2.1}_{- 1.8}$  &    13.3$^{+ 3.3}_{- 2.3}$  & 1.82$^{+0.53}_{-0.40}$  \\ 
 7962-12703    &    19.0$^{+ 1.1}_{- 1.3}$  &    12.2$^{+ 1.7}_{- 1.6}$  & 0.79$^{+0.22}_{-0.17}$  &    19.2$^{+ 1.1}_{- 1.4}$  &    12.1$^{+ 1.8}_{- 1.5}$  & 0.79$^{+0.23}_{-0.17}$  \\ 
 7990-3704    &    22.6$^{+ 9.0}_{- 6.6}$  &     4.1$^{+ 1.9}_{- 1.2}$  & 1.12$^{+0.54}_{-0.37}$  &    21.4$^{+ 5.5}_{- 5.2}$  &     4.4$^{+ 1.6}_{- 1.0}$  & 1.19$^{+0.50}_{-0.33}$  \\ 
 7990-9101    &    10.8$^{+ 2.7}_{- 5.4}$  &    10.7$^{+ 9.7}_{- 2.6}$  & 1.75$^{+1.41}_{-0.62}$  &     9.0$^{+ 3.6}_{- 4.4}$  &    12.7$^{+11.5}_{- 3.9}$  & 2.07$^{+1.69}_{-0.81}$  \\ 
 7992-6104    &    10.4$^{+ 1.3}_{- 1.6}$  &     8.9$^{+ 1.9}_{- 1.5}$  & 1.01$^{+0.29}_{-0.23}$  &    10.1$^{+ 1.3}_{- 1.9}$  &     9.2$^{+ 2.3}_{- 1.6}$  & 1.04$^{+0.33}_{-0.24}$  \\ 
 8082-6102    &    -19.4$^{+ 7.3}_{- 5.5}$  &     8.1$^{+ 4.9}_{- 2.0}$  & 1.12$^{+0.64}_{-0.33}$  &    -20.6$^{+ 8.8}_{- 5.4}$  &     7.6$^{+ 5.6}_{- 1.8}$  & 1.06$^{+0.71}_{-0.32}$  \\ 
 8083-6102    &    -15.4$^{+30.3}_{- 8.3}$  &    14.5$^{+24.1}_{- 5.0}$  & 2.26$^{+3.20}_{-0.99}$  &    -6.8$^{+20.6}_{- 8.9}$  &    23.7$^{+36.7}_{-10.9}$  & 3.65$^{+4.94}_{-1.89}$  \\ 
 8083-12704    &     2.2$^{+ 6.5}_{- 9.2}$  &    12.3$^{+25.4}_{- 6.3}$  & 2.03$^{+3.50}_{-1.19}$  &     3.8$^{+ 2.0}_{- 2.0}$  &    17.5$^{+18.5}_{- 6.3}$  & 2.75$^{+2.62}_{-1.21}$  \\ 
 8133-3701    &    -24.0$^{+ 9.4}_{- 8.8}$  &     4.1$^{+ 2.7}_{- 1.2}$  & 1.07$^{+0.74}_{-0.41}$  &    -25.6$^{+ 9.7}_{-12.8}$  &     3.8$^{+ 2.3}_{- 1.3}$  & 0.97$^{+0.63}_{-0.39}$  \\ 
 8134-6102    &    16.3$^{+12.0}_{- 7.3}$  &    13.7$^{+11.3}_{- 5.7}$  & 1.21$^{+0.95}_{-0.58}$  &    16.6$^{+10.3}_{- 6.6}$  &    13.3$^{+ 9.4}_{- 5.1}$  & 1.14$^{+0.80}_{-0.48}$  \\ 
 8137-9102    &    -14.4$^{+10.3}_{- 5.5}$  &     6.6$^{+11.2}_{- 2.1}$  & 0.65$^{+0.95}_{-0.25}$  &    -15.0$^{+12.4}_{- 6.8}$  &     6.2$^{+12.4}_{- 2.1}$  & 0.64$^{+1.09}_{-0.28}$  \\ 
 8140-12701    &    11.1$^{+ 2.9}_{- 2.3}$  &     9.4$^{+ 2.9}_{- 2.1}$  & 0.98$^{+0.33}_{-0.25}$  &    11.4$^{+ 2.4}_{- 2.1}$  &     9.2$^{+ 2.4}_{- 1.8}$  & 0.96$^{+0.29}_{-0.22}$  \\ 
 8140-12703    &    -15.7$^{+ 3.8}_{- 2.6}$  &    10.9$^{+ 3.5}_{- 2.1}$  & 1.05$^{+0.40}_{-0.28}$  &    -17.1$^{+ 3.6}_{- 2.7}$  &     9.9$^{+ 2.8}_{- 1.8}$  & 0.95$^{+0.36}_{-0.24}$  \\ 
 8243-6103    &    -26.4$^{+ 3.1}_{- 2.4}$  &     9.9$^{+ 1.7}_{- 1.5}$  & 1.42$^{+0.30}_{-0.26}$  &    -27.4$^{+ 2.7}_{- 2.4}$  &     9.5$^{+ 1.5}_{- 1.4}$  & 1.34$^{+0.30}_{-0.23}$  \\ 
 8244-3703    &    37.8$^{+ 8.0}_{- 6.0}$  &     3.8$^{+ 1.0}_{- 0.7}$  & 0.97$^{+0.29}_{-0.22}$  &    33.5$^{+ 6.2}_{- 3.3}$  &     4.3$^{+ 0.8}_{- 0.8}$  & 1.08$^{+0.28}_{-0.23}$  \\ 
 8247-3701    &    -4.4$^{+ 2.2}_{- 3.1}$  &    17.0$^{+17.3}_{- 6.9}$  & 4.15$^{+4.01}_{-1.99}$  &    -5.1$^{+ 1.8}_{- 3.6}$  &    14.4$^{+ 8.9}_{- 5.8}$  & 3.35$^{+2.27}_{-1.48}$  \\ 
 8249-6101    &    -12.9$^{+ 1.1}_{- 1.2}$  &     7.8$^{+ 1.2}_{- 1.1}$  & 0.61$^{+0.13}_{-0.10}$  &    -12.6$^{+ 0.9}_{- 1.3}$  &     7.9$^{+ 1.2}_{- 1.2}$  & 0.62$^{+0.13}_{-0.11}$  \\ 
 8254-9101    &    -18.5$^{+ 3.9}_{- 2.4}$  &    12.0$^{+ 3.3}_{- 2.1}$  & 0.94$^{+0.26}_{-0.18}$  &    -19.5$^{+ 3.4}_{- 2.0}$  &    11.4$^{+ 2.5}_{- 1.9}$  & 0.89$^{+0.21}_{-0.17}$  \\ 
 8256-6101    &    24.7$^{+ 1.0}_{- 1.6}$  &     6.3$^{+ 0.9}_{- 0.8}$  & 0.66$^{+0.13}_{-0.11}$  &    23.6$^{+ 1.1}_{- 3.0}$  &     6.8$^{+ 1.1}_{- 1.0}$  & 0.71$^{+0.15}_{-0.13}$  \\ 
 8257-3703    &    21.3$^{+ 2.2}_{- 2.8}$  &     7.8$^{+ 1.5}_{- 1.2}$  & 1.05$^{+0.41}_{-0.27}$  &    20.7$^{+ 2.4}_{- 3.4}$  &     8.1$^{+ 1.8}_{- 1.3}$  & 1.10$^{+0.44}_{-0.29}$  \\ 
 8257-6101    &    -21.0$^{+ 8.5}_{- 9.4}$  &     5.6$^{+ 3.9}_{- 1.8}$  & 1.42$^{+0.93}_{-0.50}$  &    -19.1$^{+ 4.2}_{- 3.9}$  &     6.3$^{+ 1.9}_{- 1.3}$  & 1.58$^{+0.49}_{-0.37}$  \\ 
 8274-6101    &    23.8$^{+ 0.9}_{- 4.2}$  &     6.9$^{+ 1.4}_{- 1.0}$  & 0.67$^{+0.15}_{-0.12}$  &    23.2$^{+ 1.5}_{- 5.7}$  &     7.2$^{+ 1.9}_{- 1.2}$  & 0.70$^{+0.19}_{-0.14}$  \\ 
 8312-12702    &    -9.7$^{+ 1.9}_{- 1.3}$  &    10.0$^{+ 2.6}_{- 1.7}$  & 1.11$^{+0.41}_{-0.26}$  &    -7.9$^{+ 3.0}_{- 2.1}$  &    12.5$^{+ 7.4}_{- 3.1}$  & 1.46$^{+0.87}_{-0.48}$  \\ 
 8312-12704    &    -11.9$^{+ 4.5}_{- 2.7}$  &     7.6$^{+ 4.5}_{- 1.8}$  & 0.73$^{+0.45}_{-0.24}$  &    -11.8$^{+ 4.4}_{- 2.8}$  &     7.7$^{+ 4.4}_{- 1.8}$  & 0.74$^{+0.44}_{-0.23}$  \\ 
 8313-9101    &    -13.5$^{+ 3.8}_{- 3.5}$  &    10.5$^{+ 4.2}_{- 2.5}$  & 2.11$^{+1.18}_{-0.66}$  &    -16.2$^{+ 2.0}_{- 2.3}$  &     8.6$^{+ 1.7}_{- 1.4}$  & 1.66$^{+0.75}_{-0.38}$  \\ 
 8317-12704    &    22.9$^{+ 3.9}_{- 4.5}$  &    13.8$^{+ 3.8}_{- 2.6}$  & 1.34$^{+0.39}_{-0.30}$  &    23.3$^{+ 3.5}_{- 4.0}$  &    13.6$^{+ 3.3}_{- 2.4}$  & 1.31$^{+0.37}_{-0.27}$  \\ 
 8318-12703    &    18.0$^{+ 5.2}_{- 8.3}$  &    11.9$^{+ 9.8}_{- 3.2}$  & 1.86$^{+1.42}_{-0.72}$  &    13.0$^{+ 7.4}_{-11.5}$  &    15.9$^{+31.9}_{- 5.9}$  & 2.65$^{+4.30}_{-1.29}$  \\ 
 8320-6101    &    15.0$^{+ 3.4}_{- 4.0}$  &     9.7$^{+ 3.5}_{- 2.1}$  & 1.46$^{+0.56}_{-0.39}$  &    14.0$^{+ 3.9}_{- 5.0}$  &    10.4$^{+ 5.8}_{- 2.6}$  & 1.59$^{+0.85}_{-0.49}$  \\ 
 8326-3704    &    10.6$^{+ 4.5}_{- 3.6}$  &     8.6$^{+ 4.5}_{- 2.7}$  & 1.16$^{+0.64}_{-0.41}$  &    11.4$^{+ 4.0}_{- 3.3}$  &     8.0$^{+ 3.6}_{- 2.1}$  & 1.09$^{+0.50}_{-0.35}$  \\ 
 8326-6102    &    -41.7$^{+16.8}_{-11.7}$  &     4.5$^{+ 2.9}_{- 1.1}$  & 0.91$^{+0.53}_{-0.29}$  &    -55.1$^{+ 9.8}_{-15.9}$  &     3.3$^{+ 0.9}_{- 0.7}$  & 0.63$^{+0.20}_{-0.15}$  \\ 
 8330-12703    &    11.5$^{+ 2.4}_{- 3.3}$  &     8.7$^{+ 3.6}_{- 1.9}$  & 0.92$^{+0.39}_{-0.24}$  &     5.3$^{+ 3.8}_{- 6.4}$  &    17.5$^{+36.6}_{- 6.9}$  & 1.97$^{+3.54}_{-0.90}$  \\ 
 8335-12701    &    21.1$^{+ 5.2}_{- 4.0}$  &    10.4$^{+ 2.9}_{- 2.2}$  & 1.15$^{+0.72}_{-0.39}$  &    25.9$^{+ 3.4}_{- 2.6}$  &     8.5$^{+ 1.5}_{- 1.3}$  & 0.92$^{+0.54}_{-0.27}$  \\ 
 8439-6102    &    -25.3$^{+ 2.7}_{- 3.8}$  &     6.0$^{+ 1.1}_{- 1.0}$  & 0.83$^{+0.37}_{-0.23}$  &    -25.9$^{+ 3.9}_{-11.5}$  &     5.7$^{+ 1.5}_{- 1.7}$  & 0.77$^{+0.39}_{-0.26}$  \\ 
 8439-12702    &    15.2$^{+ 2.5}_{- 1.6}$  &    12.7$^{+ 2.3}_{- 2.2}$  & 1.18$^{+0.27}_{-0.22}$  &    16.0$^{+ 2.8}_{- 1.5}$  &    12.0$^{+ 2.2}_{- 2.1}$  & 1.12$^{+0.24}_{-0.22}$  \\ 
 8440-12704    &    18.6$^{+ 3.0}_{- 2.3}$  &     9.7$^{+ 1.9}_{- 1.7}$  & 1.74$^{+0.56}_{-0.43}$  &    17.2$^{+ 3.2}_{- 3.3}$  &    10.6$^{+ 2.9}_{- 2.0}$  & 1.92$^{+0.69}_{-0.48}$  \\ 
 8447-6101    &    -53.4$^{+10.1}_{-13.9}$  &     6.0$^{+ 1.7}_{- 1.3}$  & 0.67$^{+0.20}_{-0.16}$  &    -57.1$^{+10.7}_{-15.4}$  &     5.6$^{+ 1.7}_{- 1.3}$  & 0.62$^{+0.20}_{-0.15}$  \\ 
 8452-3704    &    -33.7$^{+21.8}_{-29.5}$  &     4.0$^{+ 6.3}_{- 1.9}$  & 1.12$^{+1.58}_{-0.65}$  &    -26.8$^{+12.1}_{-13.2}$  &     5.0$^{+ 4.1}_{- 1.7}$  & 1.32$^{+1.17}_{-0.60}$  \\ 
 8452-12703    &    17.9$^{+ 2.4}_{- 3.2}$  &     8.7$^{+ 2.1}_{- 1.6}$  & 1.33$^{+0.59}_{-0.34}$  &    11.6$^{+ 5.3}_{- 9.3}$  &    13.3$^{+26.1}_{- 4.5}$  & 2.44$^{+3.86}_{-1.17}$  \\ 
 8481-12701    &    -34.2$^{+ 5.1}_{-11.3}$  &     5.6$^{+ 1.5}_{- 1.3}$  & 0.99$^{+0.32}_{-0.25}$  &    -38.8$^{+13.9}_{-17.4}$  &     5.1$^{+ 2.8}_{- 1.6}$  & 0.92$^{+0.51}_{-0.32}$  \\ 
 8482-9102    &    -15.4$^{+ 9.9}_{- 5.3}$  &    13.1$^{+19.8}_{- 4.0}$  & 2.68$^{+3.62}_{-1.01}$  &    -8.4$^{+ 5.7}_{- 7.1}$  &    22.9$^{+36.0}_{-10.5}$  & 4.52$^{+6.29}_{-2.27}$  \\ 
 8482-12703    &     3.0$^{+ 5.0}_{- 7.5}$  &    22.1$^{+41.6}_{-10.3}$  & 4.43$^{+7.30}_{-2.34}$  &     7.3$^{+ 3.4}_{- 3.7}$  &    15.0$^{+15.4}_{- 4.9}$  & 2.87$^{+2.69}_{-1.17}$  \\ 
 8482-12705    &     9.9$^{+ 4.8}_{- 6.1}$  &    21.7$^{+31.0}_{- 7.6}$  & 2.61$^{+3.23}_{-1.09}$  &     4.6$^{+ 6.9}_{-13.1}$  &    27.6$^{+48.0}_{-13.0}$  & 3.30$^{+5.08}_{-1.67}$  \\ 
 8486-6101    &    18.5$^{+ 4.4}_{- 3.2}$  &     6.6$^{+ 1.7}_{- 1.4}$  & 1.44$^{+0.80}_{-0.43}$  &    14.7$^{+ 5.9}_{- 4.7}$  &     8.2$^{+ 4.3}_{- 2.5}$  & 1.87$^{+1.30}_{-0.73}$  \\ 
 8548-6102    &    14.3$^{+10.2}_{- 9.6}$  &     9.1$^{+15.2}_{- 3.8}$  & 1.48$^{+2.16}_{-0.75}$  &    14.3$^{+ 8.8}_{-12.4}$  &     9.1$^{+17.9}_{- 3.5}$  & 1.55$^{+2.49}_{-0.77}$  \\ 
 8548-6104    &    -18.9$^{+30.9}_{-20.8}$  &     6.5$^{+14.0}_{- 2.9}$  & 1.55$^{+2.90}_{-0.80}$  &    -17.1$^{+16.3}_{-10.2}$  &     9.2$^{+18.9}_{- 3.6}$  & 2.23$^{+3.89}_{-1.08}$  \\ 
 8549-12702    &    -28.2$^{+ 4.7}_{- 9.5}$  &     6.9$^{+ 1.9}_{- 1.7}$  & 1.17$^{+0.37}_{-0.31}$  &    -25.9$^{+ 4.4}_{- 6.5}$  &     7.6$^{+ 1.9}_{- 1.7}$  & 1.29$^{+0.37}_{-0.32}$  \\ 
 8588-3701    &    -32.9$^{+ 8.4}_{- 3.4}$  &     5.4$^{+ 1.8}_{- 0.9}$  & 1.17$^{+0.40}_{-0.25}$  &    -33.2$^{+ 6.2}_{- 3.4}$  &     5.3$^{+ 1.3}_{- 0.9}$  & 1.14$^{+0.30}_{-0.24}$  \\ 
 8601-12705    &     6.8$^{+ 6.1}_{- 9.8}$  &    21.4$^{+47.1}_{- 9.1}$  & 3.78$^{+7.30}_{-1.89}$  &     6.7$^{+ 4.6}_{- 4.1}$  &    24.8$^{+35.4}_{-10.3}$  & 4.17$^{+5.44}_{-1.96}$  \\ 
 8603-12701    &    -32.8$^{+28.1}_{-30.4}$  &     5.1$^{+10.5}_{- 2.4}$  & 1.20$^{+2.19}_{-0.62}$  &    -47.5$^{+27.0}_{-12.8}$  &     3.7$^{+ 4.4}_{- 0.9}$  & 0.85$^{+0.88}_{-0.27}$  \\ 
 8603-12703    &     2.1$^{+ 3.5}_{- 6.0}$  &    34.0$^{+61.8}_{-15.9}$  & 2.73$^{+4.69}_{-1.32}$  &     0.0$^{+ 4.7}_{- 9.3}$  &    29.5$^{+70.2}_{-16.7}$  & 2.44$^{+5.29}_{-1.41}$  \\ 
 8604-12703    &    14.6$^{+ 4.5}_{- 4.0}$  &    11.8$^{+ 4.9}_{- 3.0}$  & 1.21$^{+0.54}_{-0.36}$  &    14.3$^{+ 3.1}_{- 3.4}$  &    12.2$^{+ 4.2}_{- 2.6}$  & 1.25$^{+0.48}_{-0.34}$  \\ 
 8612-6104    &    -11.4$^{+ 4.9}_{- 4.4}$  &    11.1$^{+ 8.4}_{- 3.3}$  & 1.43$^{+1.09}_{-0.54}$  &    -9.9$^{+ 4.9}_{- 4.5}$  &    12.8$^{+12.4}_{- 4.2}$  & 1.69$^{+1.50}_{-0.71}$  \\ 
 8612-12702    &    18.7$^{+ 6.5}_{- 4.5}$  &     8.7$^{+ 3.1}_{- 2.3}$  & 1.58$^{+0.66}_{-0.47}$  &    19.3$^{+ 8.5}_{- 2.2}$  &     8.0$^{+ 2.0}_{- 2.2}$  & 1.42$^{+0.50}_{-0.42}$  \\   
\hline
\end{tabular}
\begin{tablenotes}
\small
\item Note: Same as Table \ref{tab: op_pap} but for results measured using the kinematic PAs.
\end{tablenotes}
\end{table*}

\appendix
\section{Tests on bar parameter determinations using simulations}
\subsection{Simulation data}
\label{ap: gtr116}
The simulated barred galaxy we used for our investigations was chosen from a series of dynamical N-body simulations, which are configured to study the effect of gas and of halo shape on the growth, evolution and properties of the bar \citep{athanassoula2013}. Compared to previous simulations, these have several advantages. The halo is live and is represented by one million particles, a number which is sufficient for an adequate description of the resonances and therefore of the angular momentum exchange. They have also used a large number of gas particles, in all standard cases with a mass of $M_{\rm gas}= 5\times 10^{4} M_{\odot}$ per particle, while the standard mass resolution for `STARS` is $M_{\rm stars}= 2.5\times 10^{4} M_{\odot}$. Contrary to most previous dynamical studies of bar formation and evolution, the gas has both cold and hot phases and is modelled including star formation, feedback and cooling. Furthermore, a high spatial resolution is used with a gravitational softening of 50 pc. See \cite{athanassoula2013} for more details.

\begin{figure*}
\includegraphics[width = 0.9\textwidth, height=0.3\textwidth]{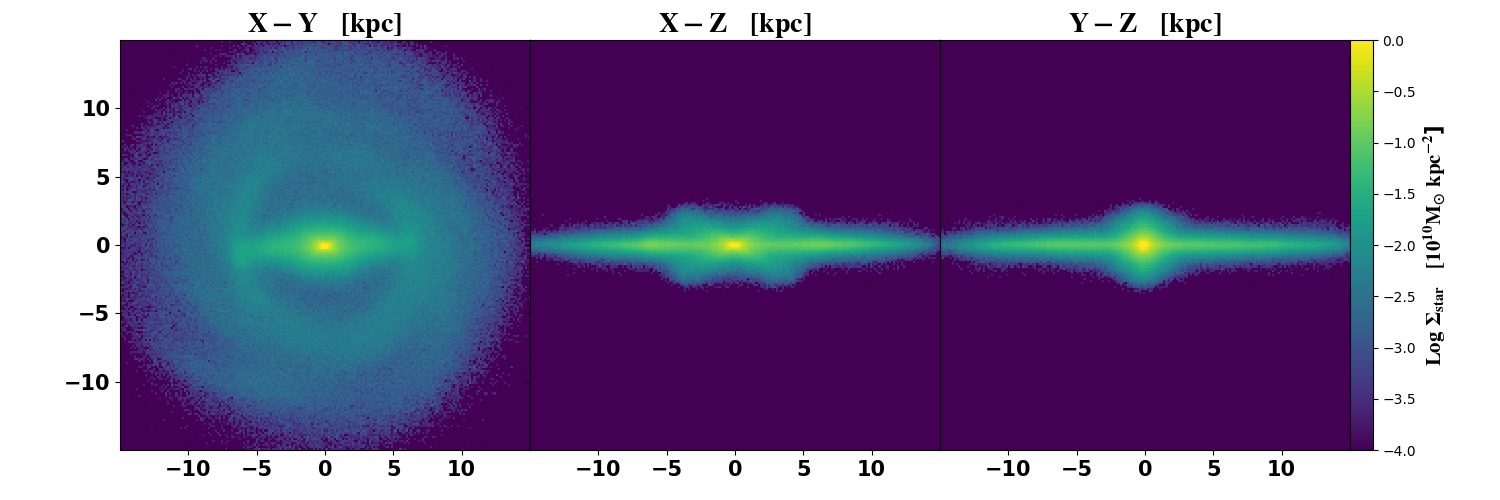}
\caption{Projection of star particles along three directions for simulation gtr116. The box size is 15 kpc. This galaxy has an evolution time of 7.99 Gyr.}
\label{sim_gal}
\end{figure*}

We chose a galaxy from Run 116 (hereafter gtr116) at a snapshot with t= 7.99 Gyr. The halo shape parameter and gas fraction can be seen in Table 1 of \cite{athanassoula2013}. This is a galaxy which has a disk about 15 kpc in radius. A visual inspection gives a bar length about 6.5 kpc. The pattern speed of this galaxy at this snapshot is 18.63 km/s/kpc, calculated from the change of the bar orientation between different snapshots. Three projections of star particles of this galaxy are shown in Fig.~\ref{sim_gal}.

\subsection{Influence of the slit length to the TW method}
\label{ap: slit}
In the weighted averaged velocity and position of Eq.~\ref{eq_tw}, the integrals of X should be over $-\infty < X < +\infty$ in theory. However the spatial coverage of the IFU is limited and the data quality is poor outside. Consequently the integrations are always limited to $-X_{0} < X,Y < +X_{0}$. But how large should $X_{0}$, i.e. the slit length, be for an adequate measurement of the pattern speed? To answer this question, we perform some tests using simulation gtr116 to check the influence of the slit length on the estimation of the bar pattern speed, with different inclinations and PA differences between the disc and the bar.

\begin{figure*}
\includegraphics[width = 1.0\textwidth, height=0.5\textwidth]{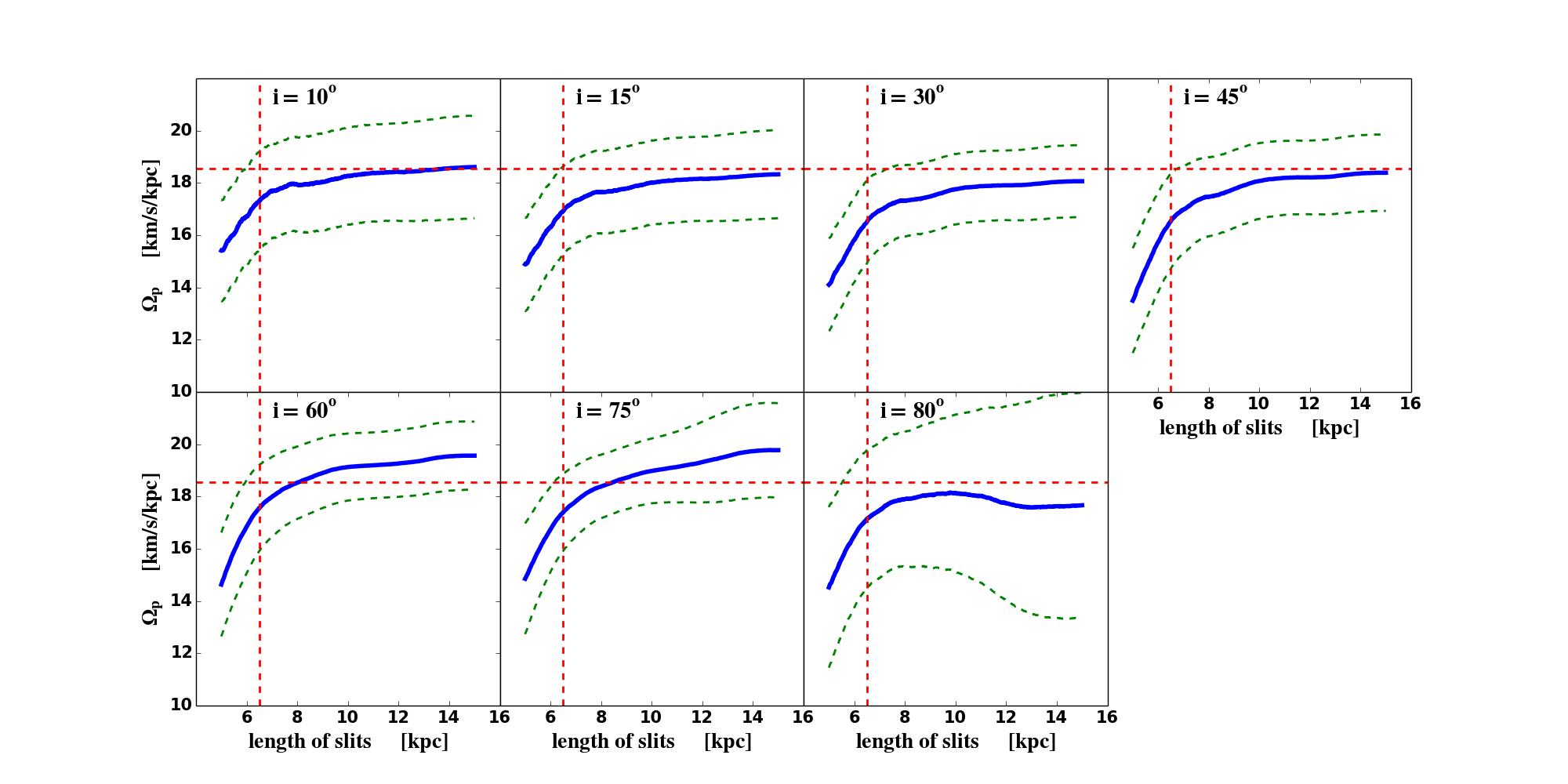}
\caption{Influence of the slit length on the bar pattern speed for different inclinations. The difference of the disc PA and the bar PA is set to 45 degrees. In each panel, the horizontal dashed line indicates the true pattern speed of the simulated galaxy, while the vertical dashed line indicates the bar length. The green dashed lines indicate the linear fitting error of $\langle V \rangle$ vs. $\langle X \rangle$.}
\label{sim_sl_i}
\end{figure*}

\begin{figure*}
\includegraphics[width = 1.0\textwidth, height=0.5\textwidth]{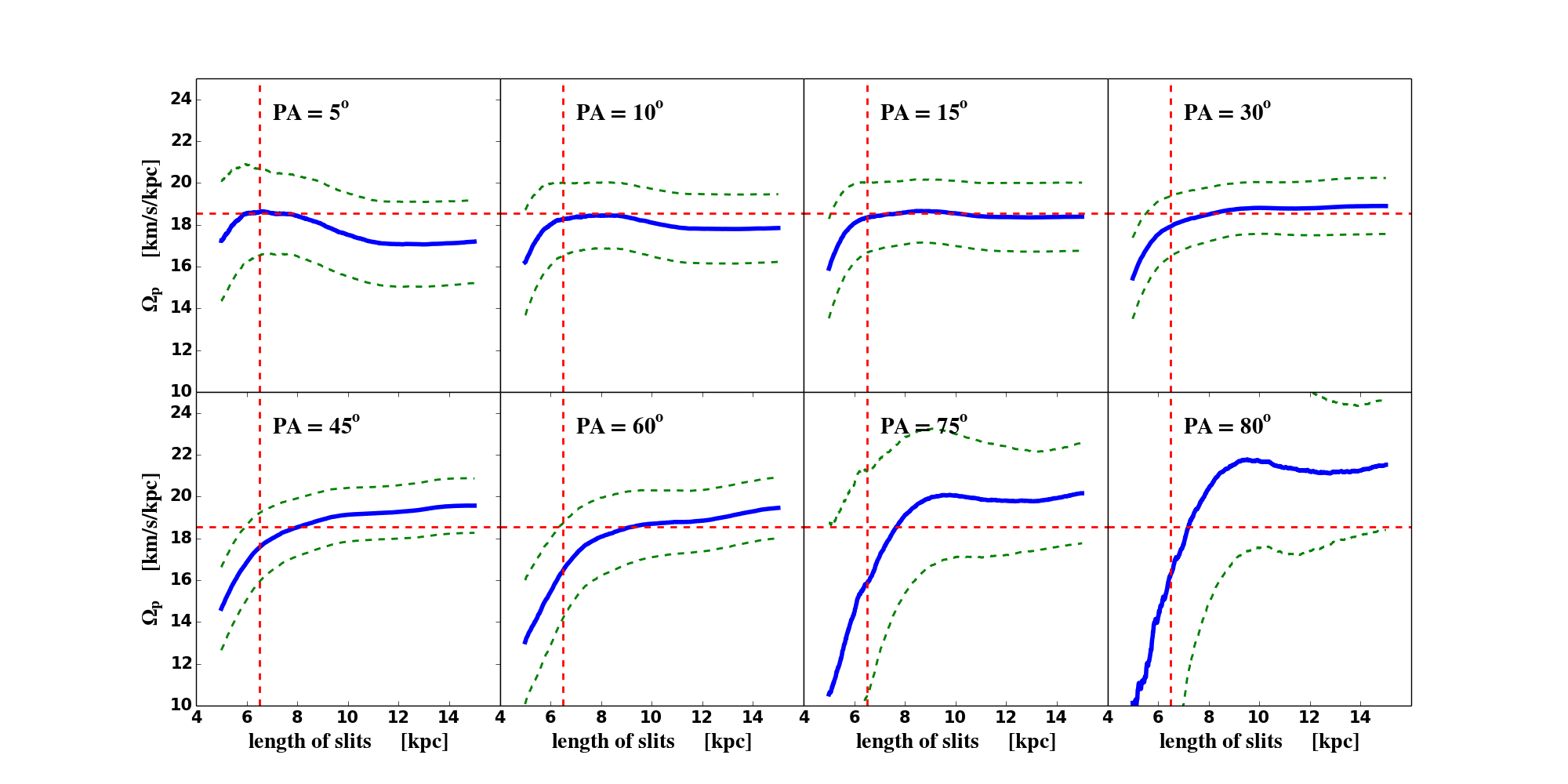}
\caption{Same as Fig. \ref{sim_sl_i} but for different PA differences between the disc and the bar. The inclination angle here is set to 60 degrees.}
\label{sim_sl_dpa}
\end{figure*}

As shown in Fig.~\ref{sim_sl_i}, the profiles of the slit length versus the pattern speed show similar patterns in that the measured pattern speed increases as the slit length increases to a length a bit larger than the bar length, after which it is nearly flat to the edge of the disc. For different inclinations, the patterns are similar. For larger inclinations, the flat pattern speeds are slightly larger than true values, and the errors are larger too. In Fig.~\ref{sim_sl_dpa}, the patterns of profiles for different PA differences are similar to those of Fig.~\ref{sim_sl_dpa}. We can see a clear trend in that as the PA difference increases from 5$^{\circ}$ to 80$^{\circ}$, i.e. from being parallel to being perpendicular to the disc major axis, the flat pattern speed increases from smaller than to larger than the true pattern speed. Though in our work we use simple cuts in inclination ($0.3< b/a <0.8$) and in PA difference ($10^{\circ} < |PA_{\rm d} - PA_{\rm b}| < 80^{\circ}$), a better sample selection criterion can be made by a two dimensional plot of the performance of the TW method with different inclinations and PA differences.

\subsection{Twisting effect of the bar to the kinematic PA measurement}
\label{ap: pak}
The kinematic PA is measured from the galaxy velocity map using a Python program {\tt fit\_kinematic\_pa.py} written by Michele Cappellari. It basically finds the best angle with the lowest difference between the observed velocity map and the symmetrized velocity map. For this method, the symmetrization of the velocity map is important, which is to some extent influenced by the bar torque. Thus we use gtr116 to test the effect of the bar on the kinematic PA measurement.

\begin{figure*}
\includegraphics[width = 0.9\textwidth, height=0.3\textwidth]{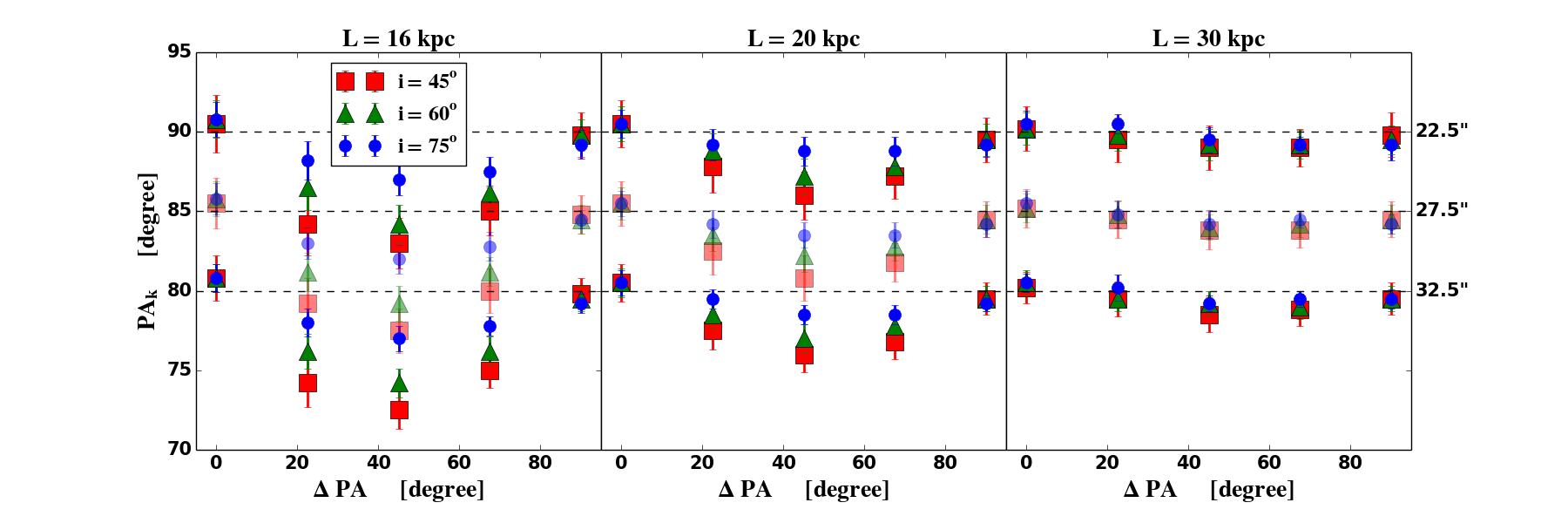}
\caption{Kinematic PAs measured with different IFU coverages, field of views, inclinations and PA differences between the disc and the bar. The left, middle and right panels are for coverages with hexagon diameters of 16 kpc, 20 kpc and 30 kpc, respectively. The red, green and blue correspond to inclinations of 45, 60, 75 degrees, respectively. The $x$ axis in each panel shows 5 PA differences between the disc and the bar, from 0 to 90$^{\circ}$ with equal spacing. The true kinematic PA is 90$^{\circ}$. The three IFU bundle diameters, i.e. field of views, are $22.5''$, $27.5''$, $32.5''$ (for example, IFU coverage of 16 kpc and a field of view of $22.5''$ mean $16\  {\rm kpc} = 22.5''$), vertically shifted in each panel by 0, $-$5$^{\circ}$ and $-$10$^{\circ}$, respectively. The $-$5$^{\circ}$ shift in each panel is fainter for distinction.}
\label{sim_pak}
\end{figure*}

We first create the mock IFU velocity maps, all with the same pixel size of $0.5''$. We also add a 2.5$''$ PSF by randomly moving the particles in the X and Y directions. The measured kinematic PAs are shown in Fig.~\ref{sim_pak}. We use three IFU coverages (i.e. physical hexagon diameters of 16 kpc, 20 kpc and 30 kpc), three IFU field of views (hexagon diameters of $22.5''$, $27.5''$ and $32.5''$), three inclinations (45, 60, 75 degrees) and five PA differences between the disc and the bar (0, 22.5, 45, 67.5, 90 degrees) to make mock IFUs. (For example, physical hexagon diameter of 16 kpc and a field of view of $22.5''$ mean $16\ {\rm kpc}= 22.5''$.) As seen from the three panels in Fig.~\ref{sim_pak}, a larger IFU coverage has smaller kinematic PA errors, because the outer part is more symmetric. For the lowest coverage, a physical hexagon side length of 8 kpc, which is slightly larger than the bar length, the error can reach to 8 degrees. The three spatial resolutions, i.e. three rows in each panel, show markedly less differences. For different inclinations, face-on cases have larger discrepancies than edge-on cases, and this phenomenon is more obvious in lower IFU physical sizes compared to the bar size. Finally, for different PA differences between the disc and the bar, a 45 degree PA difference gives the largest measurement error, which is reasonable because at this angle the bar torque twists the velocity field most seriously. However, the range of the PA difference between the disc and bar in our sample selection (see Section \ref{ssec_sample}) is 10 to 80 degrees. A PA difference of 45 degrees is better for choosing pseudo slits and applying the TW method. The influence of the bar twisting effect on our bar pattern speeds measured using kinematic PAs depends on the difference between the disc PA $PA_{\rm d,k}$ and the bar PA $PA_{\rm b}$, the galaxies' IFU coverages and inclinations in our sample. This effect may contribute most to the difference in the ${\cal R}$ ratios measured using kinematic PAs and photometric PAs. Nevertheless, it will not effect our main results, as discussed in Section \ref{sssec_operr}.

\subsection{$V_{\rm c}$ recovery of JAM using gtr116}
\label{ap: vc}
In our work, we use the circular velocity $V_{\rm c}$ from the JAM method, which has been tested using cosmologically simulated galaxies in \cite{Li2016}. They found that the total mass of a galaxy is well constrained (1$\sigma$ error $\sim$ 10-18\%). Taking the 0.1 dex M*/L into consideration, we give a 12\% systematic error for our circular velocities. In order to check the potential influence of a strong bar, we use the strongly barred galaxy gtr116 to test JAM's performance in recovering $V_{\rm c}$.

\begin{figure}
\includegraphics[width = \columnwidth]{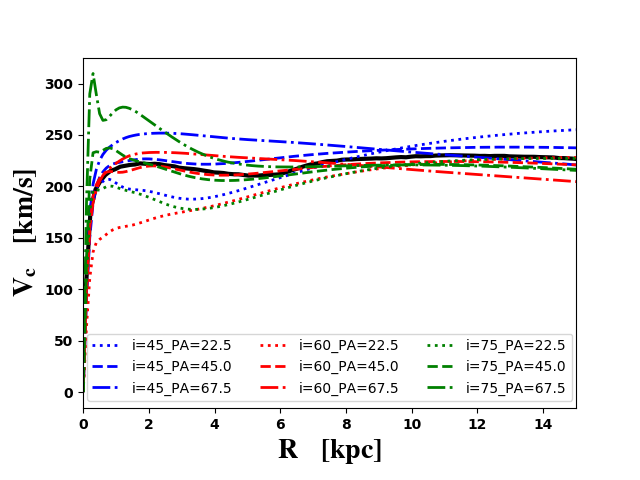}
\caption{Rotation curves recovered using the JAM method. The blue, red and green lines stand for inclinations of 45, 60 and 75 degrees, respectively. The dotted, dashed, dot-dashed lines are for 22.5, 45.0, 67.5 degrees PA differences between the disc and the bar, respectively. The black line is the true rotation curve of the simulated galaxy.}
\label{sim_vc}
\end{figure}

We put the galaxy at 150 Mpc away, and the angular IFU bundle size is 32.5$''$, i.e. the largest bundle containing 127 fibres. Thus the IFU coverage of the galaxy is a hexagon with a side length about 12 kpc. We choose three inclinations ($i=$ 45, 60, 75 degrees) and three PA differences between the bar and the disc ($|PA_{\rm d} - PA_{\rm b}|= 22.5, 45.0, 67.5$ degrees) to make mock IFU data. The IFU pixel size is $1''$, which is slightly larger than the MaNGA pixel size by taking the PSF into consideration. The mock image resolution is $0.5''$, which is higher than that of the IFU. The recovered circular velocities are shown in Fig.~\ref{sim_vc}. Though there are some discrepancies in the inner 5 kpc region, the outer flat circular velocities are recovered with about 10\% dispersion. This result reinforces our decision to use JAM circular velocities in our work.

\bsp	
\label{lastpage}
\end{document}